\documentclass[10pt,preprint]{aastex}

\usepackage{booktabs}
\usepackage{float}

\usepackage{natbib}
\usepackage[pdftex,plainpages=false,pdfpagelabels=true,breaklinks=true,pagebackref=true]{hyperref}

\hypersetup{%
 bookmarksnumbered = true,
 bookmarksopen=False,
 pdfborder=0 0 0,         
 pdffitwindow=true,      
 pdfnewwindow=true,  
 colorlinks=true,            
 linkcolor=blue,             
 citecolor=magenta,     
 filecolor=magenta,      
 urlcolor=cyan               
}

\usepackage{algorithm}
\usepackage{algorithmic}

\newcommand{\genasis}{GenASiS}

\newcommand{\f}[2]{\frac{#1}{#2}}
\newcommand{\vect}[1]{\mathbf{#1}}
\newcommand{\pderiv}[2]{\frac{\partial #1}{\partial #2}}
\newcommand{\divergence}[1]{\mathbf{\nabla} \cdot #1}
\newcommand{\gradient}[1]{\mathbf{\nabla} #1}

\newcommand\lar{\leftarrow}
\newcommand\rar{\rightarrow}
\newcommand\lrar{\leftrightarrow}

\newcommand\qlar{{\lar q}}
\newcommand\qrar{{q\rar}}
\newcommand\qminus{{-q \lrar}}
\newcommand\qplus{{\lrar q+}}
\newcommand\qlarPlus{{ \qlar +}}
\newcommand\qrarMinus{{- \qrar }}

\newcommand\yplus{{\lrar y+}}
\newcommand\yminus{{-y \lrar}}

\newcommand\leftState{{\mbox{\tiny L}}}
\newcommand\rightState{{\mbox{\tiny R}}}
\newcommand\HLL{{\mbox{\tiny HLL}}}
\newcommand\HLLC{{\mbox{\tiny HLLC}}}

\slugcomment{To be submitted to ApJS}

\shorttitle{GenASiS: Mesh and Hydrodynamics}
\shortauthors{Cardall et al.}

\begin{document}


\title{GENASIS: \emph{GEN}ERAL \emph{A}STROPHYSICAL \emph{SI}MULATION \emph{S}YSTEM. \\ I. REFINABLE MESH AND NONRELATIVISTIC HYDRODYNAMICS}

%
%
%
%
%
%

\author{Christian Y. Cardall\altaffilmark{1,2}, Reuben D. Budiardja\altaffilmark{1,2,3,4}, Eirik Endeve\altaffilmark{5}, and Anthony Mezzacappa\altaffilmark{1,2,5}}
\email{cardallcy@ornl.gov}

\altaffiltext{1}{Physics Division, Oak Ridge National Laboratory, Oak Ridge, TN 37831-6354, USA}
\altaffiltext{2}{Department of Physics and Astronomy, University of Tennessee, Knoxville, TN 37996-1200, USA}
\altaffiltext{3}{Joint Institute for Heavy Ion Research, Oak Ridge National Laboratory, Oak Ridge, TN 37831-6374, USA}
\altaffiltext{4}{National Institute for Computational Sciences, University of Tennessee, Knoxville, TN 37996, USA}
\altaffiltext{5}{Computer Science and Mathematics Division, Oak Ridge National Laboratory, Oak Ridge, TN 37831-6354, USA}


%
%

\begin{abstract}
\genasis\ ({\em Gen}eral {\em A}strophysical {\em Si}mulation {\em S}ystem) is a new code being developed initially and primarily, though by no means exclusively, for the simulation of core-collapse supernovae on the world's leading capability supercomputers. 
This paper---the first in a series---demonstrates a centrally refined coordinate patch suitable for gravitational collapse and documents methods for compressible nonrelativistic hydrodynamics.
We benchmark the hydrodynamics capabilities of GenASiS against many standard test problems;  
the results illustrate the basic competence of our implementation, demonstrate the strengths and limitations of the HLLC relative to the HLL Riemann solver in a number of interesting cases, and provide preliminary indications of the code's ability to scale and to function with cell-by-cell fixed-mesh refinement.  
\end{abstract}


\sloppy



\section{INTRODUCTION}
\label{sec:Introduction}

Many problems in astrophysics and cosmology will press against the boundaries of supercomputer software and hardware development for some time to come. 
Among the most challenging are time-dependent systems that should be treated in three position space dimensions, plus up to three momentum space dimensions for those that involve radiation transport. 
Several types of physics---and their couplings---must be addressed. 
Taken together, physics such as self-gravity, turbulent cascades, reactions that may or may not be in equilibrium, and the operation of both microscopic and macroscopic processes often imply the simultaneous relevance of multiple scales in space and time, typically requiring some kind of spatial adaptivity and multiple solvers (at least some of which may be time-implicit). 
Solvers deployed with good resolution in as much of phase space as possible fill the memory and churn through the cycles of the largest supercomputers available. 
That the relevant theory is described by time-dependent partial (and sometimes integro-)differential equations implies the need for synchronous evolution and communication between different regions of position space (and sometimes momentum space). 
These latter aspects seem ever more difficult to address due to the fact that increases in computing capability seem only to come with such additional burdens as distributed memory and distributed (and, more recently, heterogeneous) processing capacity. 
From the perspective of working astrophysicists and cosmologists, it can seem that the physics itself recedes ever further away as their efforts are channeled towards developing and tailoring codes to the specific features of these increasingly complex high-performance machines. 
In this environment, the availability of well-designed codes with broadly applicable physics capabilities is increasingly valuable to researchers.

One such computationally demanding problem is the elucidation of the explosion mechanism of core-collapse supernovae \citep{Mezzacappa2005,Woosley2005,Kotake2006,JANKA2007,Kotake2012,Kotake2012b,Janka2012,Burrows2013,Janka2012b}. 
A star with mass greater than $\sim 8\, M_\odot$ develops a degenerate core during its final burning stage. 
Once the core becomes sufficiently large---roughly the Chandrasekhar mass---it becomes unstable and undergoes catastrophic collapse on a nearly free-fall time scale. 
Collapse of an inner subsonic portion of the core halts around nuclear density (where nucleons begin to overlap), leading to the development of a shock wave at the interface between this inner core and the supersonically infalling outer portion of the core. 
The shock eventually will disrupt the entire star and give rise to the luminous supernova, but it stagnates shortly after its formation due to the endothermic reduction of heavy nuclei into their constituent nucleons and enervating neutrino losses. 
The mechanism of shock revival---that is, the explosion mechanism---remains to be fully elucidated by numerical simulations, but is expected to involve some combination of heating by intense neutrino fluxes streaming from the nascent neutron star, fluid instabilities, rotation, and magnetic fields. 
In fact, the relative contributions of these phenomena may vary from event to event: pre-supernova stars differ in properties such as mass and rotation, leading to explosions that may or may not be jet-like with an associated gamma ray burst, and either a black hole or a neutron star as a compact remnant.

Even this brief description conveys an initial sense of the multiphysics and multiscale nature of core-collapse supernova explosion mechanism simulations. 
The problem manifestly requires self-gravity, which must be general relativistic to treat the more rare and extreme case of black hole formation, and ideally would be at least approximately relativistic even when the compact remnant is a neutron star. 
The treatment of hydrodynamics---and ideally magnetohydrodynamics, especially in connection with black hole formation---must be able to handle shocks. At high density the equation of state must describe neutron-rich nuclear matter at finite temperature, and at low density it is desirable to track nuclear composition with a reaction network spanning a wide range of nuclear species. 
Neutrino transport must span diffusive, decoupling, and free-streaming regimes, and include several species and their interactions with each other and with various fluid constituents. 
Gravitational collapse, a steepening density cliff at the surface of the nascent neutron star, and regions of turbulence strongly recommend some form of spatial adaptivity.
The stiff equations of neutrino transport and a nuclear network normally require time-implicit evolution. 
At the present time computational limitations still require dimensional reduction of phase space; simple estimates suggest that at least exascale resources will be required for the full transport problem. 
There is fairly wide agreement that retention of at least energy dependence in full neutrino transport is important, and that all three position space dimensions should be included, 
but only a couple of simulations of this type have been reported \citep{Takiwaki2012,Hanke2013}, with others in progress \citep{Bruenn2009}.

\genasis\ ({\em Gen}eral {\em A}strophysical {\em Si}mulation {\em S}ystem) is a new code being developed, at least initially and primarily, for the simulation of core-collapse supernovae on the world's leading capability supercomputers. 
`General' denotes the capacity of the code to include and refer to multiple algorithms, solvers, and physics and numerics choices with the same abstracted names and/or interfaces. 
In \genasis\ this is accomplished with features of Fortran 2003 that support the object-oriented programming paradigm \citep[e.g.][]{Reid2007,Adams2008}.
`Astrophysical' roughly suggests---over-broadly, at least initially---the types of systems at which the code is aimed, and the kinds of physics and solvers it makes available. 
`Simulation System' indicates that the code is not a single program, but a collection of modules, structured as classes, that can be invoked by a suitable driver program set up to characterize and initialize a particular problem.

One fundamental characteristic of a simulation code that underpins almost everything else is the nature---or even existence, when one considers particle methods---of its meshing, as this is the stage upon which the physics plays out. 
Smoothed-particle hydrodynamics has been widely used in the broader astrophysics community as an efficient way to get to three dimensions, but its accuracy has been controversial \citep[e.g.][]{Agertz2006,Price2008,Springel2010}.  
Unstructured meshes are useful for the complex geometries of engineering contexts, but their high overhead is probably not justified for the more simple geometries of most astrophysics problems.
(On the other hand, the relatively new use of adaptive Voronoi tesselations \citep{Springel2010a} is an interesting new approach that merits further consideration.)   
Moving patches may be useful for following compact objects in orbit \citep[e.g.][]{Scheel2006}, but introduce additional complicated source terms associated with non-inertial reference frames, and are not an obvious fit for more centrally-condensed problems like core-collapse supernovae.
When more structured grid-based approaches are considered, one major choice is whether to use adaptive mesh refinement (AMR) in an effort to deploy computational resources only where needed, and if so, what type of AMR should be used. 

Most existing core-collapse supernova codes use strategies other than AMR to handle gravitational collapse. 
One code uses smoothed-particle hydrodynamics \citep{Fryer2006}. Others use an at least relatively high resolution radial mesh, usually in only one \citep{Rampp2002,Thompson2003,Liebendorfer2004,Sumiyoshi2005} or two \citep{Buras2006a,Livne2007,Bruenn2009} position space dimensions, often with Lagrangian coordinates (in spherical symmetry) or a moving radial mesh to follow the infall. 
Use of a radial mesh and associated spherical coordinates imposes severe time step limitations at coordinate singularities unless special measures are taken, such as suppression of lateral fluid motion in the few cells nearest the origin \citep[e.g.][]{swesty2009}; use of an unstructured mesh to morph to a different coordinate system at low radius \citep[e.g.][]{Livne2007}; an overlap of a radial mesh with a Cartesian mesh at low radius \citep[e.g.][]{Scheidegger2008}; or an overlap of two separate radial meshes in a so-called `yin-yang' configuration \citep[e.g.][]{Wongwathanarat2010b}. 
While these approaches may give satisfactory gravitational collapse, with a moving radial mesh also reasonably handling the steepening and moving density cliff at the neutron star surface, they cannot address regions of turbulence as well as AMR can. 
Those codes that do use (or intend to use) AMR use the block-structured variety \citep{Almgren2010,Couch2012}.  

The computational demands of the multiphysics nature of core-collapse supernovae motivate us to explore cell-by-cell AMR \citep[e.g.][]{Khokhlov1998} in developing \genasis.
Block-structured approaches \citep{Berger1984,Berger1989,MacNeice2000} are more common, and certainly there are efficiencies associated with the use of predictable basic building blocks. 
Nevertheless, it remains possible that block-structured AMR might not prove optimal in a multiphysics context, for at least two reasons.

One basic consideration with the potential to favor cell-by-cell AMR is that the computational cost per spatial cell becomes very high if one aims (at least eventually, if not initially) towards the high dimensionality of the full neutrino phase space (position space plus momentum space)---and, for that matter, towards large nuclear reaction networks. 
In this case cell-by-cell AMR could be advantageous if it requires a smaller number of total cells for a given accuracy than block-structured AMR.
This may prove true in part because it allows more fine-grained control over cell division and placement, and also because the use of many blocks at a given level of refinement in block-structured AMR can lead to a larger ratio of ghost to computational cells.
  
Moreover, cell-by-cell AMR also may turn out to have some advantages with respect to the implementation of the types of solvers needed in a multiphysics context.  
One initial motivation for block-structured AMR is the ability to deploy existing `unigrid' time-explicit hydrodynamics solvers on individual regular cell blocks. 
While this is convenient when hydrodynamics is the only physics involved, the fact that cell-by-cell AMR is not conceptualized around local time-explicit solvers raises the possibility that it might be more amenable to elliptic and other global solvers, including those that are time-implicit.
It is of course possible to develop global solvers in block-structured AMR, for instance for gravity \citep[e.g.][]{OShea2005,Ricker2008,Almgren2010} and radiation \citep[e.g.][]{Rijkhorst2006,Wise2011,Zhang2011}; but until alternatives are more fully explored, it is not obvious that this is the most natural environment imaginable for them.

While a handful of other codes used in astrophysics or cosmology do use cell-by-cell AMR \citep[e.g.][]{Khokhlov1998,Teyssier2002,Gittings2008}, our approach in \genasis\ has at least one notable difference. 
In arranging storage and writing solvers, rather than addressing the oct-tree as a single mesh consisting of the union of leaf cells at all levels of refinement, \genasis\ has a more explicit level-by-level orientation. 
Relative to the union-of-leaf-cells cell-by-cell perspective, we expect the level-by-level perspective to facilitate multigrid approaches to elliptic solves, and possibly time-implicit global solves. 
Writing solvers for single levels, with interactions between levels handled separately, restores some of the flavor of simplicity of the independent solves on individual blocks featured in block-structured AMR; at the same time, treating entire levels at once eliminates the drawbacks of having to stitch together results from multiple blocks at the same level (via additional V-cycles, corner iteration, etc.).  
The potential benefits for multiphysics solvers of our level-by-level approach to cell-by-cell AMR remain to be tested and reported in future papers in this series that address gravity and neutrino transport.  

Besides a mesh capable of handling gravitational collapse, another basic requirement for the simulation of core-collapse supernovae is a hydrodynamics solver.
The equations of hydrodynamics can be categorized as a system of hyperbolic balance equations, for which a vast mathematical literature on different solvers exists \citep[e.g.][and references therein]{Shu1998,LeVeque2002,Toro2009}.  
The multiphysics nature of core-collapse supernova explosion mechanism simulations requires that the hydrodynamics solver accommodate a non-ideal equation of state and other physical ingredients (e.g. magnetic fields, gravity, neutrino-matter interactions, etc.) in a robust manner.  
Core-collapse supernovae are ultimately general relativistic systems, and the hydrodynamics solvers should be generalizable to a relativistic description.  
Moreover, the solver must be able to accurately describe the formation and evolution of shocks and other discontinuities.  
Our choice to use AMR to at least partially deal with the multiscale nature of core-collapse supernovae also puts constraints on the choice of hydrodynamics solver.  

Finite volume methods based on approximate Riemann solvers are good candidates under these various considerations.  
The hydrodynamics solvers implemented so far in \genasis\ are second-order accurate, are based on the integral formulation of the underlying hyperbolic system, and use the method of lines approach to the solution of partial differential equations \citep[e.g.][]{Shu1998,Kurganov2001}.  
Second-order spatial accuracy is achieved with monotonic linear spatial interpolation \citep[e.g.][]{Kurganov2000}. We employ the so-called HLL family of Riemann solvers \citep[][]{Harten1983,Einfeldt1988,Toro1994,Kurganov2001} to compute intercell fluxes.  
Second-order temporal accuracy is achieved with a Total Variation Diminishing (TVD) Runge-Kutta method \citep[e.g.][]{Shu1998}.  
In particular, schemes based on HLL-type Riemann solvers rely on minimal information about the eigenstructure of the underlying hyperbolic system, and have been promoted as general black-box solvers for conservation laws and related equations \citep{Kurganov2000}.  
Indeed, HLL-type Riemann solvers have been designed for a range of hyperbolic equations, including nonrelativistic hydrodynamics and MHD \citep[e.g.][]{Toro1994,Batten1997,Linde2002,Londrillo2004,Miyoshi2005}, and special \emph{and} general relativistic hydrodynamics and MHD \citep[e.g.][]{DelZanna2002,DelZanna2003,Gammie2003,Duez2005,Mignone2005,DelZanna2007,Mignone2009}.  
HLL-type Riemann solvers have also been used for general relativistic neutrino radiation-hydrodynamics simulations of core-collapse supernovae \citep{Muller2010,Muller2012}.  

This paper---the first in a series---describes some baseline capabilities of GenASiS that will be needed in core-collapse supernova simulations.
In particular, we explain some concepts underlying the refinable discretized spaces on which calculations are to be performed (Section~\ref{sec:RefinableMesh});
document methods for compressible nonrelativistic hydrodynamics (Section~\ref{sec:Solvers});
and benchmark the hydrodynamics capabilities of GenASiS against many standard test problems (Section~\ref{sec:Tests}), including an example with a fixed centrally refined coordinate patch of a type suitable for gravitational collapse.

\section{REFINABLE MESH}
\label{sec:RefinableMesh}

In this section we briefly explain some concepts underlying the refinable discretized spaces on which calculations are to be performed with \genasis, and exhibit a centrally refined coordinate patch like that we intend to use for gravitational collapse.

We regard a physical space as a manifold covered by one or more coordinate patches.
A sufficiently simple manifold can be described with a single coordinate patch; this is the case with the example in this section, and in all the test problems in Section~\ref{sec:Tests}. 
But one can imagine many reasons to use multiple coordinate patches. 
For example, a `yin-yang' manifold includes two overlapping coordinate patches that---like the two pieces of leather that form the surface of a baseball---cover a three-dimensional space with separate spherical coordinates in such a way as to avoid a coordinate singularity on the polar axis \citep[e.g.][]{Wongwathanarat2010b}. 
One could efficiently handle both spherically symmetric radial gravitational collapse and multidimensional phenomena at smaller radius by marrying a central three-dimensional Cartesian coordinate patch to a one-dimensional radial coordinate patch that begins near the surface of the Cartesian box and extends to much larger radius \citep{Scheidegger2008}. 
Choices of algorithm and approximation might suggest that different pieces of physics be treated on separate coordinate patches, with some facility for interpolation between them \citep{Rampp2002,Buras2006a}. 
A binary stellar system could be handled by covering the two bodies with separate coordinate patches that move within a larger coordinate patch to follow the orbital motion \citep{Scheel2006}. 
Some manifolds, such as some spaces described by general relativity, may be sufficiently complicated as to require multiple coordinate patches for a reasonable description. 
Another example is the phase space---position space plus momentum space---needed for relativistic kinetic theory, which in mathematical idealization contains an infinite number of coordinate patches: this is a `tangent bundle,' consisting of a (curved, in the general case) base space, i.e position space, together with a flat tangent space, i.e. a momentum space, at every point of the base space. 

In representing a single coordinate patch in \genasis, we approximate the mathematical ideal of continuity with a finite sequence of meshes which provide, as necessary, increasing refinements of the coarsest (top-level) mesh. 
Our coordinate patches can be  one-, two-, or three-dimensional. 
The refinable structure that underlies our approximate representation of a three-dimensional continuous coordinate patch is an oct-tree (or, in restricted use in two dimensions or even one dimension, a quad- or binary tree respectively) that enables cell-by-cell refinement. 
The fundamental unit of this structure is a class \citep[in the object-oriented sense; in Fortran 2003, see e.g.][]{Reid2007,Adams2008} representing a single finite `cell,' which is a segment, quadrilateral, or cuboid that can be split into two, four, or eight cells in one, two, or three dimensions respectively.
Several constructs provide increasingly comprehensive interfaces to the oct-tree structure underlying a single refinable coordinate patch.

The first type of construct providing an interface to a portion of the oct-tree is a `cell list,' or linked list of cells. 
This can be used to, for example, create lists of selected parent cells in order to facilitate loops over frequently addressed subsets of cells on a particular level of the tree.

\begin{figure}
  \epsscale{1.0}
  \plottwo{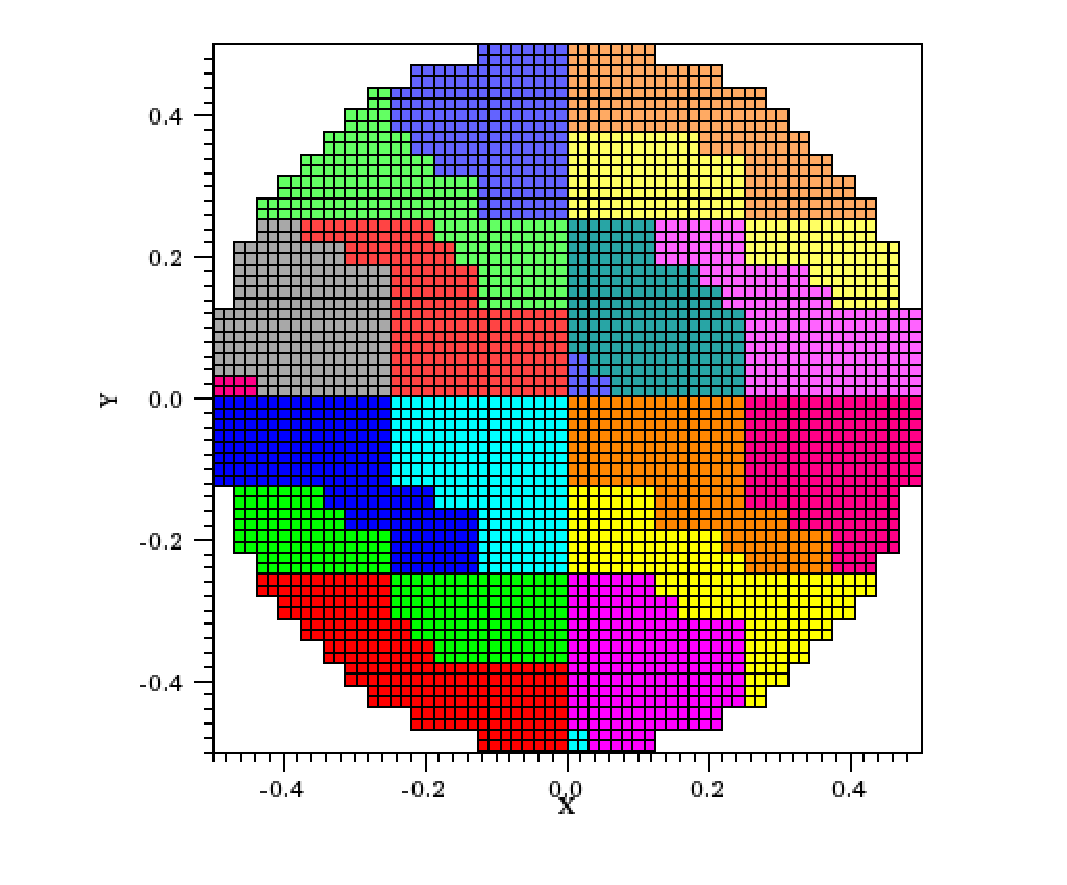}{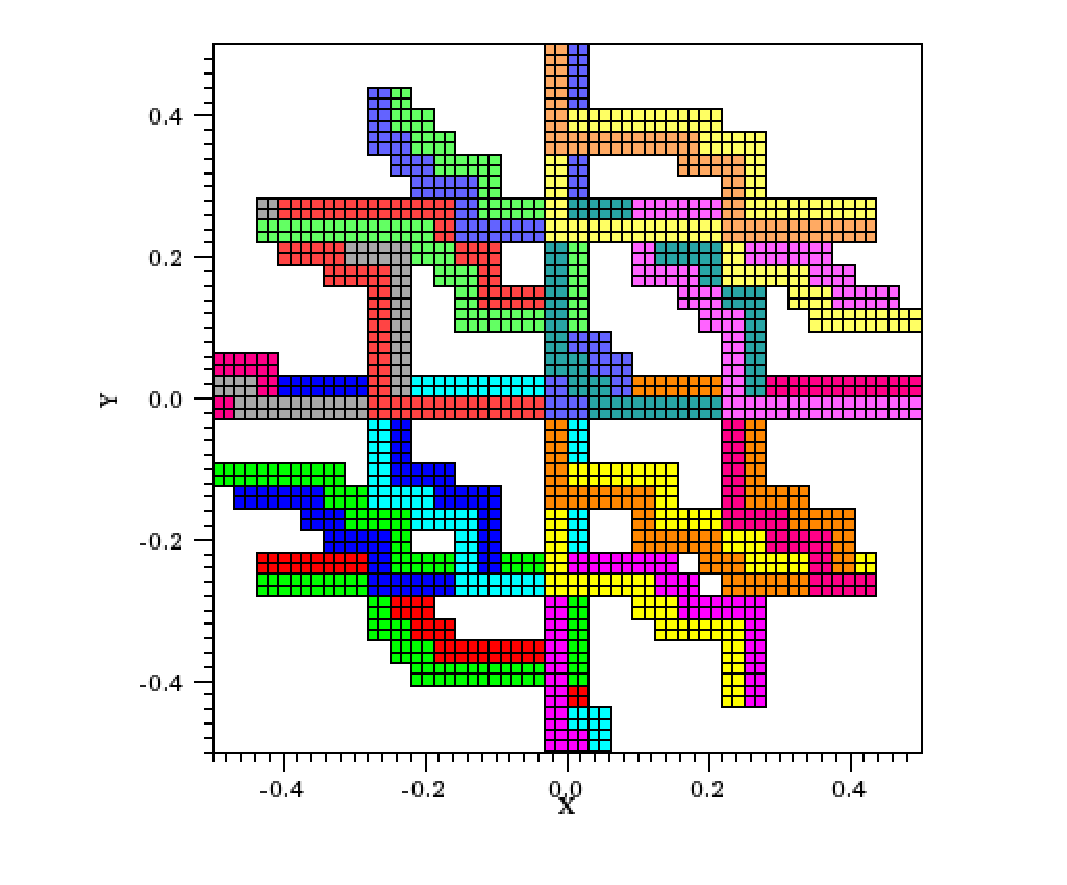}
  \caption{Domain decomposition of a spherical submesh, illustrating proper cells (left) and the ghost cells partially bounding them (right).
The ghost cells of the different processes partially overlap.   
Each individual cell is pictured; two dimensions, low resolution, and a small number of processes (16) are used for clarity.}
\label{fig:Decomposition}
\end{figure}

The next layer of interface to the oct-tree---a `submesh'---is a grid in its own right. 
It consists of a subset of cells at a single level of the oct-tree, whose combined arrangement may be irregular in shape and even consist of multiple disconnected pieces.   
Among the members of a submesh are two cell lists, for `proper' and `ghost' cells, which allow the submesh to be domain-decomposed for parallel processing in a distributed-memory environment.
(Proper cells are typically the normal working computational cells assigned to a particular process. 
Ghost cells typically compose a partial boundary layer around the proper cells; 
their data is obtained through message passing with the neighboring processes that own those cells. 
See for instance Figure~\ref{fig:Decomposition}.)

\begin{figure}
  \epsscale{1.0}
  \plottwo{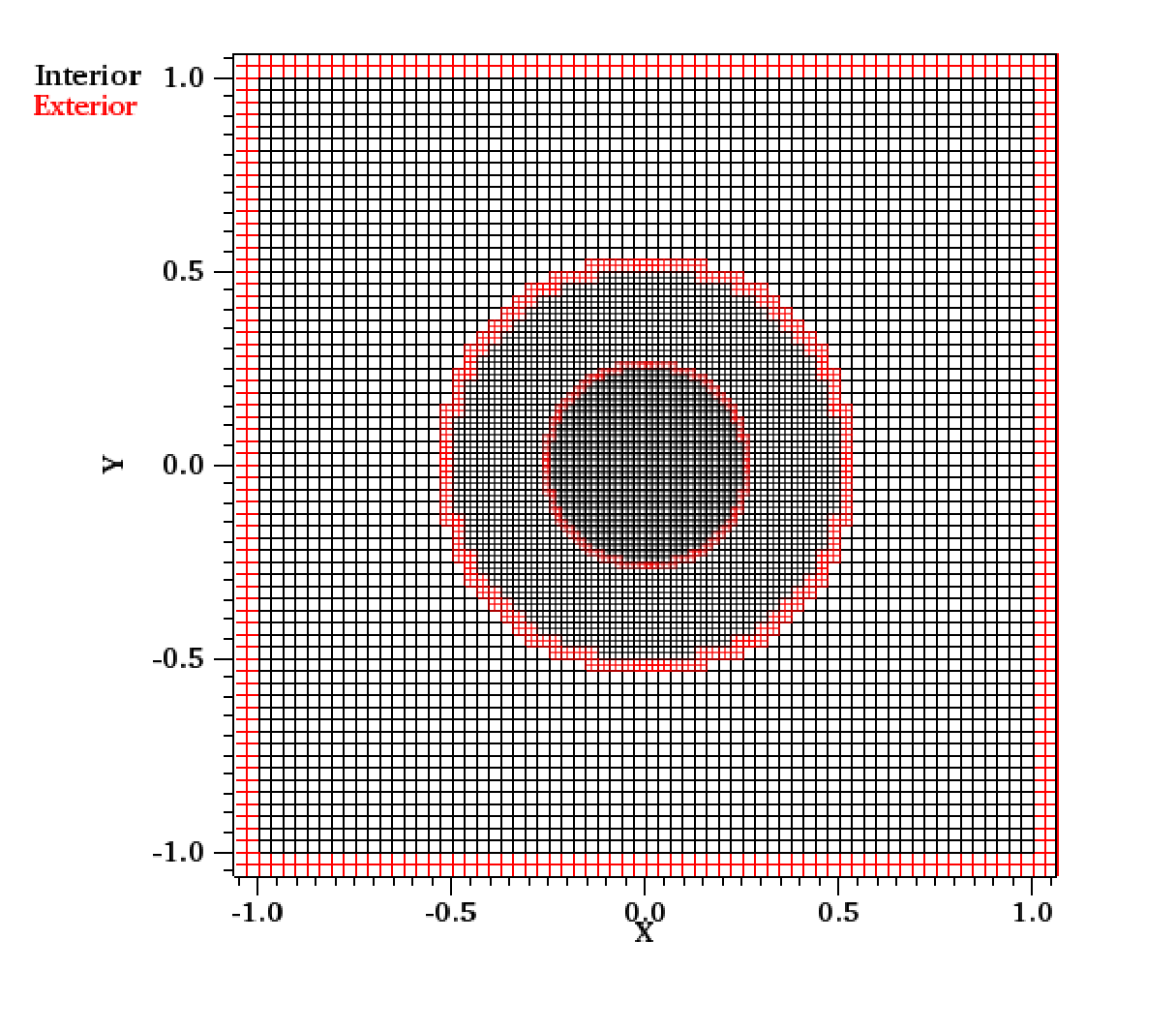}{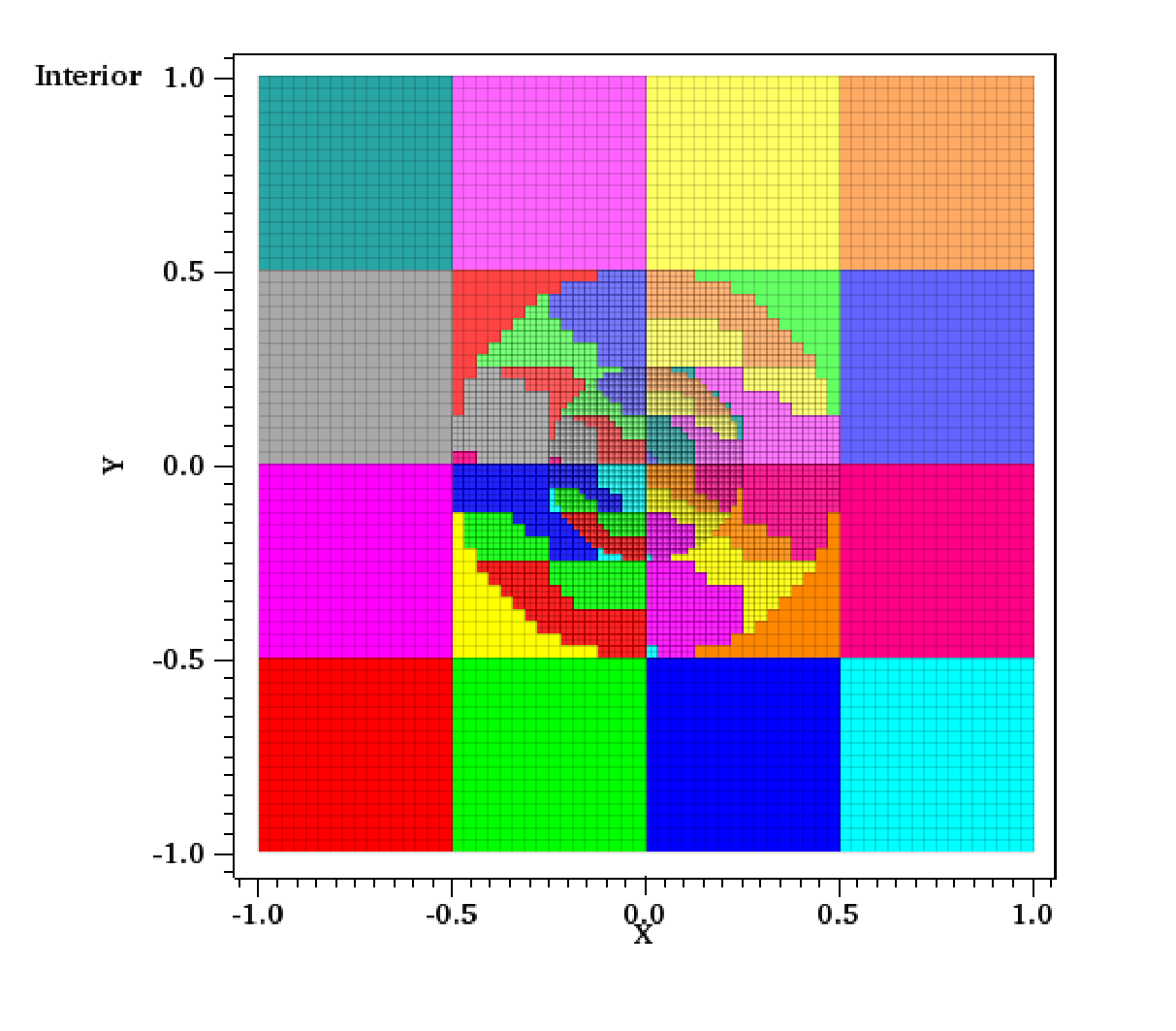}
\caption{A centrally refined coordinate patch with three `meshes,' i.e. three levels. 
Shown for each mesh are its Interior and Exterior `submeshes' (left). 
Together these comprise all the cells at a particular level of the oct-tree, with the Interior (black) consisting of the normal computational cells and the Exterior (red) constituting a boundary layer.
The Interior of each mesh or level is independently domain-decomposed (right).
Children and Parents submeshes are not shown, as there is no permanent physical field data storage associated with them (and therefore no geometry to display); 
but their proper and ghost cell lists contain domain decomposition information needed for the communications required by prolongation and restriction.
Each individual cell is pictured; two dimensions, low resolution, and a small number of processes (16) are used for clarity.
The domain decompositions of Levels 1 (coarsest) and 2 (middle) are partially hidden under the display of Levels 2 and 3 respectively.}
  \label{fig:Submeshes}
\end{figure}

A `mesh' includes four submeshes comprising all the cells at a given level of the oct-tree, and also provides links to adjacent refinement levels.
As illustrated in the left panel of Figure~\ref{fig:Submeshes},
two of these submeshes together comprise all the cells at a particular level of the oct-tree: the `Interior' submesh (or simply `Interior') contains all the normal computational cells, and the `Exterior' submesh includes all the cells that form a boundary layer---either the  boundary of the coordinate patch as a whole, or a coarse/fine boundary at the edge of a particular level---surrounding the Interior. 
As shown in the right panel of Figure~\ref{fig:Submeshes}, the Interior submesh on each level is independently domain-decomposed. 
As further discussed below, the need to exchange data between adjacent levels with independent domain decompositions gives rise to two additional submeshes at each level: 
the Level $i$ `Children' submesh provides a link to the Level $i+1$ Interior, while the Level $i$ `Parents' submesh connects with the Level $i-1$ Interior.

Major data storage for physical fields is organized around the `mesh' concept.
Groups of related field variables are stored in rank-two arrays.
Each cell at a given level of the oct-tree is assigned a number, which corresponds to the row number (first index, in Fortran) of this array;
the second dimension of the array indexes different physical fields.
The first dimension of the array is subdivided into sections reserved for data associated with the cells of the Interior and Exterior submeshes described in the previous paragraph.
To avoid unnecessary (and inconveniently distributed) memory usage, there is no permanent physical field data storage associated with the Children and Parents submeshes, which would be redundant with the data associated with the Interior and Exterior submeshes on adjacent levels. 
As discussed further below, rather than being associated with permanent field data storage, the purpose of the proper and ghost cell lists of the Children and Parents submeshes is to facilitate the communications needed to exchange data between adjacent levels with independent domain decompositions.

The class implementing the `mesh' concept also has members with connectivity information.
This involves lists of cell numbers and corresponding sibling cell numbers, in order to facilitate certain operations on particular sets of selected cells without having to walk through the oct-tree, access sibling pointers, etc. 
The application of boundary conditions is one example of an operation that uses this sort of connectivity information. 
Another is the overlapping of work and communication in time-explicit solves requiring only near-neighbor information (e.g. in hydrodynamics): `exchange' cells (those whose data must be communicated to the ghost cells of other message-passing processes) can be updated, and non-blocking communication initiated, before `non-exchange' cells are updated. 
Connectivity information for these two categories of cells allows data to be loaded from their non-contiguous loci in mesh-oriented storage (described in the previous paragraph) to contiguous storage in a `packed' rank-two array.
This is useful for efficient execution of intra-cell operations, that is, those that do not relate values in different cells.
Some examples of the use of `unpacked,' mesh-oriented, connectivity-informed  storage on the one hand; and `packed,' mesh-agnostic storage on the other, will be seen in the hydrodynamics algorithms described in Section~\ref{subsec:Updates}.

\begin{figure}
  \epsscale{1.0}
  \plotone{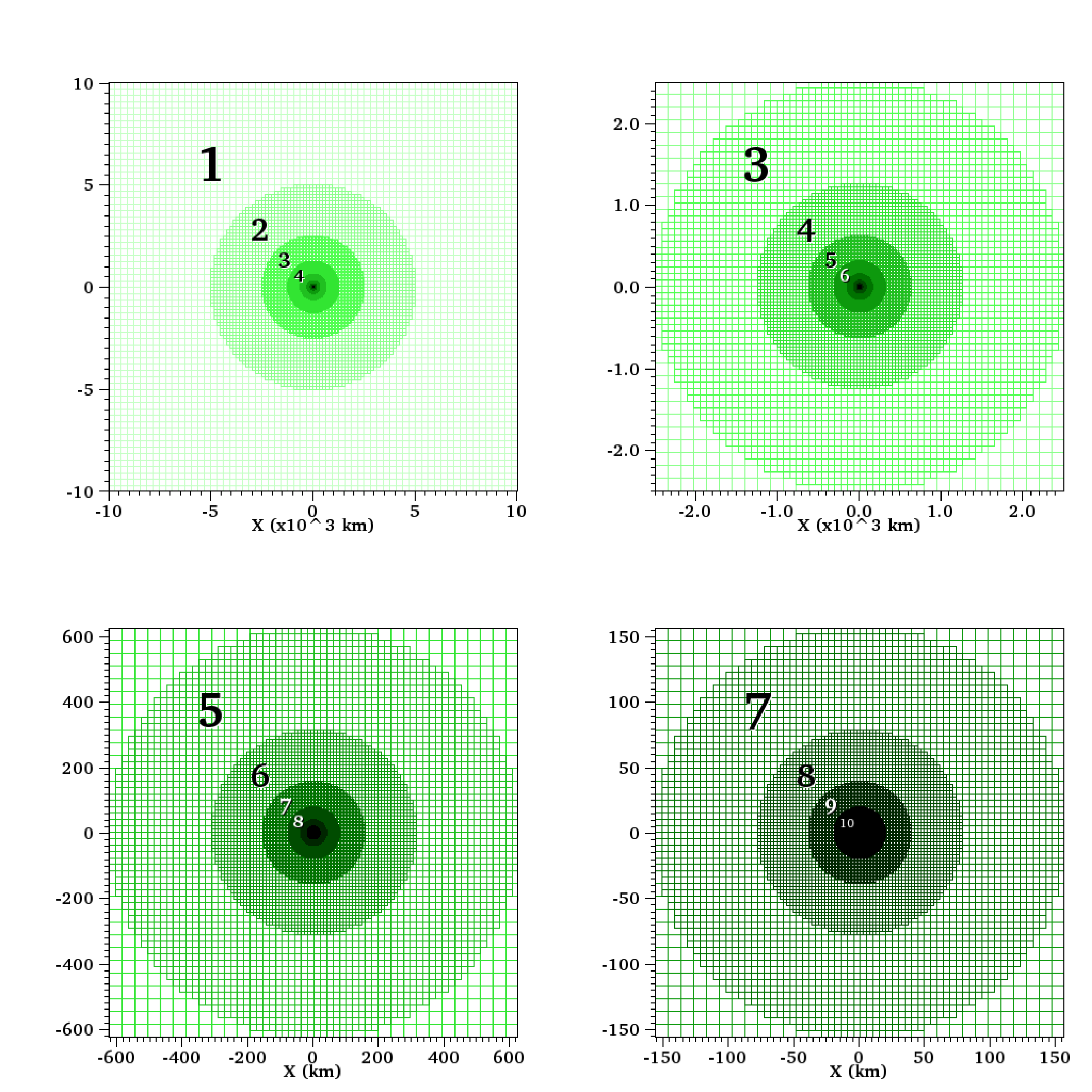}
  \caption{Levels of a two-dimensional coordinate patch with 10 levels, illustrating the dynamic range in length scales accessed during the gravitational collapse of the core of a massive star.
Outlines of $2\times 2$ cell arrays, rather than individual cells, are displayed.  
Level~1---the coarsest mesh---encompasses a square domain of width 20,000~km
with $128\times 128$ cells of width $1.56\times 10^2$ km.
Successive levels are refined by a factor of two in each dimension, labeled by their level number, and outlined in increasingly dark shades of green.
The upper left, upper right, lower left, and lower right panels zoom in to smaller and smaller regions surrounding the origin. 
Level~10---the finest mesh---has cells of width 0.305 km.}
  \label{fig:Chart2D}
\end{figure}

Finally, a class representing a coordinate patch has among its members an array of `meshes,' each element of which corresponds to a (potential) level of the refinable oct-tree.  
As mentioned above and illustrated already in Figure~\ref{fig:Submeshes}, each successive level is a refinement of the previous level, allowing the coordinate patch to approximate the ideal of continuity as needed.
Another example is shown in Figure~\ref{fig:Chart2D}.
It consists of 10 levels and illustrates the dynamic range in length scales accessed during the gravitational collapse of the core of a massive star.

\begin{figure}
  \epsscale{1.0}
  \plottwo{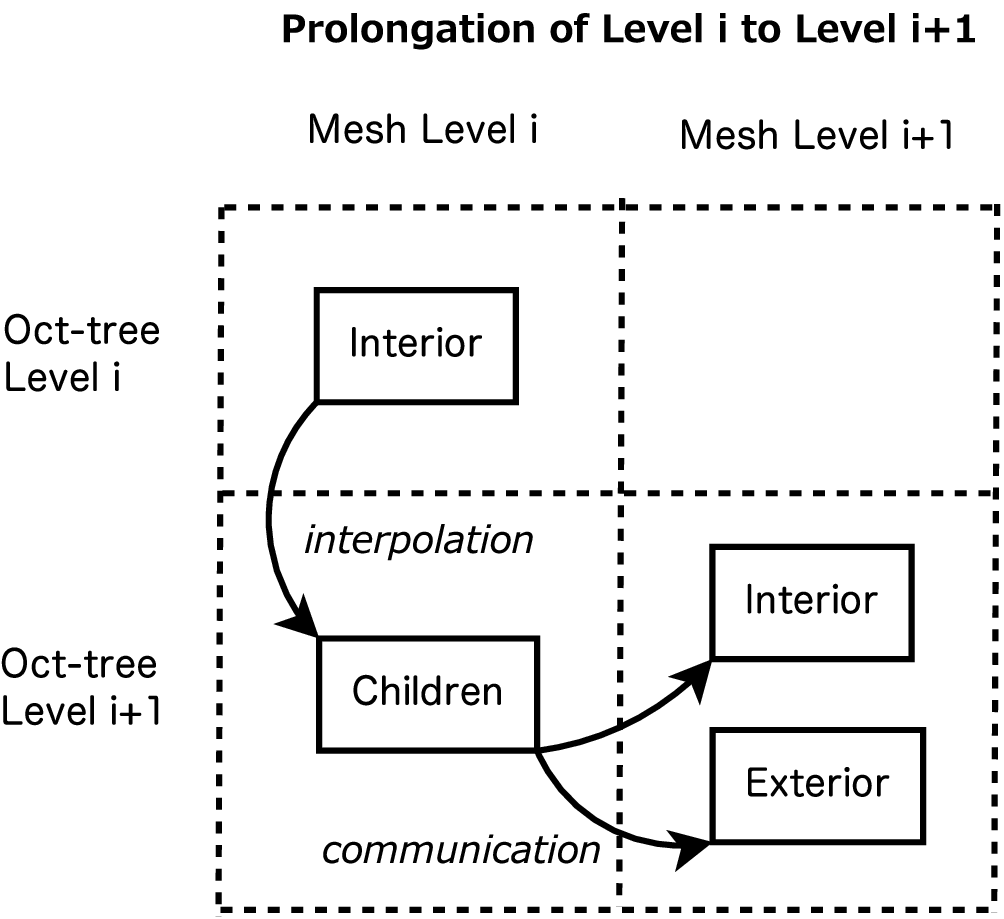}{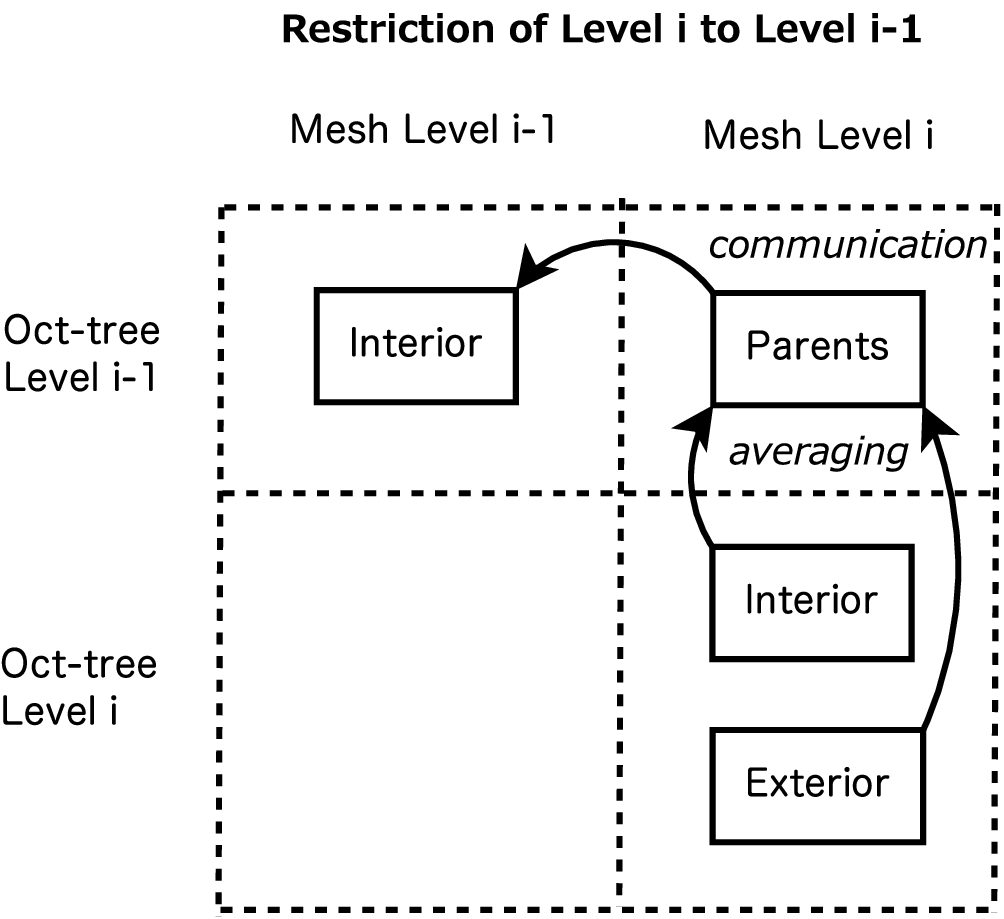}
  \caption{{\em Left: } Prolongation of Level~$i$ to Level~$i+1$. 
  This involves the Interior and Children submeshes of the Level~$i$ mesh (left column), as well as the Interior and/or Exterior submeshes of the Level~$i+1$ mesh (right column). 
  The Level~$i$ Interior submesh consists of cells at Level $i$ of the oct-tree (upper row), while the Level $i$ Children and Level $i+1$ Interior and Exterior submeshes consist of cells at Level $i+1$ of the oct-tree (lower row). 
  Interpolation to the Children is followed by communication to the finer-level Interior and/or Exterior.
  {\em Right: } Restriction of Level~$i$ to Level~$i-1$. 
  This involves the Interior, Exterior, and Parents submeshes of the Level~$i$ mesh (right column), as well as the Interior submesh of the Level~$i-1$ mesh (left column). The Level~$i$ Interior and Exterior submeshes consist of cells at Level $i$ of the oct-tree (lower row), while the Level $i$ Parents and Level $i-1$ Interior submeshes consist of cells at Level $i-1$ of the oct-tree (upper row). Averaging to the Parents is followed by communication to the coarser-level Interior.}
  \label{fig:ProlongationRestriction}
\end{figure}

The class representing a coordinate patch also has methods for `prolongation' (interpolation of Level~$i$ data to Level~$i+1$) and `restriction' (averaging of Level~$i$ data to Level~$i-1$). 
These operations are illustrated in Figure~\ref{fig:ProlongationRestriction}.
The first step of prolongation (left panel in Figure~\ref{fig:ProlongationRestriction}) is to interpolate data from the Level~$i$ Interior submesh to the Level~$i$ Children submesh.\footnote{Gradients used in the interpolation optionally can be computed with a slope limiter in order to respect discontinuities.} 
The Level~$i$ Children submesh is refined relative to the Level~$i$ Interior with which it is associated.
But in a distributed-memory message passing environment, the Level~$i$ Children follows the domain decomposition of the Level~$i$ Interior, as mentioned above, making the interpolation step a local operation. 
This is why both the Interior and Children submeshes are part of the same Level~$i$ `mesh.'
But the distribution of the Level~$i+1$ Interior cells among processes in general may differ from that of the Level~$i$ Interior, as the Interior submeshes on each level are independently domain-decomposed (right panel in Figure~\ref{fig:Submeshes}). 
Thus the second and final step of prolongation is communication, in general involving message passing, between from the Level~$i$ Children to the Level~$i+1$ Interior and/or Exterior. 
Similarly, restriction (right panel in Figure~\ref{fig:ProlongationRestriction}) involves first an averaging of Level~$i$ Interior and/or Exterior data to the Level~$i$ Parents (a local operation), followed by communication from the Level~$i$ Parents to the Level~$i-1$ Interior (which, again, in general involves message passing).

\section{HYDRODYNAMICS METHODS}
\label{sec:Solvers}

The equations of hydrodynamics are hyperbolic balance equations of the form
\begin{equation}
  \pderiv{\mathcal{U}}{t}+\divergence{\mathcal{F}(\mathcal{U})}
  =\mathcal{S}(\mathcal{U}). \label{eq:BalanceEquation}
\end{equation}
Here $\mathcal{U}$ is a vector of `conserved' or `balanced' variables, whose rate of change is determined by the divergence of the fluxes $\mathcal{F}(\mathcal{U})$, and by the sources $\mathcal{S}(\mathcal{U})$. 
In the absence of source terms, a balance equation reduces to a `conservation law': for an infinitesimal volume $dV$, the rate of change of $\mathcal{U}\, dV$ is equal to the integrated flux $-\oint_{S} \mathcal{F}(\mathcal{U})\cdot dS$ flowing in through the closed surface $S$ surrounding $dV$.  
The system described by Equation (\ref{eq:BalanceEquation}) is said to be hyperbolic if the Jacobian matrix associated with the flux divergence is diagonalizable, with eigenvalues all real \citep[e.g.][]{LeVeque2002}.  
Associated with $\mathcal{U}$ are two other sets of variables: `primitive' variables $\mathcal{W}$, often equal in number to the number of balanced variables; and some additional `auxiliary' variables $\mathcal{A}$, typically determined by one or more closure relations involving the primitive variables (e.g. an `equation of state' in the case of hydrodynamics). 

The divergence structure of hyperbolic balance equations naturally lends itself to a finite-volume approach \citep[e.g.][]{LeVeque2002}. 
Spatial discretization involves taking the volume average of Equation (\ref{eq:BalanceEquation}) over each of the cuboid cells in our multilevel grid structure:
\begin{equation}
  \pderiv{\langle\mathcal{U}\rangle}{t}
  = - \frac{1}{V}\sum_q \left[\left( A_q \langle\mathcal{F}^q \rangle \right)_\qrar - \left( A_q \langle\mathcal{F}^q \rangle \right)_\qlar\right]
  + \langle\mathcal{S}\rangle. \label{eq:BalanceEquationAverage}
\end{equation}
The sum is over dimensions $q$. 
Angle brackets of quantities associated with arrowed subscripts denote an area average over an outer ($\rar$) or inner ($\lar$) face of the cell, and angle brackets without such arrowed subscripts indicate a cell volume average. 
The cell volume and face areas are $V$ and $A_q$, respectively. 
While discretized in space, Equation (\ref{eq:BalanceEquationAverage}) remains continuous in time. 
(It is also exact, until numerical approximations to volume-averaged sources and area-averaged fluxes on the faces are taken.)
Once fully discretized in space, it is viewed as a system of ordinary differential equations, which then can be discretized in time and integrated using standard explicit techniques (e.g. Runge-Kutta methods). 
This approach is frequently used to design higher order methods for hyperbolic systems \citep[e.g.][]{Shu1998}. 
(We only consider second-order methods in the work presented here, however.)  
The spatial order of accuracy can be different from the temporal order of accuracy, although the overall formal accuracy of the scheme is limited to the lower of the two.  
This general approach---in which all dimensions but one are discretized, so as to allow application of methods for ordinary differential equations---is called the method of lines. 
It has also been called a semi-discrete method \citep[e.g.][]{Kurganov2000}.  

\subsection{Reconstruction}

Before the face-averaged fluxes in Equation (\ref{eq:BalanceEquationAverage}) can be computed, variable values on the faces must be obtained through a `reconstruction' of the spatial dependence of the variables within each cell. 
In \genasis\ we first obtain approximate values for the primitive variables $\mathcal{W}$ \citep[as opposed to the characteristic variables; e.g.][]{Shu1998,LeVeque2002} on the cell faces, and then use them to compute the conserved and auxiliary variables $\mathcal{U}$ and $\mathcal{A}$. 
In the work presented here we take the primitive variables to be represented by a linear expansion within a cell.  
(For cells of equal size, linear reconstruction results in a second-order spatial scheme.) 
For some point inside the cell---which we shall call the cell center---the values of the primitive variables are equal to the cell volume averages: 
\begin{equation}
  \mathcal{W}_\lrar = \langle\mathcal{W}\rangle,
\end{equation}
where the double-headed arrow ($\lrar$) denotes evaluation at the cell center. 
(In Cartesian coordinates---which we use in all the test problems in this paper---the cell center defined in this way coincides with the geometric center, i.e. the point equidistant from the centers of all cell faces.)
Denoting the first derivative of $\mathcal{W}$ in the $q$ dimension by $\mathcal{D}_q\left[ \mathcal{W}_\lrar \right]$, the values at the inner and outer faces are
\begin{eqnarray}
\mathcal{W}_\qlar &=& \mathcal{W}_\lrar - \mathcal{D}_q\left[ \mathcal{W}_\lrar \right] \left( q_\lrar - q_\qlar \right), \\
\mathcal{W}_\qrar &=& \mathcal{W}_\lrar + \mathcal{D}_q\left[ \mathcal{W}_\lrar \right] \left( q_\qrar - q_\lrar \right),
\end{eqnarray}
where the last factors are the coordinate distances between the faces and the cell center. Spurious oscillations near discontinuities can be reduced by using a slope limiter that enforces monotonic reconstruction; we use \citep[e.g.][]{Kurganov2000}
\begin{equation}
\mathcal{D}_q\left[ \mathcal{W}_\lrar \right]
= \mbox{MM}
  \left[
    \vartheta\left(\f{\mathcal{W}_\lrar - \mathcal{W}_\qminus}{q_\lrar - q_\qminus}\right),\, 
    \left(\f{\mathcal{W}_\qplus - \mathcal{W}_\qminus}{q_\qplus - q_\qminus}\right),\, 
    \vartheta\left(\f{\mathcal{W}_\qplus - \mathcal{W}_\lrar}{q_\qplus - q_\lrar}\right)
  \right]. \label{eq:MinMod}
\end{equation}
Center values in left and right (previous and next) neighboring cells in the $q$ dimension are labeled by subscripts $\qminus$ and $\qplus$ respectively.
The minmod function $\mbox{MM}[\cdot]$ compares its arguments and chooses the one with the smallest magnitude.  
If the arguments do not all have the same sign, the minmod function returns zero.  
The three arguments in Equation (\ref{eq:MinMod}) are left, centered, and right slopes, with the left and right slopes multiplied by the slope limiter parameter $\vartheta\in[1,2]$. Higher values of $\vartheta$ promote the centered difference, and are therefore less diffusive and more accurate for smooth flows, but also more prone to oscillations in the presence of discontinuities. The value $\vartheta=1$ is equivalent to the traditional minmod that selects between only the left and right derivatives; this is generally unacceptably diffusive. We often use $\vartheta = 1.4$.

\subsection{Riemann Solvers}
\label{sec:RiemannSolvers}

Because variables on cell faces are reconstructed independently for each cell, the values obtained from the cells on the left and right sides of an interface do not generally match up, and must be resolved to obtain a single value of the flux to be applied to both cells. 
In the finite-volume approach it is natural to consider the two states on the immediate left and right side of the interface as constituting a `Riemann problem' consisting of two regions of constant data, separated by a single discontinuity, and governed by a one-dimensional version of Equation (\ref{eq:BalanceEquation}): 
\begin{equation}
  \pderiv{\mathcal{U}}{t}+\pderiv{\mathcal{F}^q(\mathcal{U})}{q}
  =\mathcal{S}(\mathcal{U}). \label{eq:BalanceEquation1D}
\end{equation}
(For second-order methods, the extension to the multidimensional case is simple, and is achieved by applying the one-dimensional spatial discretization prescription separately in each dimension.)
The initial conditions in the two regions are commonly referred to as the `left' and `right' states; as applied to cell interfaces, these are
\begin{eqnarray}
(\mathcal{U}_\leftState)_\qlar &=& \mathcal{U}_\qrarMinus, \label{eq:conservedLeft}\\
(\mathcal{U}_\rightState)_\qlar &=& \mathcal{U}_\qlar
\end{eqnarray}
at a cell's inner face, and
\begin{eqnarray}
(\mathcal{U}_\leftState)_\qrar &=& \mathcal{U}_\qrar, \\
(\mathcal{U}_\rightState)_\qrar &=& \mathcal{U}_\qlarPlus \label{eq:conservedRight}
\end{eqnarray}
at a cell's outer face, where the subscripts $\qrarMinus$ and $\qlarPlus$ respectively denote outer face values of the previous cell and inner face values of the next cell.
As a Riemann problem evolves, several waves propagate away from the initial discontinuity into the left and right states, with velocities given by the eigenvalues $\lambda_q$ of the Jacobian matrix $\partial \mathcal{F}^q / \partial {\mathcal{U}}$.

A full solution of the Riemann problem consists of finding the `characteristics,' or trajectories of these several waves, and determining the (not necessarily constant) values of the variables in the regions bounded by them \citep[e.g.][]{LeVeque2002};  
but in practical applications, it is often desirable to work with only approximate solutions to the Riemann problem, and here we use two variants of the HLL-type approximate Riemann solvers \citep[][]{Harten1983,Einfeldt1988,Toro1994}.
These determine fluxes from the jump conditions 
\begin{equation}
  \lambda_{d}\left(\mathcal{U}_- - \mathcal{U}_+\right)
  = \mathcal{F}^q \left(\mathcal{U}_-\right)-\mathcal{F}^q \left(\mathcal{U}_+\right), \label{eq:JumpConditions}  
\end{equation}
that obtain at the boundaries between the regions separated by the propagating waves ($\mathcal{U}_-$ and $\mathcal{U}_+$ are the states immediately to the left and right of the discontinuity).

The first variant we use, which we simply label `HLL,' was devised by \cite{Harten1983}.  
The Riemann problem is approximated as consisting of only three constant states---$\mathcal{U}_\leftState$, $\mathcal{U}_*$, and $\mathcal{U}_\rightState$---separated by two waves.\footnote{Here we assume that at least two wave speeds can be associated with the hyperbolic system represented by Equation (\ref{eq:BalanceEquation1D}).}  
Application of Equation (\ref{eq:JumpConditions}) across the two waves gives
\begin{eqnarray}
  -\alpha^q_- \left(\mathcal{U}_\leftState - \mathcal{U}_*\right)
  &=& \mathcal{F}^q \left(\mathcal{U}_\leftState\right)-\mathcal{F}^q \left(\mathcal{U}_* \right), \label{eq:JumpConditionHLLMinus}\\  
  \alpha^q_+ \left(\mathcal{U}_* - \mathcal{U}_\rightState\right)
  &=& \mathcal{F}^q \left(\mathcal{U}_*\right)-\mathcal{F}^q \left(\mathcal{U}_\rightState\right). \label{eq:JumpConditionHLLPlus}  
\end{eqnarray}
Here $\mathcal{U}_\leftState$ and $\mathcal{U}_\rightState$ are the reconstructed face values on either side of a cell interface, as in Equations (\ref{eq:conservedLeft})-(\ref{eq:conservedRight}), 
and $\alpha^q_\pm = \mbox{max}\left[ 0, \pm \lambda^q_\pm \left(\mathcal{U}_\leftState \right), \pm \lambda^q_\pm \left(\mathcal{U}_\rightState \right) \right] $
are the fastest left- ($-$) and right- ($+$) moving hyperbolic wave speeds (magnitudes of eigenvalues of $\partial \mathcal{F}^q / \partial {\mathcal{U}}$). 
As the expression for $\alpha^q_\pm$ indicates, the wave speed estimates are computed from values on the left and right side of the interface, and the largest of the two is used in Equations (\ref{eq:JumpConditionHLLMinus}) and (\ref{eq:JumpConditionHLLPlus}) \citep[cf.][]{Davis1988}.  
Note that $\alpha^q_{\pm} \ge 0$. 
Solving Equations (\ref{eq:JumpConditionHLLMinus}) and (\ref{eq:JumpConditionHLLPlus}) for $\mathcal{U}_\HLL = \mathcal{U}_*$ and $\mathcal{F}^q_\HLL = \mathcal{F}^q\left( \mathcal{U}_* \right)$ yields
\begin{eqnarray}
\mathcal{U}_\HLL &=&
	\frac{\alpha^q_+ \,\mathcal{U}_\rightState + \alpha^q_- \,\mathcal{U}_\leftState - \left[\mathcal{F}^q \left(\mathcal{U}_\rightState\right) - \mathcal{F}^q \left(\mathcal{U}_\leftState\right) \right] }
	{\alpha^q_+ + \alpha^q_-}, \label{eq:BalancedHLL} \\
\mathcal{F}^q_\HLL &=&
	\frac{\alpha^q_+ \,\mathcal{F}^q \left(\mathcal{U}_\leftState\right) + \alpha^q_- \, \mathcal{F}^q \left(\mathcal{U}_\rightState\right) - \alpha^q_+ \alpha^q_- \left( \mathcal{U}_\rightState - \mathcal{U}_\leftState\right) }
	{\alpha^q_+ + \alpha^q_-}. \label{eq:FluxHLL}
\end{eqnarray}
When this solver is selected, it is $\mathcal{F}^q_\HLL$ that goes into the cell-averaged balance equation in Equation (\ref{eq:BalanceEquationAverage}).
For `supersonic' flow to the left ($\alpha^q_{+}=0$) or to the right ($\alpha^q_{-}=0$) the HLL flux reduces to a pure upwind flux (i.e., information from only one side of the interface is used to compute the flux).  
For `subsonic' flows, the third term on the right-hand-side of (\ref{eq:FluxHLL}) acts as a diffusion term which damps out grid-scale oscillations.  

The second variant we report here is known as the HLLC solver \citep{Toro1994}. 
In this case a third wave, which we here call the `middle wave,' is included in addition to the fastest left- and right-moving waves entering into the HLL solver.  
Thus the Riemann problem has four constant states: $\mathcal{U}_\leftState$, $\mathcal{U}_{*\leftState}$, $\mathcal{U}_{*\rightState}$, and $\mathcal{U}_\rightState$, with the middle wave separating the middle states $\mathcal{U}_{*\leftState}$ and $\mathcal{U}_{*\rightState}$. 
The jump conditions at the three waves are
\begin{eqnarray}
  -\alpha^q_- \left(\mathcal{U}_\leftState - \mathcal{U}_{*\leftState}\right)
  &=& \mathcal{F}^q \left(\mathcal{U}_\leftState\right)-\mathcal{F}^q \left(\mathcal{U}_{*\leftState} \right), \label{eq:JumpConditionHLLCMinus}\\  
  \alpha^q_m \left(\mathcal{U}_{*\leftState} - \mathcal{U}_{*\rightState}\right)
  &=& \mathcal{F}^q \left(\mathcal{U}_{*\leftState}\right)-\mathcal{F}^q \left(\mathcal{U}_{*\rightState}\right), \label{eq:JumpConditionHLLCMiddle}  \\
  \alpha^q_+ \left(\mathcal{U}_{*\rightState} - \mathcal{U}_\rightState\right)
  &=& \mathcal{F}^q \left(\mathcal{U}_{*\rightState}\right)-\mathcal{F}^q \left(\mathcal{U}_\rightState\right). \label{eq:JumpConditionHLLCPlus}  
\end{eqnarray}
Here $\alpha^q_\pm$ are the same as in the HLL case, and $\alpha^q_m$ is the middle wave speed estimate, which can be positive or negative (unlike $\alpha^q_\pm$, which are restricted to non-negative values). 
Unlike the system of Equations (\ref{eq:JumpConditionHLLMinus}) and (\ref{eq:JumpConditionHLLPlus}) in the HLL case, the present system of Equations (\ref{eq:JumpConditionHLLCMinus})-(\ref{eq:JumpConditionHLLCPlus}) is underdetermined: there are now four (sets of) unknowns $\left[\mathcal{U}_{*\leftState}, \mathcal{U}_{*\rightState}, \mathcal{F}^q \left(\mathcal{U}_{*\leftState} \right), \mathcal{F}^q \left(\mathcal{U}_{*\rightState} \right)\right]$ but only three (sets of) equations given by the jump conditions. 
Additional relations must be introduced, the details of which necessarily depend upon the particular system. 
(A concrete example is given in the case of hydrodynamics discussed in Section \ref{subsec:Fluids}, in which the starting point is to use the HLL states of Equation (\ref{eq:BalancedHLL}) to estimate the middle wave speed $\alpha_m^q$.)
Once this has been done and $\mathcal{F}^q \left(\mathcal{U}_{*\leftState} \right)$ and $\mathcal{F}^q \left(\mathcal{U}_{*\rightState} \right)$ have been obtained, the HLLC flux is given by 
\begin{equation}
 \mathcal{F}_\HLLC^q
  =\left\{
  \begin{array}{rl}
    \mathcal{F}^q \left(\mathcal{U}_{\leftState} \right) & \mathrm{if\ } \alpha^q_{-}=0,  \\
   \mathcal{F}^q \left(\mathcal{U}_{*\leftState} \right)  & \mathrm{if\ } \alpha^q_m \ge 0, \\
 \mathcal{F}^q \left(\mathcal{U}_{*\rightState} \right)  & \mathrm{if\ } \alpha^q_m < 0, \\
  \mathcal{F}^q \left(\mathcal{U}_{\rightState} \right) & \mathrm{if\ } \alpha^q_{+}=0.
  \end{array}
  \right. \label{eq:FluxHLLC}
\end{equation}
When this solver is selected, it is $\mathcal{F}^q_\HLLC$ that goes into the cell-averaged balance equation in Equation (\ref{eq:BalanceEquationAverage}).
As in the HLL case, the HLLC flux reduces to an upwind scheme for supersonic flows to the left ($\alpha^q_{-}=0$) or to the right ($\alpha^q_{+}=0$).  

The
HLLC solver is subject to so-called `odd-even decoupling' \citep[e.g.][]{Quirk1994} when a shock propagates parallel to a coordinate axis. 
To prevent this, we have implemented a shock detection scheme that automatically switches to the HLL solver in the immediate vicinity of a shock and in directions transverse to its propagation direction. 
Our tests show that the greater diffusivity of the HLL solver suppresses the odd-even decoupling instability without sacrificing the superior results of the HLLC solver on fluid instability tests in Section~\ref{sec:FluidInstabilityTests}.

%

\subsection{Updates}
\label{subsec:Updates}

Our implementation of reconstruction, flux computation (Riemann solve), and assembly of the right-hand side of Equation (\ref{eq:BalanceEquationAverage}) is written to address one level of our multilevel grid structure at a time; that is, one `mesh' in a `coordinate patch' at a time (see Section~\ref{sec:RefinableMesh}). 
There are however two ways in which better information from a next finer level is utilized.  
One of these is applied in the present layer of implementation: for cells on the coarse side of a coarse/fine boundary with the next level, flux updates from faces on the coarse/fine boundary computed by and restricted from the finer level replace those computed at the present (coarser) level. 
Aside from making use of more highly resolved data, this ensures conservative evolution on the union of leaf cells of all levels (modulo source terms). 
The second use of information from the next finer level is applied at a higher layer of coding discussed in more detail later: non-leaf cells have their updated balanced variables replaced by restriction from the next finer level (see Section~\ref{sec:Evolution}). 
This restriction does not have direct consequences for global conservation checks, because these are tallied only from leaf cells; nevertheless this step is performed for the sake of consistency between levels.  

In pursuit of efficient performance we seek to both (a) overlap work and communication in a message-passing parallel environment, and (b) work on data stored in contiguous memory when possible.
To accomplish these goals we identify two types of cells.
`Exchange' cells are cells near process boundaries whose data must be communicated (or exchanged) to fill the ghost cells on neighboring processes in a `ghost exchange' during the update of the hyperbolic balance equations.
(The ghost cells of a given process are exchange cells on neighboring processes; see Section~\ref{sec:RefinableMesh}.)
`Non-exchange' cells, in contrast, do not have to communicate their data to neighboring processes.
Goals (a) and (b) above can be accomplished by processing exchange cells and non-exchange cells separately in some instances, and by packing data from these two sets of cells into contiguous memory before certain operations.  
Storage we refer to as `unpacked' is the normal mesh-oriented storage discussed in Section~\ref{sec:RefinableMesh}. 
It is this type of storage in which connectivity information culled from the oct-tree is available, i.e. location in the mesh and knowledge of siblings. 
Operations that need this information must be performed with unpacked storage. 
These include exchange of ghost cell data, and operations across cell faces that require data from sibling cells, in particular calculations of differences and fluxes.
`Packed' storage, on the other hand, has no connectivity information. 
It provides for efficient intra-cell operations by furnishing sequential memory access to sets of data---in particular, data only from exchange cells or data only from non-exchange cells---that are discontiguous in unpacked storage. 

Algorithm \ref{alg:BalanceEquation} outlines a balance equation update on a single level. 
Uses of packed and unpacked storage are indicated by (P) and (U) respectively. 
One major loop is explicitly shown, over cell types (exchange vs. non-exchange): exchange cells are operated on first, non-blocking communication is initiated, and then operations are performed on non-exchange cells while communication continues in the background.  
In Algorithm \ref{alg:BalanceEquation} and later listings, right-arrows ($\rightarrow$) indicate the flow of various operations (e.g. {\em Compute}, {\em Pack}, and {\em Unpack}), with the input on the left and the result on the right.\footnote{For example, packed finite differences are computed from unpacked primitive variables in line 2;
primitive variables in unpacked storage are copied into packed storage in line 3;  
packed reconstructed primitive variables are computed from packed primitive variables and finite differences in line 4; and so on.}  
Application of the updates to obtain new values of the variables is deferred to the time stepper (see Section \ref{subsec:Steps}).
The slowest parts of the algorithm are those that involve both packed and unpacked storage, as these necessarily involve indirect indexing of the unpacked storage. 
These include computations across cell faces (differences and fluxes) and moving data between packed and unpacked storage. Ideally, the efficiency of operations in packed storage makes up for the overhead of data movement between the two storage types.

\begin{algorithm}
\begin{algorithmic}[1]
\FOR{iCells = EXCHANGE, NON-EXCHANGE}
\STATE {\em Compute: } Primitive (U) $\rightarrow$ Differences (P)
\STATE {\em Pack: } Primitive (U) $\rightarrow$ Primitive (P)
\STATE {\em Compute: } Primitive (P), Differences (P) $\rightarrow$ Reconstructed Primitive (P)
\STATE {\em Compute: } Reconstructed Primitive (P) $\rightarrow$ Reconstructed Balanced, Reconstructed Auxiliary (P)
\STATE {\em Unpack: } Reconstructed (P) $\rightarrow$ Reconstructed (U)
\STATE {\em Apply Boundary Conditions: } Reconstructed Interior (U) $\rightarrow$ Reconstructed Exterior (U)
\STATE {\em Compute: } Reconstructed (U) $\rightarrow$ Riemann Solver Input (U)
\STATE {\em Compute: } Riemann Solver Input (U) $\rightarrow$ Fluxes (P)
\STATE {\em Compute: } Fluxes (P), Sources (P) $\rightarrow$ Updates (P)
\STATE {\em Unpack: } Updates (P) $\rightarrow$ Updates (U)
	\IF{iCells == EXCHANGE} 
		\STATE {\em Begin Exchange: } Updates (U) 
	\ENDIF
\ENDFOR
\STATE {\em Finish Exchange: } Updates (U)
\caption{Single-level balance equation update}
\label{alg:BalanceEquation}
\end{algorithmic}
\end{algorithm}

\subsection{Steps}
\label{subsec:Steps}

As discussed in connection with Equation (\ref{eq:BalanceEquationAverage}), a spatially discretized system with a continuous time derivative can be considered a system of time-dependent ordinary differential equations. We write this as
\begin{equation}
  \frac{d\langle \mathcal{U} \rangle}{dt} = \mathcal{L}\left[\langle\mathcal{U}\rangle\right],
  \label{eq:OrdinaryDifferentialEquation}
\end{equation}
where (for example) in the case of balance equations $\mathcal{L}\left[\langle\mathcal{U}\rangle\right]$ is given by the right-hand side of Equation (\ref{eq:BalanceEquationAverage}).
The system consists of one equation---or rather, one set of equations, since $\mathcal{U}$ is a vector---for $\langle \mathcal{U} \rangle$ in every cell. 
(We suppress indexing not only over the components of $\mathcal{U}$, but also over cells, taking this to be implied by the volume-average-denoting angle brackets.)
Equation (\ref{eq:OrdinaryDifferentialEquation}) is amenable to discretization and integration in time using standard explicit techniques. (By `explicit' we mean that the solution at time $t^{n+1}$ depends only on values known at $t^n$, where the sans-parentheses superscript $n$ denotes a value at the beginning of the $n$th time step.) 

A basic building block of many such integration algorithms is a single update
\begin{equation}
\mathcal{K}^{(i)} = \Delta t \, \mathcal{L}\left[\langle\mathcal{U}\rangle^{(i-1)}\right],
\label{eq:Update}
\end{equation}
where $\Delta t = t^{n+1} - t^n$ is the full time step.
Stable and accurate schemes typically involve multiple such updates or `substeps,' indexed by the parenthetical superscript $(i)$, with higher-order schemes requiring more updates. 
Typically $\langle\mathcal{U}\rangle^{(0)} = \langle\mathcal{U}\rangle^{n}$ on the right-hand side of Equation (\ref{eq:Update}) for the first update $\mathcal{K}^{(1)}$. 
Subsequent intermediate values $\langle\mathcal{U}\rangle^{(i)}$ then depend on prior updates.

In all the test problems in this paper we use a second-order Total Variation Diminishing (TVD) Runge-Kutta step \citep[e.g.][also known as Heun's method]{Shu1998} consisting of two substeps. 
The two updates $\mathcal{K}^{(1)}$ and $\mathcal{K}^{(2)}$ are obtained by respectively using 
\begin{eqnarray}
  \langle\mathcal{U}\rangle^{(0)} &=& \langle\mathcal{U}\rangle^{n}, \label{eq:RungeKutta1} \\
  \langle\mathcal{U}\rangle^{(1)} &=& \langle\mathcal{U}\rangle^{n} + \mathcal{K}^{(1)} \label{eq:RungeKutta2}
\end{eqnarray}
on the right-hand side of Equation (\ref{eq:Update}). 
Then
\begin{equation}
  \langle\mathcal{U}\rangle^{n+1} = \langle\mathcal{U}\rangle^{n} + \frac{1}{2}\left( \mathcal{K}^{(1)} + \mathcal{K}^{(2)} \right) 
  \label{eq:RungeKutta3}
\end{equation}
is the solution at $t^{n+1}$. 

There are a couple of points to make in connection with balance equations. 
First, in this context $\langle\mathcal{U}\rangle^{n+1}$ are the balanced variables, from which the primitive and auxiliary variables $\langle\mathcal{W}\rangle^{n+1}$ and $\langle\mathcal{A}\rangle^{n+1}$ are subsequently obtained. These operations, and also the computations in Equations (\ref{eq:RungeKutta1})-(\ref{eq:RungeKutta3}), are all performed on all proper and ghost cells of the Interior submesh: 
Algorithm \ref{alg:BalanceEquation} implementing computation of an update $\mathcal{K}$ based on Equation (\ref{eq:BalanceEquationAverage}) already includes a ghost exchange of $\mathcal{K}$. 
(See Section~\ref{sec:RefinableMesh} for more on proper and ghost cells and the Interior submesh.) 
Thus no additional communication is needed at this point to complete the step. Moreover, being performed over all Interior cells, application of the updates works on contiguous memory even though it is `unpacked' storage.
Second, flux updates at faces on the coarse/fine boundary with a coarser level are recorded in a manner consistent with the update scheme described above, in order that they can be utilized when the next coarser level is evolved (as discussed at the beginning of Section \ref{subsec:Updates}).

Time step determination, addressing of multiple levels, and looping over many steps are implemented in a higher layer of coding and are further discussed later (Section \ref{sec:Evolution}).

\subsection{Nonrelativistic Fluid}
\label{subsec:Fluids}


We now discuss a particular example of a system governed by balance equations of the form of Equation~(\ref{eq:BalanceEquation}): a nonrelativistic fluid.
We describe only ideal fluids; dissipative processes (e.g. viscosity and heat conduction) are not included, and the equations describe adiabatic flows: $\partial s/\partial t+\vect{v}\cdot\gradient{s}\equiv ds/dt=0$, where $s$ is the entropy per baryon \citep[see for example][for an introduction to fluid mechanics]{Landau1959}.

First we specify the most basic fluid variables and the fluxes appearing in the balance equations.
The balanced variables $\mathcal{U}$ are the conserved baryon density~$D$, the momentum density~$\mathbf{S}$, and the balanced energy density $G$.
The primitive variables $\mathcal{W}$ are the comoving baryon density $n$, the three-velocity~$\mathbf{v}$, and the internal energy density $e$.
The auxiliary variables $\mathcal{A}$ are the average baryon mass $m$ and the pressure $p$.  
The relations between these variables are
\begin{eqnarray}
  D &=& n, \label{eq:BalancedBaryonDensity} \\
  \mathbf{S} &=& m n \mathbf{v}. \label{eq:MomentumDensity} \\
  G &=& e + \frac{1}{2} m n |\mathbf{v}|^2.
  \label{eq:balancedEnergy}
\end{eqnarray}
The baryon, momentum, and energy fluxes $\mathcal{F}(\mathcal{U})$ are
\begin{eqnarray}
  \mathcal{F}(D) &=& n\mathbf{v}, \label{eq:FluxBaryon} \\
  \mathcal{F}(\mathbf{S}) &=& mn\mathbf{v}\mathbf{v} + p \mathbf{I}, \label{eq:FluxMomentumFluid} \\
  \mathcal{F}(G) &=& \left( e + p + \frac{1}{2} m n |\mathbf{v}|^2 \right) \mathbf{v}, 
\end{eqnarray}
where $\mathbf{I}$ is the rank-two unit tensor.

Next we discuss aspects needed for the Riemann solvers discussed in Section~\ref{sec:RiemannSolvers}.
There are three distinct wave velocities: $\lambda_-^q = v^q - c_s$, $\lambda_m^q = v^q$, and $\lambda_+^q = v^q + c_s$, where the adiabatic sound speed $c_s = \sqrt{(\partial p / \partial \rho )_s} = \sqrt{(\Gamma_s p / \rho)}$, with adiabatic index $\Gamma_s \equiv (\partial \ln p / \partial \ln \rho )_s$ and mass density $\rho=mn$.  
The eigenvalues $\lambda_-^q$ and $\lambda_+^q$ are propagation velocities associated with shock or rarefaction waves, while $\lambda_m^q$ is the propagation velocity of contact and shear waves.  
(The contact wave is also referred to as the entropy wave.)  
The HLL solver uses only the largest-magnitude wave velocities $\lambda_\pm^q$.  
Moreover, recall (Section \ref{sec:RiemannSolvers}) that in the supersonic case (either $\alpha_+^q = 0$ or $\alpha_-^q = 0$) the HLLC solver yields the same result as the HLL solver, namely, upwind fluxes computed solely from the reconstructed values on either the left or right side of the interface (see Equation (\ref{eq:FluxHLLC})).  
For subsonic flows the HLLC solver uses, in addition, an estimate of the middle wave velocity $\alpha_m^q = \lambda_m^q$.
We follow \citet{Batten1997} and form the middle wave speed from the mass density and momentum density of the HLL average state (cf. our Equations (\ref{eq:BalancedHLL}), (\ref{eq:BalancedBaryonDensity}), (\ref{eq:MomentumDensity})):
\begin{equation}
  \alpha_m^q = \frac{S_\HLL^q}{m D_\HLL} 
	= \frac{S_\rightState^q \left(\alpha_+^q - v_\rightState^q \right)
		+ S_\leftState^q \left(\alpha_-^q + v_\leftState^q \right)
		- \left(p_\rightState - p_\leftState \right)}
		{m D_\rightState \left(\alpha_+^q - v_\rightState^q \right)
		+ m D_\leftState \left(\alpha_-^q + v_\leftState^q \right)}. \label{eq:MiddleWaveSpeed}
\end{equation}
Recall that the three waves in the HLLC framework separate four constant states: the reconstructed values $\mathcal{U}_\leftState$ and $\mathcal{U}_\rightState$, and the middle states $\mathcal{U}_{*\leftState}$ and $\mathcal{U}_{*\rightState}$ separated by the middle wave. 
In the subsonic case the fluxes are constructed from either $\mathcal{U}_{*\leftState}$ or $\mathcal{U}_{*\rightState}$, depending on the sign of $\alpha_m^q$ (Equation (\ref{eq:FluxHLLC})). 
Obtaining $\mathcal{U}_{*\leftState}$ or $\mathcal{U}_{*\rightState}$ in terms of the known reconstructed values $\mathcal{U}_\leftState$ or $\mathcal{U}_\rightState$ is accomplished by making a key assumption followed by use of the jump conditions of Equations (\ref{eq:JumpConditionHLLCMinus}) or (\ref{eq:JumpConditionHLLCPlus}) across the outer waves.
The key assumption is to set the normal velocity component equal to the estimated middle wave speed,
\begin{equation}
  v_{*\leftState/\rightState}^q = \alpha_m^q, 
  \label{eq:NormalVelocityMiddle}
\end{equation}
for both the left and right middle states.
The density jump conditions give
\begin{equation}
D_{*\leftState/\rightState} 
	= \frac{\alpha_\mp^q \pm v_{\leftState/\rightState}^q}
	{\alpha_\mp^q \pm \alpha_m^q} D_{\leftState/\rightState}, 
\end{equation}
with the upper and lower signs corresponding to the left ($\leftState$) and right ($\rightState$) versions of the equation respectively. 
Using this in conjunction with the transverse components of the momentum jump conditions yields
\begin{equation}
  v_{*\leftState/\rightState}^{r\ne q} = v_{\leftState/\rightState}^{r\ne q},
\end{equation}
while the normal component produces
\begin{equation}
  p_{*\leftState/\rightState} = p_{\leftState/\rightState} 
	- D_{\leftState/\rightState}
	\left( \alpha_\mp^q \pm v_{\leftState/\rightState}^q \right)
	\left( \alpha_m^q \pm v_{\leftState/\rightState}^q  \right). 
\label{eq:PressureMiddle}
\end{equation}
The energy jump conditions give
\begin{equation}
  G_{*\leftState/\rightState} =
	\frac{\left( \alpha_\mp^q \pm v_{\leftState/\rightState}^q \right) 
		G_{*\leftState/\rightState}
		\pm  v_{\leftState/\rightState}^q  p_{\leftState/\rightState}
		\mp  \alpha_m^q  p_{*\leftState/\rightState} }
	{\alpha_\mp^q \pm \alpha_m^q}. \label{eq:EnergyMiddle}
\end{equation}
Equations (\ref{eq:MiddleWaveSpeed})-(\ref{eq:EnergyMiddle}) provide the necessary and sufficient information needed to build the HLLC flux.  
As an aside, it is not immediately apparent from Equation (\ref{eq:PressureMiddle}), but it turns out that enforcing the normal velocity to be constant across the middle wave in the Riemann fan (cf. Equation (\ref{eq:NormalVelocityMiddle})) implies that the pressure is also constant across the middle wave. This is because the normal component of the momentum jump condition across the middle wave (Equation (\ref{eq:JumpConditionHLLCMiddle})) yields
\begin{equation}
p_{*\leftState} 
	+ m D_{*\leftState} v_{*\leftState}^q
		\left( v_{*\leftState}^q - \alpha_m^q \right)
= p_{*\rightState} 
	+ m D_{*\rightState} v_{*\rightState}^q
		\left( v_{*\rightState}^q - \alpha_m^q \right).
\end{equation}
It then follows from Equation (\ref{eq:NormalVelocityMiddle}) that $p_{*\leftState} = p_{*\rightState}$.

An equation of state of the form
\begin{equation}
  p=\kappa n^{\gamma}
\end{equation}
closes the system of hydrodynamics equations for all the test problems in this paper.
From the first law of thermodynamics for an ideal fluid,
\begin{equation}
  de = \frac{e + p}{n}\, dn,
\end{equation}
it follows that $p = (\gamma-1)\, e$ and that the adiabatic index $\Gamma_s = \gamma$. 
The adiabatic index $\gamma$ is a constant parameter. In the absence of energy source terms, the `polytropic constant' $\kappa$ remains constant unless shocks generate dissipation (captured automatically by the finite-volume approach with HLL or HLLC solvers).

\subsection{Evolution}
\label{sec:Evolution}

We now discuss the evolution of a free fluid,
which in basic outline is simply a loop over individual time steps covering the interval from parameters {\tt StartTime} to {\tt EndTime}; see the routine {\tt Evolve} in Algorithm \ref{alg:Evolve}.
However, there are complications from multiple levels of refinement.\footnote{There are further complications in the case of a space with multiple coordinate patches. These would depend on the manner in which the patches are stitched together to form a manifold. We do not discuss this further as all the examples in this paper use a single (refinable) coordinate patch.}
Our multilevel explicit evolution features `subcycling' of deeper levels, or `refinement in time' as well as space. 
The first hint of this appears in line 3 of Algorithm \ref{alg:Evolve}: there is a global {\tt Time} for the coordinate patch as a whole, but also an array {\tt LevelTime} that tracks the time to which each level has been evolved.  
At the beginning of a global time step all elements of {\tt LevelTime} are initialized to the global {\tt Time}.
A call to {\tt EvolveLevel} for the coarsest level (line 4) recursively evolves all the deeper levels, updating the elements of {\tt LevelTime} in the process.  
All levels are synchronized by the time this call returns, so that the global {\tt Time} is updated to the element of {\tt LevelTime} corresponding to the coarsest level (line 5).  

\begin{algorithm}
  \begin{algorithmic}[1]
    \STATE {\em Set: } Time = StartTime
    \WHILE{Time $<$ EndTime}
    \STATE {\em Set: } LevelTime ( 1 : nLevels ) = Time
    \STATE {\em Call: } EvolveLevel ( LevelTime, iLevel = 1 )
    \STATE {\em Set: } Time = LevelTime ( 1 )
    \ENDWHILE
    \caption{Evolve ( )}
    \label{alg:Evolve}
  \end{algorithmic}
\end{algorithm}

The basic idea of the algorithm for {\tt EvolveLevel}, called in line~4 of Algorithm \ref{alg:Evolve}, is straightforward (see Algorithm \ref{alg:EvolveLevel}): 
the fact that two steps of Level $i$ are performed for every step of Level $i-1$ makes it convenient to have this routine simply group two successive calls to a more primitive routine {\tt StepLevel} (lines 4 and 11). 
There are also two {\tt if} blocks.
The one at the top (lines 1-3) terminates the recursion to deeper levels by returning when the deepest level has been reached.
The one after the first call to {\tt StepLevel} (lines 5-7) returns if {\tt iLevel == 1}; i.e., only a single step of the coarsest level is performed.
For {\tt iLevel~>~1}, in order to allow updated boundary data for the current level (Level~$i$) to be generated during the second call to {\tt StepLevel} in line 11, lines 8-10 approximately and temporarily evolve the coarser level (Level~$i-1$) a `half-step' to synchronize it with the current level after its first step forward in line 4.
This coarse provisional `half-step,' being for Level~$i$ boundary condition purposes only, is approximate in two time-saving ways: 
first, a restriction of updates at the coarse/fine boundary is omitted; 
and second, only the first Runge-Kutta substep, the forward-Euler Eq.~(\ref{eq:RungeKutta2}), is performed. 
Further economization is achieved by storing the Level~$i-1$ fluxes---the output of the Riemann solver---computed at this stage for later reuse.
The coarse provisional `half-step' is temporary in that, after the second step of Level~$i$, the coarser Level~$i-1$ is reset to its previous value (line 12); it subsequently will be properly evolved a full step,   
with the previously-computed fluxes being used in the first substep (Eq.~(\ref{eq:RungeKutta2})), thereby not wasting an expensive Riemann solve.
Beyond the {\tt if} blocks and the approximate coarse step, the only other wrinkle in this routine is the optional argument {\tt ForParentsThisOption}, which was not included in the top-level call in line~5 of Algorithm \ref{alg:Evolve}. 
This allows updates at coarse/fine boundaries to be accumulated between the two steps at Level~$i$ for use at Level~$i-1$.

\begin{algorithm}
  \begin{algorithmic}[1]
    \IF{iLevel $>$ nLevels} 
    \STATE {\em Return}
    \ENDIF 
    \STATE {\em Call: } StepLevel ( LevelTime, iLevel, ForParentsThisOption )
    \IF{iLevel $==$ 1} 
    \STATE {\em Return}
    \ENDIF
    \STATE {\em Set: } CoarseFluidOld = Fluid ( iLevel - 1 )
    \STATE {\em Set: } CoarseStep = LevelTime ( iLevel ) - LevelTime ( iLevel - 1 )
    \STATE {\em Call: } Stepper ( iLevel - 1 ) \% Step ( CoarseStep, ProvisionalStepOption = .true. )
    \STATE {\em Call: } StepLevel ( LevelTime, iLevel, ForParentsThisOption )
    \STATE {\em Set: } Fluid ( iLevel - 1 ) = CoarseFluidOld
    \caption{EvolveLevel ( LevelTime, iLevel, ForParentsThisOption )}
    \label{alg:EvolveLevel}
  \end{algorithmic}
\end{algorithm}

The time stepper described in Section \ref{subsec:Steps} is finally invoked for Level $i$ in the routine {\tt StepLevel} (see Algorithm~\ref{alg:StepLevel}), which is called in lines~4 and 12 of Algorithm~\ref{alg:EvolveLevel}.
Taking a step forward in time is its main purpose; but most of this routine (in terms of lines of code) is actually dedicated to interactions between adjacent levels.

Lines 1-5 of Algorithm  \ref{alg:StepLevel} show that if a deeper level exists, that level is evolved {\em before} the current level.
In line 2, the variable {\tt ForParentsNext} is allocated to hold updates at coarse/fine boundaries to be computed on Level $i+1$.
The call to {\tt EvolveLevel} $i+1$ follows on line 3.  
This call includes the optional argument containing temporary storage {\tt ForParentsNext} in which the updates at the coarse/fine boundary between Levels $i$ and $i+1$ are accumulated.  
After the call returns, these updates are restricted from Level $i+1$ to the member {\tt FromChildren} of the Level $i$ stepper (lines 4). 
(Recall that restriction is the averaging procedure that provides data for a coarse cell from the finer cells it encompasses; see Section~\ref{sec:RefinableMesh}.) 
When the stepper at Level $i$ takes a time step (line 14), these updates {\tt FromChildren} overwrite those computed from Level $i$ at the coarse/fine boundary with those from Level $i+1$, ensuring conservative evolution at this boundary with the more accurate update computed at the finer level. 

A couple of additional tasks must be accomplished before the current Level~$i$ is stepped forward in line 14 of Algorithm  \ref{alg:StepLevel}.
Boundary values (i.e. data for the Exterior submesh; see Section~\ref{sec:RefinableMesh}) of Level $i$ are filled in by prolongation from Level $i-1$ if this is not the coarsest level (lines 6-8).  
(Recall that prolongation is the interpolation operation that provides data for finer cells from coarser cells; see Section~\ref{sec:RefinableMesh}.)
Then the time step is set in lines 9-13, either in accordance with the Courant-Friedrichs-Lewy (CFL) condition for stable explicit time stepping if this is the deepest level, or to synchronize the current Level~$i$ with the more refined (and just evolved) Level~$i+1$.
In detail, the CFL-restricted time step at the deepest level depends on the cell widths $\Delta q$ and characteristic speeds $|\lambda^{-}_{q}|,|\lambda_{q}^{+}|$.
Denoting
\begin{equation}
  \Delta t_{\mbox{\tiny CFL}}
  =\min_{\mathrm{cells}}
  \left[ \min_q \left(
    \f{\Delta q}{\max(|\lambda^{-}_{q}|,|\lambda_{q}^{+}|)}
  \right) \right],
  \label{eq:courantTimeStep}
\end{equation}
we take the CFL-restricted time step to be
\begin{equation}
\Delta t = \mbox{C}_{\mbox{\tiny CFL}}\times\Delta t_{\mbox{\tiny CFL}}, 
\end{equation}
where $\mbox{C}_{\mbox{\tiny CFL}}\lesssim1/d$ for number of spatial dimensions $d$ is an appropriate `Courant factor' for Runge-Kutta evolution with the method of lines \citep[e.g.][]{Shu1988}.

The update of Level $i$ is almost complete.
After the Level $i$ step in line 14 of Algorithm  \ref{alg:StepLevel}, consistency with Level $i+1$ is completed by restriction of the fluid from Level $i+1$ to Level $i$ in cells where these levels overlap (lines 15-17).
Then the appropriate element of {\tt LevelTime} is updated in line 18 to reflect the advance by {\tt TimeStep}.
The final task is to accumulate, in the optional argument {\tt ForParentsThisOption}, the coarse/fine boundary updates stored in the Level $i$ stepper member {\tt ForParents}, which will be used later by the Level $i-1$ stepper to handle updates at the coarse/fine boundary between Levels $i-1$ and $i$ (lines 19-21).

\begin{algorithm}
  \begin{algorithmic}[1]
    \IF{iLevel + 1 $<=$ nLevels} 
    \STATE {\em Allocate: } ForParentsNext
    \STATE {\em Call: } EvolveLevel ( LevelTime, iLevel + 1, ForParentsNext )
    \STATE {\em Restrict: } ForParentsNext, Exterior $\rightarrow$ Stepper ( iLevel ) \% FromChildren, Interior
    \ENDIF 
    \IF{iLevel $>$ 1}
    \STATE {\em Prolong: } Fluid ( iLevel - 1 ), Interior $\rightarrow$ Fluid ( iLevel ), Exterior
    \ENDIF
    \IF{iLevel $==$ nLevels}
    \STATE {\em Set: } TimeStep = $\mbox{C}_{\mbox{\tiny CFL}}\times\Delta t_{\mbox{\tiny CFL}}$
    \ELSE
    \STATE {\em Set: } TimeStep = LevelTime ( iLevel + 1 ) - LevelTime ( iLevel )
    \ENDIF
    \STATE {\em Call: } Stepper ( iLevel ) \% Step ( TimeStep )
    \IF{iLevel + 1 $<=$ nLevels} 
    \STATE {\em Restrict: } Fluid ( iLevel + 1 ), Interior $\rightarrow$ Fluid ( iLevel ), Interior 
    \ENDIF 
    \STATE {\em Compute: } LevelTime ( iLevel ) = LevelTime ( iLevel ) + TimeStep
    \IF{present ( ForParentsThisOption )}
    \STATE {\em Compute: } ForParentsThisOption = ForParentsThisOption + Stepper ( iLevel ) \% ForParents
    \ENDIF
    \caption{StepLevel ( LevelTime, iLevel, ForParentsThisOption )}
    \label{alg:StepLevel}
  \end{algorithmic}
\end{algorithm}

We now summarize the recursive multilevel evolution executed by the routines {\tt Evolve}, {\tt EvolveLevel}, and {\tt StepLevel} in Algorithms~\ref{alg:Evolve}-\ref{alg:StepLevel}.
Each iteration of the loop in {\tt Evolve} takes a single global time step by calling the method {\tt EvolveLevel} for the coarsest level. 
However, the coarsest level is not immediately stepped forward. 
Instead, {\tt EvolveLevel} and {\tt StepLevel} recursively work their way down to the deepest, most refined level, which is the first to be evolved. 
From the bottom up, two steps of each Level $i+1$ are performed for each step at Level $i$. 
By construction, the evolution of the entire multilevel structure ends up fully synchronized once this process works its way up to complete a single step at the coarsest level.
The conservation (in the absence of sources, as in most of the test problems reported in this paper) implied by the divergence structure of the balance equations is ensured by using, at Level $i$, the updates at coarse/fine boundaries computed at Level $i+1$. 
Further---less critical and perhaps fastidious---consistency between levels is achieved by restriction from Level $i+1$ to Level $i$ where these overlap.  
Our algorithm for time integration of the balance equations on a multilevel grid is demonstrated in the 2D and 3D Sedov-Taylor blast wave problems (Section \ref{sec:SedovTaylor}).

\section{HYDRODYNAMICS TESTS}
\label{sec:Tests}

In this section we present results from numerical test problems demonstrating the capabilities of the numerical methods and algorithms implemented in \genasis\ to solve the equations of nonrelativistic ideal hydrodynamics.  
The test problems have been chosen to validate the correctness of our implementation and to reveal the scheme's strengths and weaknesses.  
Most are well-known in the literature \citep[e.g.][]{Toro2009,Fryxell2000,Liska2003,Stone2008}.  

By definition a test problem has an accepted solution against which program output can be compared.
For some problems an analytic solution exists;
in other cases no analytic solution exists, but `known' numerical solutions are available in the literature.  
In the cases where no analytic solution is available, it is also common practice to compare low-resolution results with a high-resolution reference solution \citep[self-convergence; e.g.][]{Stone1992c}, which we do in Section \ref{sec:interactingBlastWaves}.  
For test problems where an analytic solution $\chi$ is available, we quantify the quality of the numerical solution $\langle\chi\rangle$ produced with \genasis~by computing the relative $L_{1}$ error norm
\begin{equation}
  L_{1} \left( \chi, N \right)
  = \frac{\displaystyle{\sum_{\mbox{\tiny cells}}} \left| \langle\chi\rangle - \chi \right| }{\displaystyle{\sum_{\mbox{\tiny cells}}} \left| \chi \right| },
  \label{eq:L1ErrorNorm}
\end{equation}
where the sums extend over all $N$ cells covering the computational domain.  
For a numerical method of spatial order $p$, the error $|\langle\chi\rangle-\chi|$ decreases with the mesh spacing as $(\Delta q)^{p}\,\propto N_{q}^{-p}$, where $N_{q}$ is the number of cells in the $q$th coordinate direction (assuming a uniform mesh).  
A similar argument holds for the temporal order of the numerical method (due to the CFL condition in Equation (\ref{eq:courantTimeStep}), the ratio $\Delta t/\Delta q$ remains fixed as the mesh spacing decreases), and the temporal error decreases with decreasing mesh spacing at a rate determined by the temporal order of the scheme.  
We determine the formal order of the numerical method by computing the relative $L_{1}$ error norm in a resolution study:  
by evolving an initial condition to some specified end time with multiple grid resolutions (e.g. $N_{q,1}<N_{q,2}$), the formal order of the scheme $p$---or the rate at which the numerical solution approaches the analytic solution as the number of grid cells increases from $N_{q,1}$ to $N_{q,2}$---is determined from
\begin{equation}
  -p=\frac{\log\left[L_{1}(\chi,N_{q,1})/L_{1}(\chi,N_{q,2})\right]}{\log\left[N_{q,1}/N_{q,2}\right]}.  
  \label{eq:convergenceRate}
\end{equation}
The hydrodynamics scheme implemented in \genasis\ is designed to be second-order accurate (we use linear interpolation in space and evolve the resulting system of ODEs with a second-order Runge-Kutta time integrator), and we expect second-order convergence for smooth flows.  
For flows containing discontinuities, the scheme switches to constant spatial interpolation in the vicinity of the discontinuities (cf. Equation (\ref{eq:MinMod})), and the formal order of the scheme reduces to first order.  

We perform smooth fluid, discontinuous fluid, and fluid instability tests.
Our results illustrate the basic competence of our implementation, demonstrate the strengths and limitations of the HLLC relative to the HLL Riemann solver, and provide preliminary indications of the code's ability to scale and to function with cell-by-cell fixed-mesh refinement.
We present results problems in one (1D), two (2D), and three (3D) space dimensions.  
For the 1D and 2D test problems we set the Courant factor to $\mbox{C}_{\mbox{\tiny CFL}}=0.5$, and for the 3D tests we use $\mbox{C}_{\mbox{\tiny CFL}}=0.3$.  

\subsection{Smooth Fluid Tests}
\label{subsec:SmoothFluidTests}

Tests with smooth analytic solutions enable us to determine the order of accuracy of the hydrodynamics solvers implemented in \genasis.
We compare the numerical solutions with the analytic solution for multiple grid resolutions, compute the relative $L_{1}$ error norm (Equation (\ref{eq:L1ErrorNorm})), and calculate the rate $p$ (Equation (\ref{eq:convergenceRate})) at which the error norm decreases with increasing grid resolution.  
We demonstrate second-order accuracy with both the HLL and the HLLC solvers.  

\subsubsection{Fluid Advection}
\label{subsubsec:advection}

In this test a smooth density profile is advected with a constant velocity field.
A constant pressure field is also present, and we evolve the full system of hydrodynamics equations, not just the continuity equation;  
thus the tests in this section involve the entropy (or contact) wave.  
We present and compare results obtained with the HLL and HLLC Riemann solvers, and investigate the sensitivity of the results to the flow Mach number $\mbox{Ma}=|\vect{v}|/c_{s}$.  
It is reasonable to expect that the results obtained with the HLL solver are sensitive to $\mbox{Ma}$, since this Riemann solver considers only the acoustic waves in the Riemann fan, and ignores the entropy wave.  
This is especially true for highly subsonic flows where the entropy wave is clearly separated from the acoustic waves in the Riemann fan 
Periodic boundary conditions are used in all the tests presented in this section.  
We show 1D and 2D results.  

In the 1D tests the computational domain is confined to $x\in[0,1]$.  
The initial density is set to $\rho=\rho_{0}+\delta\rho\times\sin(2\pi x/L_{x})$, with $\rho_{0}=1.0$ and $\delta\rho=0.1$, and the only nonzero velocity component is $v_{x}=1$.  
The (constant) pressure is parametrized by the Mach number $p=\rho_{0}|\vect{v}|^{2}/(\gamma\mbox{Ma}^{2})$.  
The adiabatic index is set to $\gamma=1.4$.  
The density profile is advected across the domain five times, until $t=5.0$.  

Results from the 1D advection tests with $\mbox{Ma}=0.6$ are tabulated in Table \ref{tab:convergenceResultsAdvection1DMach06}, where we list the $L_{1}$ error norms of the density $L_{1}(\rho)$, for multiple grid resolutions, at $t=1.0$ and $t=5.0$, computed with the HLL and the HLLC Riemann solvers.  
We also list the rate at which the numerical solution converges to the true solution (the initial condition in this case) as the grid resolution is increased.  
From Table \ref{tab:convergenceResultsAdvection1DMach06} we see that the numerical scheme is second-order accurate for this test, both with the HLL and the HLLC Riemann solvers.  
The $L_{1}$ error norms are somewhat smaller (about $20\%$ for $N_{x}=16$) when the HLLC solver is used, but the rate of convergence is slightly higher with the HLL solver, and the difference is reduced to a few percent for the highest resolution runs.  

Results from the 1D advection test with $\mbox{Ma}=0.1$ are tabulated in Table \ref{tab:convergenceResultsAdvection1DMach01}.  
The difference between the HLL and HLLC results becomes more pronounced when the Mach number is reduced.  
Both Riemann solvers result in second order convergence of the $L_{1}$ error norm when $\mbox{Ma}=0.1$, but the errors obtained with the HLL solver are significantly larger---about a factor of two compared to the corresponding results listed in Table \ref{tab:convergenceResultsAdvection1DMach06}.  
Errors with the HLLC results are somewhat smaller than those in Table \ref{tab:convergenceResultsAdvection1DMach06}.  
In fact, the HLLC solver becomes exact for this test when the advection velocity is zero, as demonstrated with the isolated contact discontinuity Riemann problem in Section \ref{subsubsec:RiemannProblem}.  

In the 2D advection tests the sine wave propagates parallel to $\vect{k}=({2\pi}/{L_{x}})\hat{\vect{x}}+({2\pi}/{L_{y}})\hat{\vect{y}}$,  
that is, with an angle $\alpha=\tan^{-1}(L_{x}/L_{y})$ with respect to the $x$-axis.  
The computational domain is now restricted to $[x,y]\in[0,L_{x}]\times[1,L_{y}]$, the initial density is set to $\rho=\rho_{0}+\delta\rho\times\sin(2\pi [x/L_{x}+y/L_{y}])$, and the velocity vector is $\vect{v}=[v_{x},v_{y},v_{z}]=[\cos\alpha,\sin\alpha,0]=[L_{x}^{-1},L_{y}^{-1},0]/\left(L_{x}^{-2}+L_{y}^{-2}\right)^{1/2}$.  
With $L_{x}=2L_{y}$ and $L_{y}=\sqrt{5}/2$ the propagation angle is $\alpha\approx63.4^{\circ}$, and the sine wave returns to its initial position for $t=1.0$.  
We set $\mbox{Ma}=0.6$ and evolve until $t=5.0$.  
All other parameters are the same as in the 1D tests.

Results from the 2D advection tests at $t=1$ and $t=5$, computed with the HLL and HLLC Riemann solvers for different grid resolutions, are tabulated in Table \ref{tab:convergenceResultsAdvection2DMach06}.  
The results reveal a trend similar to that seen in the 1D tests: with both Riemann solvers, the numerical solution eventually converges to the analytic solution at the expected second-order rate.  
The errors decrease at a somewhat faster rate when the HLL solver is used, but the $L_{1}$ error norms obtained with the HLLC solver are smaller than those obtained with the HLL solver (about $7\%$ smaller at $t=5$ with $N_{x}=512$).  

\subsubsection{Linear Fluid Waves}
\label{sec:linearWaves}

Our linear wave tests are similar to those in \citet{Stone2008}. 
We initialize a 1D periodic domain $x\in[0,1]$ with a background state $\rho_{0}=1$, $p_{0}=3/5$.  
The adiabatic index is set to $\gamma=5/3$, so that the background sound speed is $c_{s,0}=1$.  
The background medium is at rest for the sound wave test, while for the shear wave tests we set $\vect{v}\cdot\hat{\vect{x}}=0.5$ (i.e., $\mbox{Ma}=0.5$).  
We initialize sound waves by setting $[\rho,v_{x},v_{y},v_{z},p]=[\rho_{0},0,0,0,p_{0}]+A\,\vect{R}_{\pm c_{s}}^{T}\,\sin(2\pi x/L_{x})$, while shear waves are initialized by setting $[\rho,v_{x},v_{y},v_{z},p]=[\rho_{0},0.5,0,0,p_{0}]+A\,(\vect{R}_{v_{y}}^{T}+\vect{R}_{v_{z}}^{T})\,\sin(2\pi x/L_{x})$.  
The amplitude of the linear waves is set to $A=10^{-6}$.  
When initializing the different wave types, $\vect{R}_{\pm c_{s}}^{T}=[c_{s,0}^{-2},\pm c_{s,0}^{-1}\rho_{0}^{-1},0,0,1]$, $\vect{R}_{v_{y}}^{T}=[0,0,1,0,0]$, and $\vect{R}_{v_{z}}^{T}=[0,0,0,1,0]$ are right eigenvectors obtained from the quasilinear form of the Euler equations in primitive variables, and are associated with left ($-$) and right ($+$) propagating sound waves, and shear waves associated with the $y$- and $z$-components of the velocity, respectively.  
We let the waves propagate a distance $L_{x}$ (until $t=1.0$ for sound waves and $t=2.0$ for shear waves) and compute the $L_{1}$ error norm.  

\begin{figure}
  \epsscale{1.0}
  \plottwo{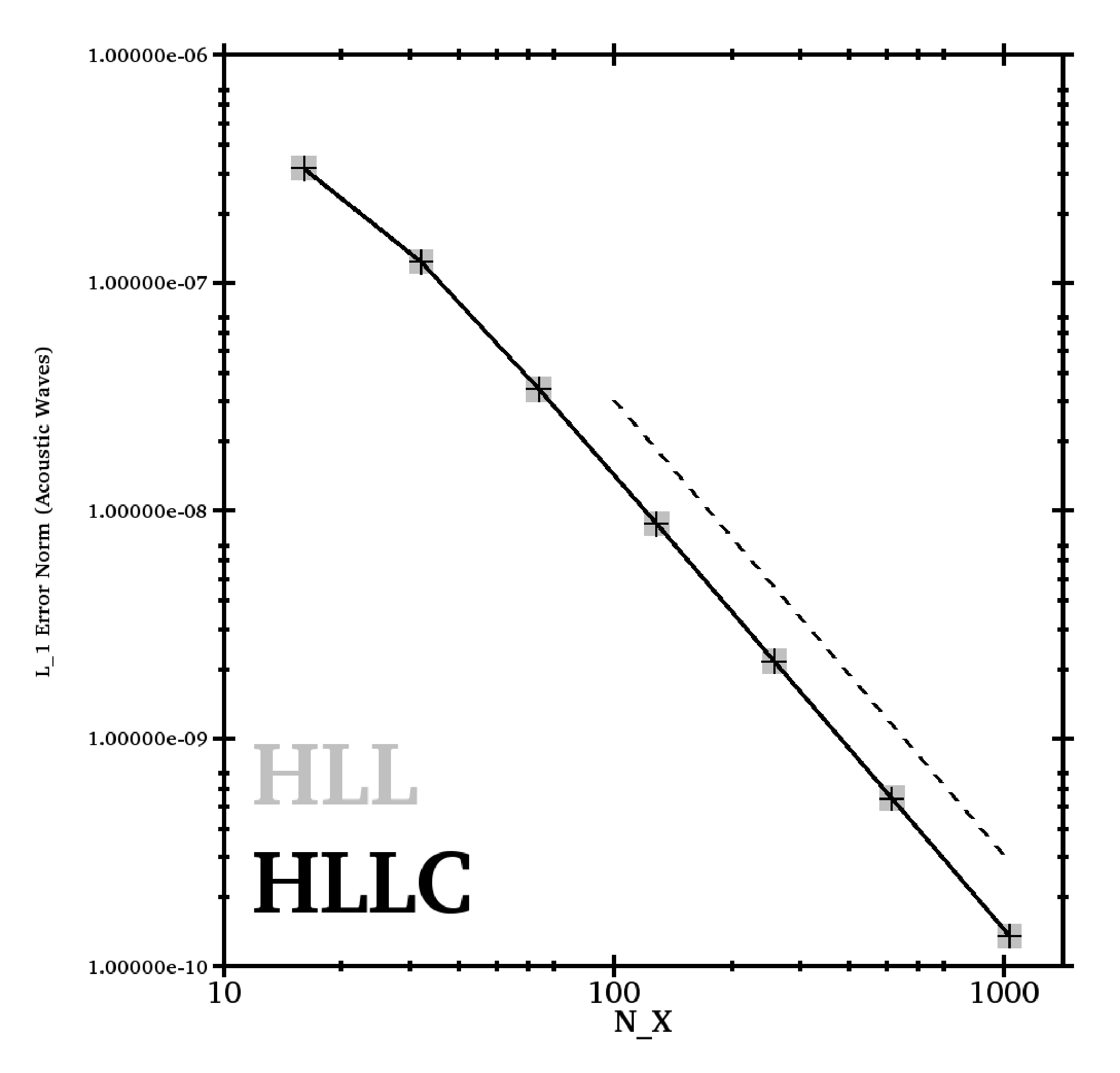}
          {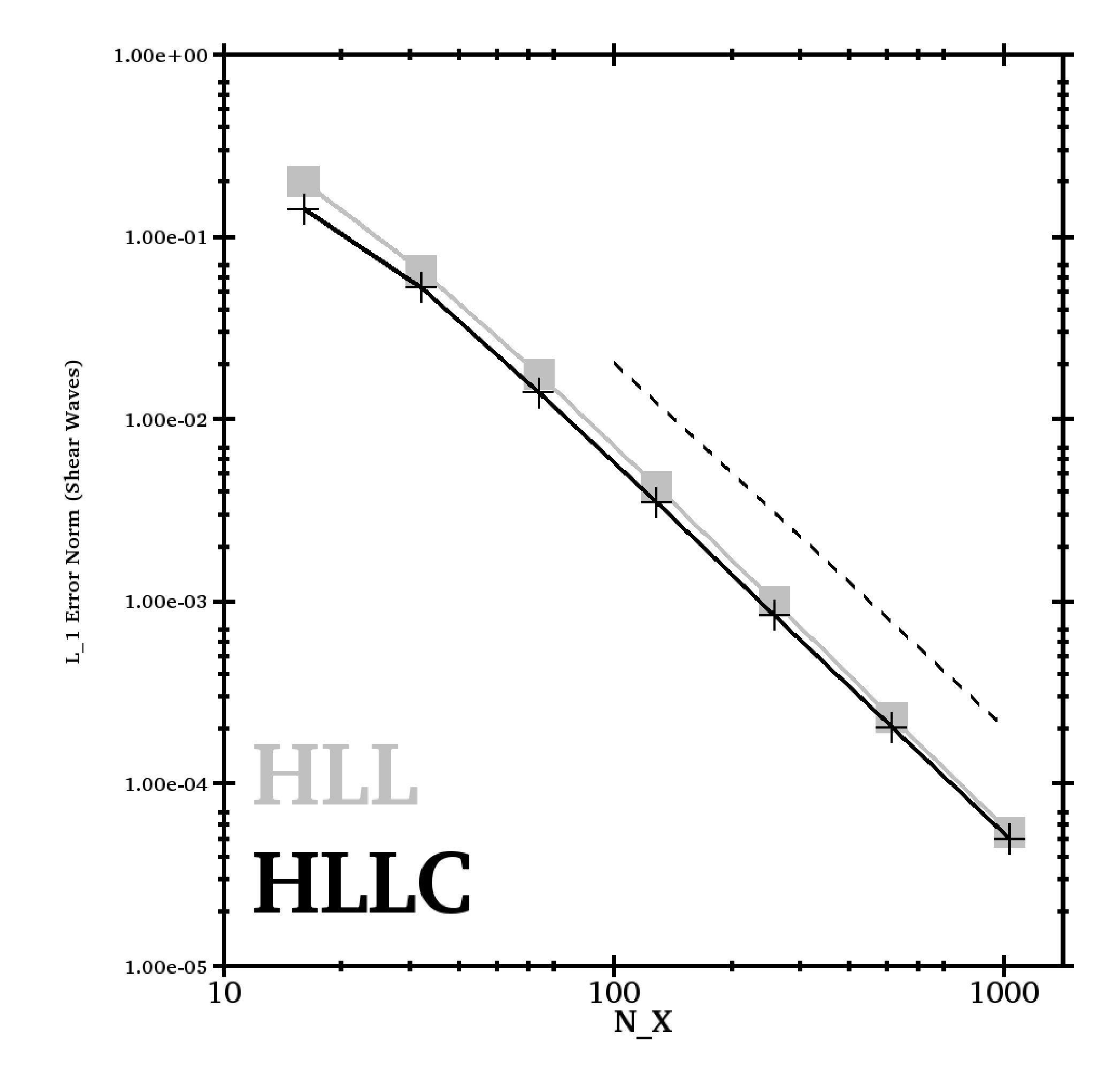}
  \caption{Relative $L_{1}$ error norms versus number of grid cells $N_{x}$ from the 1D linear wave tests.  
  The results are obtained with the HLL (grey) and HLLC (black) Riemann solvers in \genasis.  
  Error norms from the acoustic (left panel) and shear (right panel) wave tests decrease with the expected second-order rate.  
  The dashed reference lines are proportional to $N_{x}^{-2}$.}
  \label{fig:linearWavesConvergence}
\end{figure}

Results from the convergence tests are displayed in Figure \ref{fig:linearWavesConvergence}, where we plot the $L_{1}$-error norm versus $N_{x}$ for acoustic (left panel) and shear (right panel) waves.  
Results obtained with the HLL and HLLC Riemann solvers are shown in grey and black, respectively.  
We obtain second-order convergence in these tests.  
For sound waves, the results obtained with the HLL and HLLC solvers are identical, while the errors obtained with the HLLC solver are somewhat smaller for the shear wave tests (about $30\%$ for $N_{x}=16$ and about $8\%$ for $N_{x}=1024$).

\subsection{Discontinuous Fluid Tests}

Here we test the ability of \genasis\ to handle shocks and other discontinuities, which are ubiquitous in astrophysical flows.  
Finite-volume methods based on the integral formulation in Equation (\ref{eq:BalanceEquationAverage}) are particularly well suited for flows that may develop discontinuities \citep[e.g.][]{LeVeque2002}.  

\subsubsection{Riemann Problem}
\label{subsubsec:RiemannProblem}

A Riemann problem involves 
a set of piecewise constant initial data separated into left ($L$) and right ($R$) states by an initial discontinuity.  
The solution for $t>0$ typically consists of a finite set of waves propagating away from the initial location of the discontinuity.  
The adiabatic index is set to $\gamma=1.4$ in all the Riemann problems presented here.
In the 1D tests the initial discontinuity is located at $x=0.5$.  

\paragraph{Contact Discontinuity}

A 1D Riemann problem involving an isolated contact discontinuity illustrates an advantage of using the the HLLC Riemann solver as opposed to the HLL Riemann solver.  
Here we present results from the stationary and slowly moving contact discontinuity tests \citep[cf.][]{Toro2009,Liska2003} with initial conditions (left and right states)
\begin{eqnarray}
  \left[\rho,\, v,\, p\right]_{L}
  &=&\left[1.4,\, v_{\mbox{\tiny C}},\, 1.0\right] \nonumber \\
  \left[\rho,\, v,\, p\right]_{R}
  &=&\left[1.0,\, v_{\mbox{\tiny C}},\, 1.0\right],
\end{eqnarray}
where the speed of the contact discontinuity is $v_{\mbox{\tiny C}}=0.0$ and $v_{\mbox{\tiny C}}=0.1$ for the stationary and the slowly moving contact discontinuity, respectively.  
The system is evolved until $t=2.0$, at which time the moving contact discontinuity is located at $x=0.7$.  

Results from the stationary and slowly moving contact discontinuity tests for $t=2$ are plotted in Figure~\ref{fig:contactWaves} (left and right panel, respectively).  
These tests clearly demonstrate the improved resolution of the contact discontinuity when the HLLC Riemann solver is used.  
The HLLC solver is exact for the stationary contact discontinuity.  
This is because the middle wave speed estimate given by Equation (\ref{eq:MiddleWaveSpeed}) is exact in this case, and results in zero mass flux across the discontinuity.  
The diffusive part of the HLL flux (cf. the third term on the right-hand side of Equation (\ref{eq:FluxHLL})) results in a non-zero mass flux across the contact discontinuity, even as $v_{x}$ remains zero.  
Both solvers are inexact for the moving contact discontinuity test, but the HLLC solver remains superior. 

\paragraph{Sod Shock Tube}

The Sod shock tube is a well-known Riemann problem with an analytic solution.  
It was introduced by \cite{SOD1978} to benchmark algorithms for solving the Euler equations.  
The problem is initialized with left and right states given by
\begin{eqnarray}
  \left[\rho,\, v,\, P \right]_{L} &=& \left[1.0,\, 0.0,\, 1.0 \right], \nonumber \\
  \left[\rho,\, v,\, P \right]_{R} &=& \left[0.1,\, 0.0,\, 0.125\right].  
\end{eqnarray}
For $t>0$, nonlinear waves are generated and propagate away from the initial discontinuity.  
A shock wave propagates to the right and a rarefaction wave propagates to the left.  
Also, a contact discontinuity propagates to the right, between the shock and the rarefaction waves.  

Figure \ref{fig:SodShockTube_1D} shows results from this test problem for $t=0.25$.  
\genasis\ captures all the essential features of this Riemann problem with good accuracy.  
In particular, we have compared the numerical results with the analytic solution using the $L_{1}$ error norm:  Table \ref{table:SodShockTube_1D_Convergence} shows the $L_1$ error norm and the convergence rate (for mass density and pressure) obtained with both the HLL and the HLLC Riemann solvers. 
The errors obtained when using the HLLC Riemann solver are slightly smaller, especially in the mass density.  
This is because the HLLC solver better resolves the contact discontinuity, across which only the density varies.  
However, both solvers result in similar first-order convergence rate, which is expected for problems involving discontinuities.  
(We also ran this problem in two- and three-dimensional mode by letting the waves propagate along the $y$- and $z$-coordinate directions, and the results from those runs are exactly the same as the results listed in Table~\ref{table:SodShockTube_1D_Convergence}.)  

We have also used a three-dimensional version of the Sod shock tube to study the parallel scaling behavior of the hydrodynamics solver in \genasis\ (cf. Algorithm \ref{alg:BalanceEquation}).  
Figure~\ref{fig:scaling} shows results from a pure MPI weak-scaling test on Titan, the Cray XT7 machine at the Oak Ridge Leadership Computing Facility.  
The hydrodynamics algorithms implemented in \genasis\ scale well up to about $10^{5}$ MPI tasks with a single-level mesh.  

\paragraph{Double Rarefaction}

The double rarefaction problem was introduced by \cite{Einfeldt1991} to test the behavior of Riemann solvers on problems where a `vacuum' is created in a region between two strong rarefaction waves propagating away from an initial discontinuity in the velocity $v_{x}$.  
In particular, the problem was designed to reveal the lack of positivity conservation (e.g. resulting in negative pressure) by Riemann solvers based on linearization \citep[e.g. the solver provided by][]{ROE1981}, while showing that the HLLE solver (which is similar to the HLL scheme in Equation (\ref{eq:FluxHLL})) is positivity conserving (in 1D), provided that certain constraints on the wave speeds $\alpha_{\pm}^{x}$ are satisfied.  

We initialize the double rarefaction problem as described by \citet{Einfeldt1991} \citep[problem 1-2-0-3; see also][]{Toro2009}.  
The left and right states of the Riemann problem are given by
\begin{eqnarray}
  \left[\rho,\, v,\, p \right]_{L} &=& \left[1.0,\, -2.0,\, 0.4 \right] \nonumber \\
  \left[\rho,\, v,\, p \right]_{R} &=& \left[1.0,\, +2.0,\, 0.4\right].  
\end{eqnarray}
Figure \ref{fig:DoubleRarefaction_1D} shows results from running the double rarefaction problem with \genasis\ to time $t=0.15$, using $128$ cells and the HLLC Riemann solver. 
A referential solution using $1000$ cells is plotted with a solid black line.  
The referential solution was obtained with an exact Riemann solver program ({\tt e1rpexf.f}) from the NUMERICA library \citep[][see also the discussion in Chapter 4 in \citealt{Toro2009}]{Toro1999}.

The result shows that \genasis\ follows the solution obtained with the exact Riemann solver well.  
Moreover, the density and pressure remain positive.  
However, the specific internal energy $e/\rho$ shows a pathology around the location of the initial discontinuity $x=0.5$.  
(Both the density and pressure approach zero, while the specific internal energy remains finite in this problem.)  
This pathology is likely due to our use of the conservative formulation of the Euler equations (i.e., we evolve the total energy density) in a situation where the kinetic energy density is significantly larger (a factor of two initially) than the internal energy density \citep[see also results from this test (test 2) in][obtained with multiple Eulerian schemes for hydrodynamics]{Liska2003}, which can result in an inaccurate internal energy density when it is obtained by subtracting the kinetic energy density from the total (or balanced) energy density $G$ \citep[cf. Equation (\ref{eq:balancedEnergy});][]{Blondin1993}.  
For the specific internal energy, we also plot results from runs using $256$ cells (blue dashed line) and $512$ cells (green dotted line).  
The results converge to the exact solution, albeit very slowly near the center.

\paragraph{Implosion}

This is a 2D Riemann problem with initial conditions similar to the Sod shock tube.  
It was used by \citet{Liska2003} to compare several numerical schemes for solving the Euler equations \citep[see also][]{Stone2008}.  
The problem is solved on a square computational domain confined to $[x,y]\in[0,0.3]\times[0,0.3]$, with reflecting boundary conditions everywhere.  
The fluid is initially at rest, and the density and pressure are set to $\rho=0.125$ and $p=0.14$ in the region where $x+y\le0.15$, and to $\rho=1$ and $p=1$ elsewhere.  

Results obtained with \genasis\ using $400\times400$ cells and the HLLC Riemann solver are shown for select times ($t=0.045$ and $t=2.5$) in Figure \ref{fig:implosionProblem}, which is a color map of the pressure, with density contours and velocity vectors overlaid.  
(Figure \ref{fig:implosionProblem} can be compared directly with Figures 4.10 and 4.11 in \cite{Liska2003}, and Figure 17 in \cite{Stone2008}.)  
At $t=0.045$ a shock propagates diagonally towards the origin, and a rarefaction wave propagates in the opposite direction.  
A contact discontinuity is trailing the shock (cf. the density contours in the left panel in Figure \ref{fig:implosionProblem}).  
The evolution along the diagonal is similar to that of the Sod shock tube for $t=0.045$.  
At later times, wave-boundary and wave-wave interactions eventually result in a very complex flow structure.  

The results obtained with \genasis\ compare favorably with results obtained with other dimensionally unsplit codes \citep[e.g.][]{Liska2003,Stone2008}.  
In particular, the symmetry about the diagonal connecting $(0,0)$ and $(0.3,0.3)$ is perfectly preserved.  
There is no analytical solution to the implosion problem.  
However, \citet{Stone2008} argue that the production of a jet along the diagonal is part of the correct result for this test.  
The jet along the diagonal is clearly seen in the density contours in the right panel of Figure \ref{fig:implosionProblem}.  
The formation and subsequent evolution of the jet is very sensitive to the numerical scheme's ability to preserve the initial symmetry.  
We also find the formation and evolution of the jet to be sensitive to the scheme's ability to track the intermediate waves: the jet is absent when the HLL Riemann solver in \genasis\ is used, and the solution then resembles the results produced with the Positive scheme (LL) in \cite{Liska2003}.  
This is not surprising, as the jet is formed from vortices produced near the origin, which are later advected with the fluid velocity along the diagonal \citep{Stone2008}.

\subsubsection{Interacting Blast Waves}
\label{sec:interactingBlastWaves}

This problem, discussed in detail by \cite{Woodward1984}, involves multiple interactions between shocks, rarefaction waves, and contact discontinuities.  
It is considered an extremely difficult test for methods employing a uniform Eulerian mesh \citep[][]{Woodward1984}, and has been used by many authors to benchmark solvers for the Euler equations \citep[e.g.][]{Kurganov2001,Liska2003,Stone2008}.  
The problem is initialized on a computational domain confined to $x\in[0,1]$ with reflecting boundary conditions.  
The fluid is initially at rest with constant background density $\rho_{0}=1$ and pressure $p_{0}=0.01$.  
Two initial pressure jumps are introduced by setting the pressure to $p=1000$ in the region where $x<0.1$, and to $p=100$ in the region where $x>0.9$.  
For $t>0$, strong shocks, rarefactions, and contact discontinuities develop as a result of the initial pressure jumps, which later interact multiple times to create a complex flow structure.  

Figure \ref{fig:InteractingBlastWave_1D} shows results obtained with \genasis\ for $t=0.038$, using $400$ cells and the HLLC Riemann solver.  
A high-resolution reference solution, obtained by using $10^{4}$ cells, is also included in the plot (solid black line).  
The results obtained with \genasis\ are comparable to those obtained by other authors (cf. Figures 4.7 and 4.9 in \citealt{Kurganov2001}, Figure 3.10 in \citealt{Liska2003}, and Figure 9 in \citealt{Stone2008}).  
For $t=0.038$, the maximum density is about $5.4$, which is significantly lower than in the reference solution (6.4), but comparable to the results presented by \cite{Kurganov2001}, and the results obtained with many of the schemes tested by \cite{Liska2003}.  
However, our maximum density is somewhat lower than the value obtained by \citet{Stone2008}, who used third-order spatial reconstruction.  
Moreover, the contact discontinuity located around $x=0.6$ is poorly resolved, but the results obtained with \genasis\ are comparable to results obtained with other schemes based on a fixed Eulerian mesh \citep[e.g.][]{Kurganov2001,Liska2003,Stone2008}.  
Schemes based on a moving (e.g. Lagrangian) meshes perform very well on this test \cite[e.g.][]{Woodward1984,Springel2010a}.

\subsubsection{Sedov-Taylor Blast Wave}
\label{sec:SedovTaylor}

The Sedov-Taylor blast wave is a classic test in computational astrophysics.  
It has been used by many authors to benchmark multidimensional hydrodynamics algorithms \citep[e.g.][]{Fryxell2000,Almgren2010,Springel2010a,Kappeli2011}.  
It follows the self-similar evolution of a strong shock wave expanding into a uniform medium.  
We follow closely the problem setup described in \cite{Fryxell2000}, and we present results from 2D (cylindrical detonation) and 3D (spherical detonation) computations, employing both a single-level and a fixed multilevel grid.  
The problem is initialized with a fluid at rest, with uniform density $\rho_{0}=1$ and (small) pressure $p_{0}=10^{-5}$.  
The computational domain is confined to $[-0.5,0.5]$ in each coordinate dimension in these runs.  
The adiabatic index is set to $\gamma=1.4$.  
An amount of thermal energy $E_{d}=1$ is instantaneously released inside a finite detonation radius, $R_{d}$, resulting in a pressure
\begin{equation}
  p_{d}=\f{3(\gamma-1)E_{d}}{(\alpha+1)\pi R_{d}^{\alpha}}
\end{equation}
in the detonation region $r\le R_{d}$, where $\alpha=2$ and $r=\sqrt{x^{2}+y^{2}}$ for the 2D version of the test, and $\alpha=3$ and $r=\sqrt{x^{2}+y^{2}+z^{2}}$ for the 3D version.  
For $t>0$, the detonation results in the formation and expansion of a strong cylindrical (2D) or spherical (3D) shock wave.  
From dimensional arguments, the shock radius is approximately $R_{\mbox{\tiny Sh}}(t)\approx\left(E_{d}t^{2}/\rho_{0}\right)^{1/(\alpha+2)}$, and the velocity of the expanding shock wave is $\dot{R}_{\mbox{\tiny Sh}}\approx {2}{(\alpha+2)^{-1}}\left(R_{\mbox{\tiny Sh}}/t\right)$.  
At $t=0.05$ the shock has reached $r\approx0.224$ in the 2D version of the test and $r\approx0.302$ in the 3D version.  
From the shock jump conditions in Equation (\ref{eq:JumpConditions}) we find the following values for density, flow velocity, and pressure immediately behind the shock for $t=0.05$: 
$\rho_{\mbox{\tiny Sh}}^{-}\approx6.0$, $v_{\mbox{\tiny Sh}}\approx1.867$, and $p_{\mbox{\tiny Sh}}^{-}\approx4.181$ (2D), and $\rho_{\mbox{\tiny Sh}}^{-}\approx6.0$, $v_{\mbox{\tiny Sh}}\approx2.013$, and $p_{\mbox{\tiny Sh}}^{-}\approx4.864$ (3D).  

We find that the outcome of this test---in particular, the final shock position---is sensitive to the way the detonation is initiated.  
Ideally, the detonation occurs in a single point.  
However, for practical computations with a finite volume scheme employing an Eulerian Cartesian mesh, the detonation radius is typically spread out over three cells \citep{Fryxell2000}.  
To further improve the initialization of the blast wave, we subdivide each cell that is intersected by the surface of the sphere with radius $R_{d}$ into a subgrid with 20 cells in each coordinate direction \citep{Almgren2010}.  
The pressure in the intersected cells is then obtained from a volume average over the subgrid.  

Results from 2D and 3D single-level mesh calculations using the HLL Riemann solver are displayed in Figures \ref{fig:SedovTaylor2D} and \ref{fig:SedovTaylor3D}.  
(Results obtained with the HLL and the HLLC Riemann solvers are nearly identical for this test, which primarily involves an expanding strong shock, so that the contact wave captured by the HLLC solver plays a minimal role.)  
Scatter plots of the density, the velocity magnitude, and the pressure versus radius, obtained by using $256$ cells per dimension, are displayed in the two upper panels and in the lower left panel in each figure, respectively.  
Despite the fact that our results are obtained by using a single-level mesh, they compare reasonably well with the results presented in \cite{Fryxell2000}, which were obtained with an adaptively refinable mesh.  
Most noticeably, the peak values of the density and pressure just behind the shock are somewhat lower than the analytic solution, but the agreement improves with increasing spatial resolution (cf. lower right panel in Figure \ref{fig:SedovTaylor2D}).  
The shock position (relative to the analytic solution) also improves with increasing spatial resolution.  
The shock position in the 3D run overshoots the analytic value by about $3\%$ at $t=0.05$, but the agreement between the analytic and numerical results seems to improve for later times (cf. Figure \ref{fig:SedovTaylor3D_FMR}).  
The color plot of the pressure in the $xy$-plane ($z=0$) from the 3D run with $256^{3}$ cells (lower right panel in Figure \ref{fig:SedovTaylor3D}) illustrates the spherical shape of the shock surface at $t=0.05$; i.e., `grid imprints' in the shape of the shock, due to our use of Cartesian coordinates, are minimal.  

Figures~\ref{fig:SedovTaylor2D_FMR}-\ref{fig:SedovTaylor3D_FMR} display results from the Sedov-Taylor blast wave test obtained with the fixed multilevel grid capabilities in \genasis.  
The 2D results (employing seven mesh levels) are shown in Figures~\ref{fig:SedovTaylor2D_FMR} and \ref{fig:SedovTaylor2D_FMR_2}, and the 3D results (employing four mesh levels) are shown in Figure \ref{fig:SedovTaylor3D_FMR}.  
For both tests, the multilevel mesh results are compared with corresponding single-level mesh results.  
The multilevel meshes consist of concentric nested spheres (embedded in a square or cubic box at the coarsest level) with increasing resolution towards the origin.  
The spatial resolution of adjacent levels differs by a factor of two.  
This type of mesh is well suited for problems with a centrally condensed matter distribution, and we intend to employ a similar multilevel mesh in future initial simulations of core-collapse supernovae with \genasis.  
(The current multilevel mesh capabilities in \genasis\ only allow for evolution on a static mesh, but we intend to implement adaptive mesh refinement capabilities in the future.)  
For a `fair' comparison, the single-level mesh results are obtained with a resolution equal to the resolution of the deepest level in the multilevel mesh (i.e., $1024^{2}$ cells in 2D and $128^{3}$ cells in 3D).  
Of course, the multilevel mesh results become poorly resolved after the shock has crossed multiple fine-to-coarse mesh boundaries.  
The single-level mesh results are only included to show that the multilevel mesh results look reasonable in comparison (e.g. the shock positions are similar).  
Besides showing reasonable comparison with the single-level mesh results, the main purpose of including these tests is to demonstrate that the multilevel time integration algorithm, with subcycling of deeper levels and conservative flux updates across coarse-fine mesh boundaries (cf. Algorithms \ref{alg:Evolve}-\ref{alg:StepLevel}), work as intended.  

The shock has crossed one level boundary in the 2D run displayed in the left panel in Figure \ref{fig:SedovTaylor2D_FMR}.  
The shock radius is very similar in the two runs.  
(In the analytic solution, $R_{\mbox{\tiny Sh}}\approx0.1$.)  
The maximum pressure (i.e., just behind the shock) is $\sim17$ in the multilevel mesh run, and $\sim18.5$ in the single-level mesh run (versus $p_{\mbox{\tiny Sh}}^{-}\approx20.8$ in the analytic solution).  
Later ($t=0.04$), when the shock has crossed three fine-to-coarse level boundaries, the shock positions are still reasonably similar in the two runs (considering the factor of $8$ difference in spatial resolution), with $R_{\mbox{\tiny Sh}}\sim0.2$ ($R_{\mbox{\tiny Sh}}\approx0.2$ in the analytic solution).  
Just behind the shock, the pressure is $\sim3.7$ in the multilevel mesh run, and about $4.8$ in the single-level mesh run ($p_{\mbox{\tiny Sh}}^{-}\approx5.2$ in the analytic solution for $t=0.04$).  
The comparison between the multilevel and single-level mesh 3D runs in Figure \ref{fig:SedovTaylor3D_FMR} also indicates reasonable agreement (considering that the shock in the multilevel mesh run has reached level 2, where the effective resolution is only $32^{3}$, for $t=0.1$).  
Moreover, the total mass, linear momentum, and energy in the multilevel mesh runs are conserved to numerical precision.  

We have also run a larger-scale 3D Sedov-Taylor blast wave simulation, visualized in Figure~\ref{fig:Sedov_3D_Big} and in an animation available online. 
The multilevel mesh has seven levels, with six spherical meshes nested inside the coarsest rectangular mesh in a manner like that in Figure~\ref{fig:Chart2D} (and episodically visible in the online animation). 
The coarsest level has $128^3$ cells of width $\sim 7.8\times 10^{-3}$ in a domain spanning $[-0.5, 0.5]$ in each dimension.
The cells at the finest level have a width of $\sim 1.2\times 10^{-4}$, filling a domain of radius $\sim 7.8\times 10^{-3}$---the size of a single cell at the coarsest level.
(A single level mesh covering the outermost domain $[-0.5, 0.5]$ in each dimension with this resolution would have $8192^3$ cells---a problem size that would be truly heroic, if not prohibitive, even on machines the size of Titan at the the Oak Ridge Leadership Computing Facility.)
The visualization boxes in the two panels in a given row of Figure~\ref{fig:Sedov_3D_Big} have the same spatial scale;
this scale changes by a factor of four between each successive row.
Cross sections of the pressure, through the origin, are projected onto the rear walls of the visualization box in each dimension. 
The grey sphere is a contour plot of the shock surface.
Barely visible in the upper left panel at $t=0$ is a tiny spherical region of radius $\sim 8.5\times 10^{-4}$ with pressure $\sim 7.0\times 10^5$. 
At $t \sim 3.5\times 10^{-3}$ the maximum pressure immediately behind the shock is down to $\sim 5.1$ (upper right and middle left panels, visualized at different spatial scales).
At $t \sim 0.11$ the maximum pressure immediately behind the shock is down to $\sim 0.08$ (middle right and lower left panels, visualized at different spatial scales).
The final maximum pressure is $\sim 1.1\times 10^{-3}$ when the simulation ends at $t\sim 3.9$ (lower right panel).
This blast wave simulation, in which large dynamic ranges in time, space, and pressure (or energy density) have been handled successfully with modest resources (512 MPI processes), suggests that the multilevel hydrodynamics solver in GenASiS will be well suited to the hydrodynamics needs of core-collapse supernova simulations.

The weak scaling behavior of the hydrodynamics evolution on a centrally refined mesh with five levels is shown in Figure~\ref{fig:MultilevelScaling}.
The data points are for coarsest-level resolutions of $64^3$, $128^3$, and $256^3$, corresponding to effective finest-level resolutions of $1024^3$, $2048^3$, and $4096^3$.
The fact that the communications involved in the prolongations and restrictions of the multilevel evolution are not overlapped with work clearly has an impact on the parallel efficiency. 
Moreover, if one compares the time per step per cell of multilevel vs. single level simulations, there is significant overhead, and we observe numbers in the same ballpark as \citet{Teyssier2002}.
Work to reduce such overhead is worthwhile to be sure, and we will continue to investigate improvements.
But we caution that undue fixation on per-cell overhead risks obscuring a larger and more important practical point: despite the overhead, simulations of high effective resolution and large dynamic ranges, that otherwise would be out of reach (or require drastically more resources)---such as the multilevel Sedov run in the previous paragraph---become much more accessible thanks to mesh refinement and multilevel evolution.

\subsection{Fluid Instability Tests}
\label{sec:FluidInstabilityTests}


Finally, we demonstrate how two instabilities of astrophysical relevance---the Kelvin-Helmholtz and Rayleigh-Taylor instabilities---grow and develop in \genasis\ simulations.

\subsubsection{Kelvin-Helmholtz}
\label{sec:kelvinHelmholtz}

The Kelvin-Helmholtz (KH) instability is relevant to a broad range of astrophysical applications.  
For example, the shear flows induced by the spiral mode of the standing accretion shock instability \citep[SASI;][]{Blondin2003} are KH unstable.  
Energy transfer, mediated by the KH instability, from flows associated with low-order SASI modes to small-scale turbulent flows, can possibly result in the nonlinear saturation of the SASI \citep{Guilet2010,Endeve2012}.  
We include a 2D version of the KH test here to further highlight differences between simulation results obtained with the HLL and HLLC Riemann solvers in \genasis, and to compare with (and hopefully corroborate) findings reported by other authors.  
In particular, simulations of the relativistic magnetohydrodynamic KH instability \citep[e.g.][]{Mignone2009,Beckwith2011} have revealed that schemes based on approximate Riemann solvers that include the intermediate waves in the Riemann fan have significantly improved spectral resolution when compared to schemes that do not.  
Note that the inclusion of this test is \emph{not} an attempt to study in detail any aspect of the KH instability.  

Our numerical setup of the KH instability test follows closely the description detailed on the Athena website.\footnote{www.astro.princeton.edu/$\sim$jstone/Athena/tests/kh/kh.html}  
The computational domain is confined to $[x,y]\in[0,1]\times[0,1]$, with periodic boundary conditions in both spatial dimensions.  
We denote the lengths of the $x$ and $y$ sides of the computational domain $L_{x}$ and $L_{y}$ respectively.
Velocity shear layers are located at $y=0.25$ and $y=0.75$.  
The pressure is initially uniform everywhere, with $p=2.5$.  
In the region with $|y-0.5|<0.25$ we set $\rho=2$ and $v_{x}=0.5$, while in the region of the computational domain where $|y-0.5|\ge0.25$ we set $\rho=1$ and $v_{x}=-0.5$.  
For the unperturbed initial state we set $v_{y}=v_{z}=0$ everywhere.  
The adiabatic index is set to $\gamma=1.4$.  

The shear layers in the initial configuration are KH unstable to perturbations in the velocity components perpendicular to the initial flow \citep[e.g. $v_{y}$;][]{Chandrasekhar1981}.  
For our initial setup, the growth rate of a single-mode perturbation with associated wavenumber $k_{x}$ is \citep[cf. Section 101 in][]{Chandrasekhar1981}
\begin{equation}
  \Gamma_{\mbox{\tiny KH}}=k_{x}\f{\sqrt{2}}{3}\Delta_{y}v_{x}, 
\end{equation}
where $\Delta_{y}v_{x}$ is the jump in $v_{x}$ across the velocity shear layer.  
Thus the amplitude of a single mode perturbation with associated wavelength $\lambda_{x}$ grows exponentially with growth time $\Gamma_{\mbox{\tiny KH}}^{-1}\approx0.34\,(\lambda_{x}/L_{x})$;  
perturbations associated with shorter wavelengths grow at a faster rate.  
We initiate the KH instability with perturbations in the initial velocity field $\vect{v}_{0}$ by setting
\begin{equation}
  \vect{v}=\vect{v}_{0}+\delta\vect{v}, 
\end{equation}
where we use a combination of single-mode and random-mode perturbations
\begin{equation}
  \delta\vect{v}=\left[A_{\mbox{\tiny S}}\sin\left(2\pi x/L_{x}\right)+A_{\mbox{\tiny R}}\right]\times\psi(y)\,\hat{\vect{y}}.  
\end{equation}
The amplitude of the single-mode perturbation is $A_{\mbox{\tiny S}}=10^{-2}$, and $A_{\mbox{\tiny R}}(x,y)$ assumes random numbers between $-10^{-3}$ and $10^{-3}$.  
The function $\psi(y)$ concentrates the perturbations in the shear layers
\begin{equation}
  \psi(y)=\exp\left[-\f{1}{2}\left(\f{\cos\left(2\pi y/L_{y}\right)}{2\pi \sigma/L_{y}}\right)^{2}\right], 
\end{equation}
where we set $\sigma=0.1$.  
The random perturbations seed small-scale modes, whose growth may be suppressed by an excessively dissipative numerical scheme.  

We have carried out simulations using $256^{2}$ (low resolution) and $512^{2}$ (high resolution) cells, using both the HLL and HLLC Riemann solvers in \genasis.  
The simulations are evolved until $t=5.0$.  
We display the density distribution at select times in Figures~\ref{fig:kelvinHelmholtz1}-\ref{fig:kelvinHelmholtz3} ($t=0.6$, Figure~\ref{fig:kelvinHelmholtz1}; $t=1.0$, Figure \ref{fig:kelvinHelmholtz2}; $t=2.0$, Figure~\ref{fig:kelvinHelmholtz3}), which illustrate the results from the high resolution runs.  
(Results obtained with the HLL and HLLC solvers are displayed in the left and right panels, respectively.)  
The differences in the results obtained with the two Riemann solvers become apparent at an early stage.  
The run with the HLLC solver has developed small-scale `KH rolls' in the shear layer at $t=0.6$.  
These are absent in the run with the HLL solver, which seems to have only been affected by the sinusoidal part of the perturbation at this point.  
The interfaces between the low and high density fluids are clearly more diffuse in the HLL run, which is also to be expected, as this solver ignores the contact and shear waves in the Riemann fan.  
Small-scale KH rolls, although clearly affected by the more diffuse interface, have developed in the HLL run for $t=1.0$.  
The shear layers in the HLLC run are much more disturbed by the instability at this time.  
The two runs share some qualitative similarities at $t=2.0$ (i.e., the largest-scale component of the dense-fluid deformation), 
but visual comparisons reveal increasing differences
with evolving time.  
The shear layers in the HLL run are clearly populated with KH rolls of roughly equal size at $t=2.0$, while the deformations of the same layers in the HLLC run can be characterized as covering a broader spectrum of spatial scales.  

Figure~\ref{fig:kelvinHelmholtzKineticEnergy} provides additional quantitative results from the KH runs.  
In the left panel we plot the $y$-component of kinetic energy, 
\begin{equation}
  E_{\mbox{\tiny kin,y}}=\int_{V}\f{1}{2}\rho v_{y}^{2}\,dV, 
\end{equation}
versus time for the two high-resolution runs.  
The grey line is from the HLL run, while the black line is from the HLLC run.  
Note that $E_{\mbox{\tiny kin,y}}$ grows faster initially in the HLLC run than in the HLL run.  
This growth is due to the developments seeded by the small-scale random perturbations.  
In both runs, $E_{\mbox{\tiny kin,y}}$ reach similar levels at late times ($t\gtrsim2.0$), but the distribution of kinetic energy on various spatial scales remains different.  
In the right panel of Figure \ref{fig:kelvinHelmholtzKineticEnergy} we plot the spectral distribution of the kinetic energy $\hat{e}_{\mbox{\tiny kin}}(k)$ at $t=2.0$ from the low (dashed lines) and high (solid lines) resolution runs.  
(The wavenumber, $k=2\pi/\lambda_{k}$, is the magnitude of the wavevector $\vect{k}$.)  
The energy spectra are obtained from Fourier transforms of the components of $\sqrt{\rho}\vect{v}$, and satisfy
\begin{equation}
  \int_{0}^{\infty}\hat{e}_{\mbox{\tiny kin}}(k)\,dk=\int_{V}\f{1}{2}\rho |\vect{v}|^{2}\,dV
\end{equation}
\citep[cf.][]{Ryu2000}.  
Clearly, the HLLC solver results in higher spectral resolution for a given spatial resolution.  
The spectra from the high-resolution runs begin to separate already around $k=40$ (i.e., $\lambda_{k}\approx80\,\Delta x$), and the spectrum from the HLL run falls below $10^{-9}$ around $k\approx560$, while the spectrum from the HLLC run stays above $10^{-9}$ out to $k\approx920$.  
Moreover, the spectrum from the low-resolution run with the HLLC solver follows very closely the spectrum from the high-resolution run with the HLL solver.  
Thus the low-resolution run with the HLLC solver offers the same spectral resolution as the high-resolution HLL run. This is a potentially substantial computational savings, especially for 3D simulations.  

In general, the conclusions we can draw from these runs agree very well with those of other authors \citep[e.g.][]{Mignone2009,Beckwith2011}.
The inclusion of the intermediate waves in the approximate Riemann solver results in sharper contact and shear discontinuities, and unphysical suppression of the growth of KH unstable modes may be avoided, which also impacts the overall evolution of the instability---in particular its growth rate.  
Significantly improved spectral resolution of the nonlinear flows is also obtained when the HLLC solver is employed in \genasis.  

\subsubsection{Rayleigh-Taylor}

The Rayleigh-Taylor (RT) instability is another important fluid instability with relevance to astrophysical applications.  
The RT instability is extensively analyzed in \citet{Chandrasekhar1981}.  
Again, we follow closely the initial setup described on the Athena website \citep[see also][who compared results from this test computed with eight different schemes]{Liska2003}.  
The 2D computational domain is confined to $[x,y]\in[-0.25,0.25]\times[-0.75,0.75]$, and we use periodic boundary conditions at $|x|=0.25$ and reflecting boundary conditions at $|y=0.75|$.  
This test involves a uniform external force, and is the only test with nonzero sources included in this paper.  
The problem consists of a denser fluid above a less dense fluid at rest, with $\rho=\rho_{2}=2$ for $y>0$ and $\rho=\rho_{1}=1$ for $y\le0$.  
The external force acts anti-parallel to $\hat{\vect{y}}$ with a constant acceleration $g=0.1$.  
Setting $\Phi(y_{l})=0$, the corresponding potential is $\Phi=g (y-y_{l})$.  
The pressure is $p=p_{b}-\rho g y$ with $p_{b}=2.5$.  
Thus the initial configuration is in hydrostatic equilibrium $\gradient{p}+\rho\gradient{\Phi}=0$.  
The adiabatic index is set to $\gamma=1.4$.  

We specify the potential in the cell centers and compute its gradient using second-order finite differences.  
Because the analytic potential is linear in $y$, the computation of its gradient is exact.  
In particular,
\begin{equation}
  \langle\mathcal{S}\rangle
  =
  \left[
    0, 0, m\langle D\rangle, 0, \langle S^{y}\rangle
  \right]^{T}
  \times
  \left(\f{\Phi_{\yplus}-\Phi_{\yminus}}{y_{\yplus}-y_{\yminus}}\right).  
\end{equation}
are the source terms in Equation (\ref{eq:BalanceEquationAverage})---the external force and power.

In the absence of surface tension at the interface between the two fluids, the initial configuration described above is unstable to perturbations in all wavenumbers $k_{x}=2\pi/\lambda_{x}$.  
The growth rate of a single-mode perturbation is \citep[cf. \S92][]{Chandrasekhar1981}
\begin{equation}
  \Gamma_{\mbox{\tiny RT}}
  =\sqrt{gk_{x}\mbox{A}}, 
\end{equation}
where $\mbox{A}=(\rho_{2}-\rho_{1})/(\rho_{2}+\rho_{2})$ is the Atwood number, which is $1/3$ for our initial setup.  
Thus the growth time for a single-mode perturbation with wavelength $\lambda_{x}$ is $\Gamma_{\mbox{\tiny RT}}^{-1}\approx1.55\,(\lambda_{x}/L_{x})^{-1/2}$.  

We initiate the RT instability by perturbing the initial velocity field $\vect{v}=\vect{v}_{0}+\delta\vect{v}$, with $\vect{v}_{0}=0$ and a single-mode perturbation of the form
\begin{equation}
  \delta\vect{v}=\f{A_{\mbox{\tiny S}}}{4}\left[1+\cos(2\pi x/L_{x})\right]\times\left[1+\cos(3\pi y)\right]\,\hat{\vect{y}}, 
\end{equation}
with amplitude $A_{\mbox{\tiny S}}=10^{-2}$.  

Figure~\ref{fig:rayleighTaylor} displays results from the RT instability test computed with $256\times768$ cells.  
The left panel shows the density distribution at $t=8.5$ from a simulation using the HLLC Riemann solver.  
The instability has evolved well into the nonlinear stage at this time, with the characteristic rising bubble (or `mushroom cap') and falling spikes clearly displayed.  
On the top side of the bubble there are signs of emerging secondary KH instabilities at the interface between the dense and light fluids.  
In the right panel we plot contours of constant density $\rho=1.25$, comparing results obtained with the HLL (solid black) and the HLLC (dashed grey) Riemann solvers.  
The results computed with the two Riemann solvers agree qualitatively, but again the HLLC solver preserves a sharper interface between the fluids.  
This becomes particularly evident when looking at the `coil-up' of the two fluids below the mushroom cap.

\section{CONCLUSION}
\label{sec:Conclusion}

This paper is the first in a series on GenASiS ({\em Gen}eral {\em A}strophysical {\em Si}mulation {\em S}ystem), a new code ultimately aimed at state-of-the-art simulations of core-collapse supernovae and other astrophysical problems.
Distinguishing features include an object-oriented approach and cell-by-cell mesh refinement (albeit with a more explicitly level-by-level approach than in some other codes).
In this first paper we explain some concepts underlying the refinable discretized spaces on which calculations are to be performed;
document methods for compressible nonrelativistic hydrodynamics;
and benchmark the hydrodynamics capabilities of GenASiS against many standard test problems, including an example with a fixed centrally refined coordinate patch of a type suitable for gravitational collapse.

Mathematically, a continuous coordinate patch covers an individual region of a manifold. 
The class that represents a single coordinate patch in GenASiS approximates the mathematical ideal of continuity with a finite sequence of meshes which provide, as necessary, increasing refinements of the coarsest (top-level) mesh.
The refinable structure that underlies our approximate representation of a three-dimensional continuous coordinate patch is an oct-tree (or, in restricted use in two dimensions or even one dimension, a quad- or binary tree respectively) that enables cell-by-cell refinement. 
Several constructs---cell lists, submeshes, and meshes---provide increasingly comprehensive interfaces to the oct-tree structure underlying a single refinable coordinate patch.
These facilitate such tasks as the storage of physical variables associated with the oct-tree, overlapping of work and communication of ghost cell data, and the transfer of data between different levels of refinement (prolongation and restriction). 
Each level of refinement is independently domain-decomposed. 

We have developed a rather generic solver for hyperbolic balance equations, whose divergence structure naturally lends itself to a finite-volume approach naturally suited to the handling of shocks.
We have implemented second-order spatial reconstruction and the HLL and HLLC Riemann solvers, along with a second-order Runge-Kutta time stepper.
(In the immediate vicinity of a shock and when running with the HLLC option, the Riemman solver automatically switches to HLL in the directions transverse to shock propagation;
the greater diffusivity of the HLL solver successfully eliminates odd-even decoupling.) 
Our single-level solver aims for efficiency both by overlapping work and communication in a message-passing environment and by working on data stored in contiguous memory when possible.
In the context of mesh refinement, recursive, multilevel evolution works from the bottom up, with two steps at Level~$i+1$ for each step at Level~$i$. 
(After the first step at Level~$i+1$ there is a temporary step bringing Level~$i$ bringing it into synchrony with Level~$i+1$, in order that updated values for a coarse/fine boundary layer surrounding Level~$i+1$ can be obtained.) 

\begin{figure}
  \epsscale{1.0}
  \plottwo{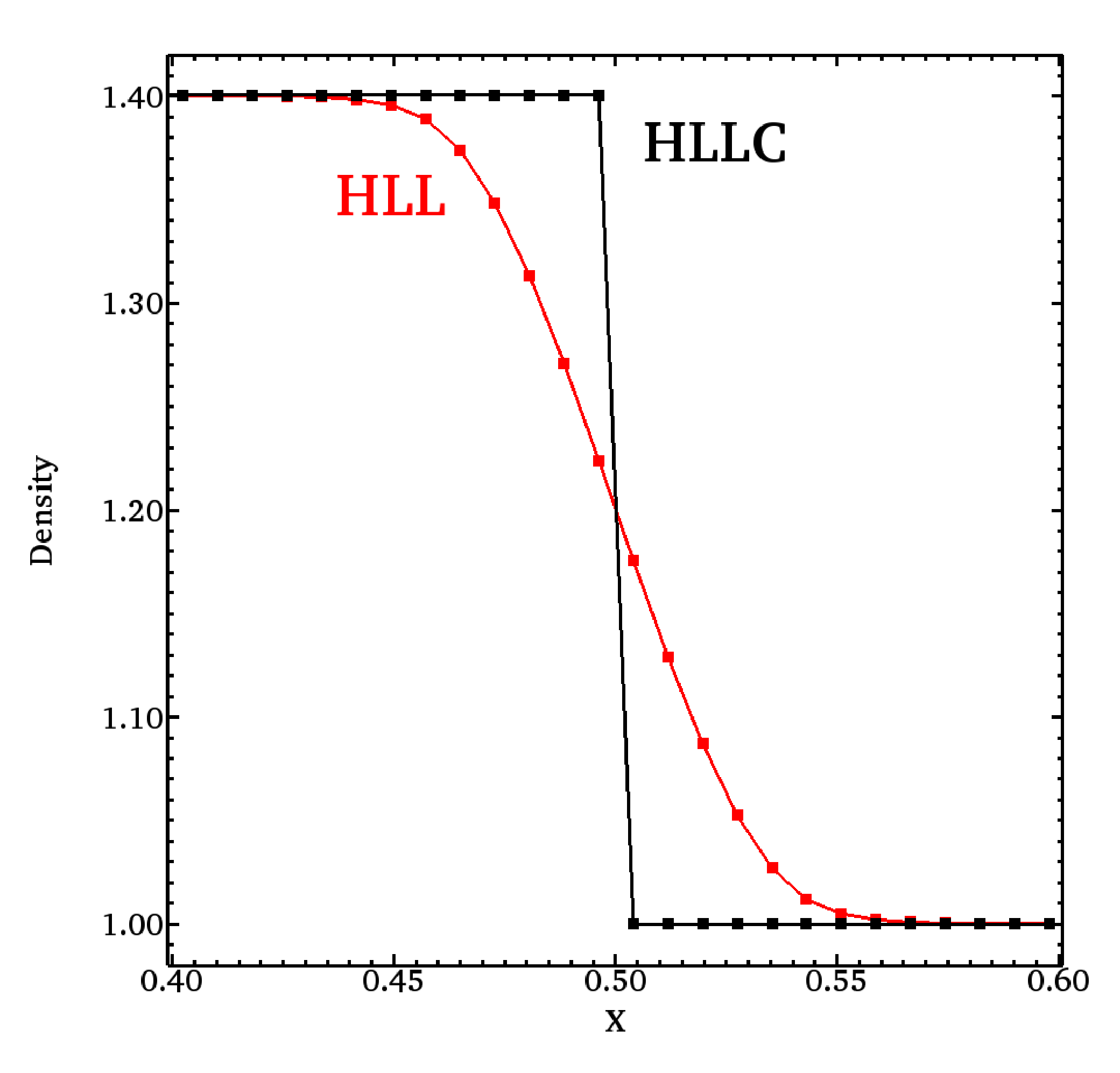}
                {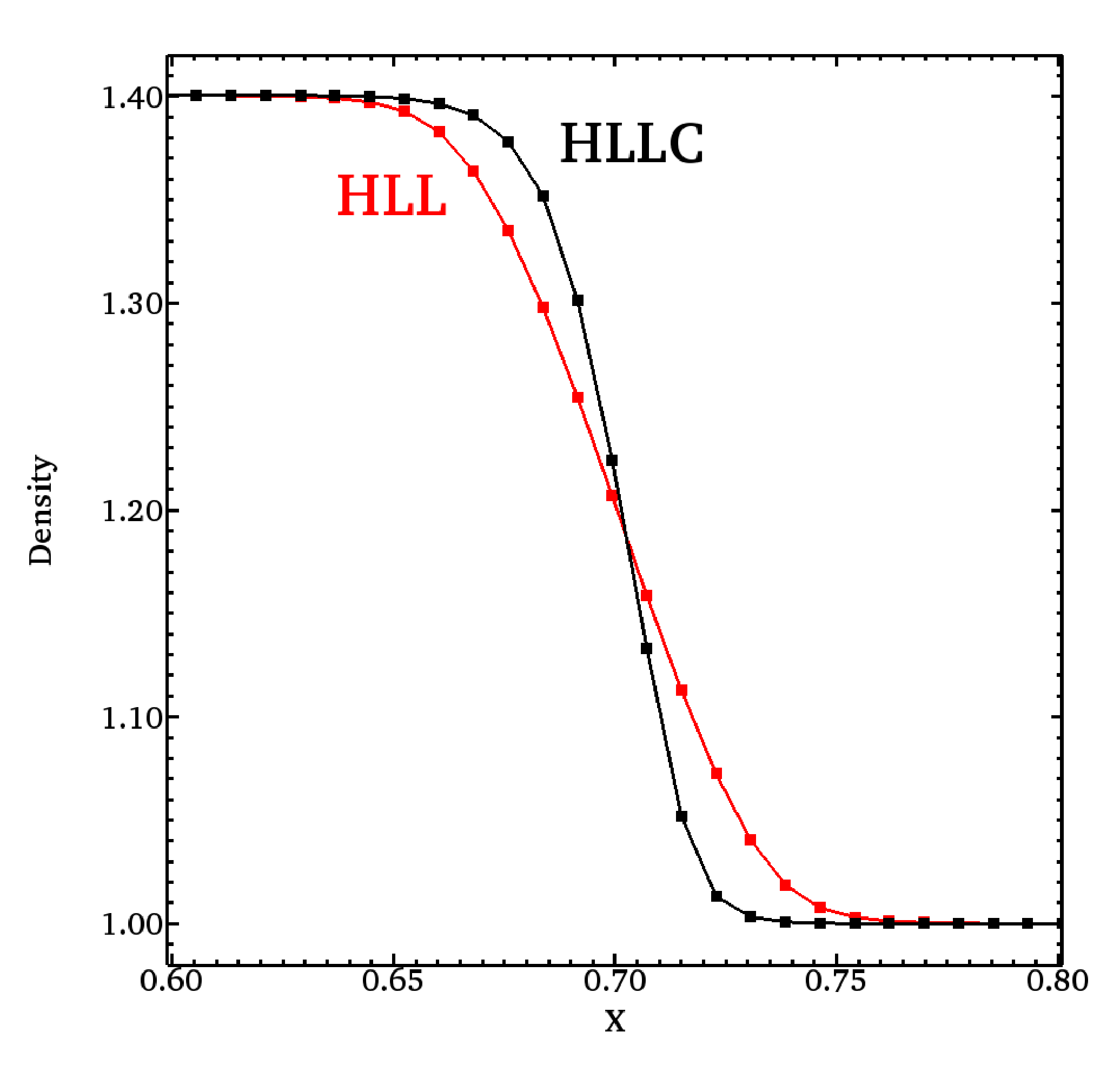}
  \caption{Results from the stationary (left) and the slowly moving (right) contact discontinuity at $t=2$, computed with a grid of $128$ zones.  
  The mass density is plotted.  
  Results computed with the HLLC Riemann solver are shown in black, and results computed with the HLL Riemann solver are shown in red.  }
  \label{fig:contactWaves}
\end{figure}

\begin{figure}
  \epsscale{1.0}
  \plottwo{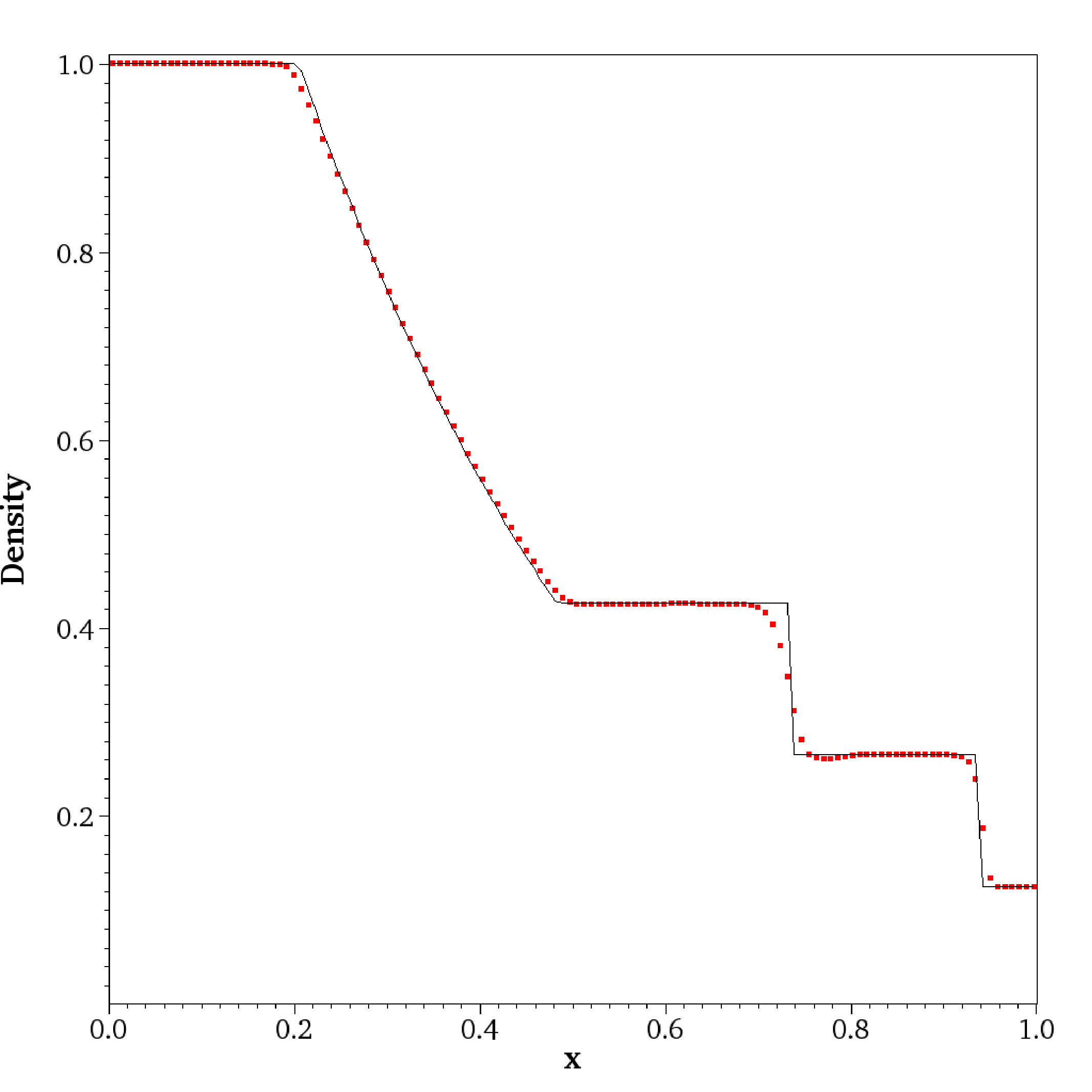}
          {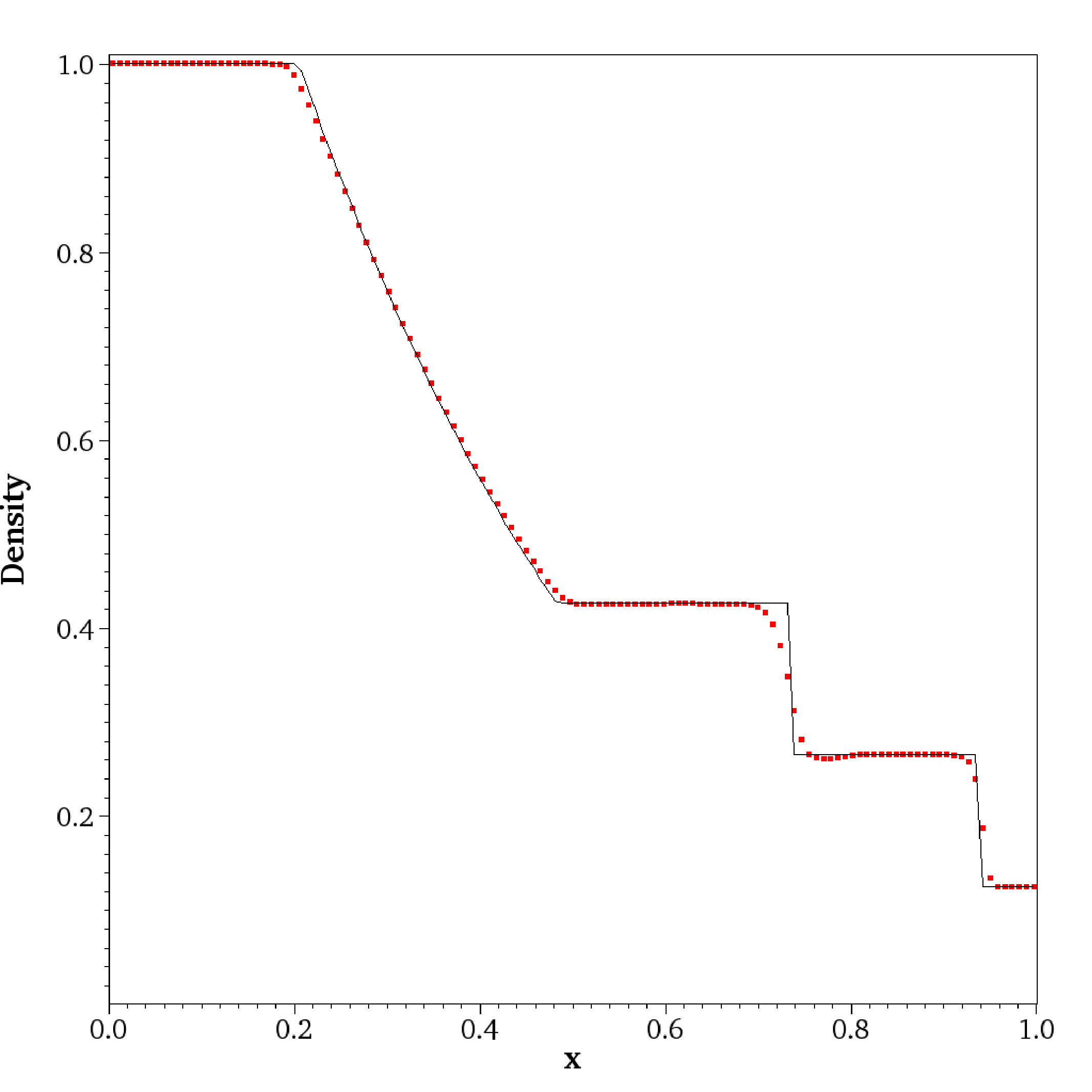} \\
  \plottwo{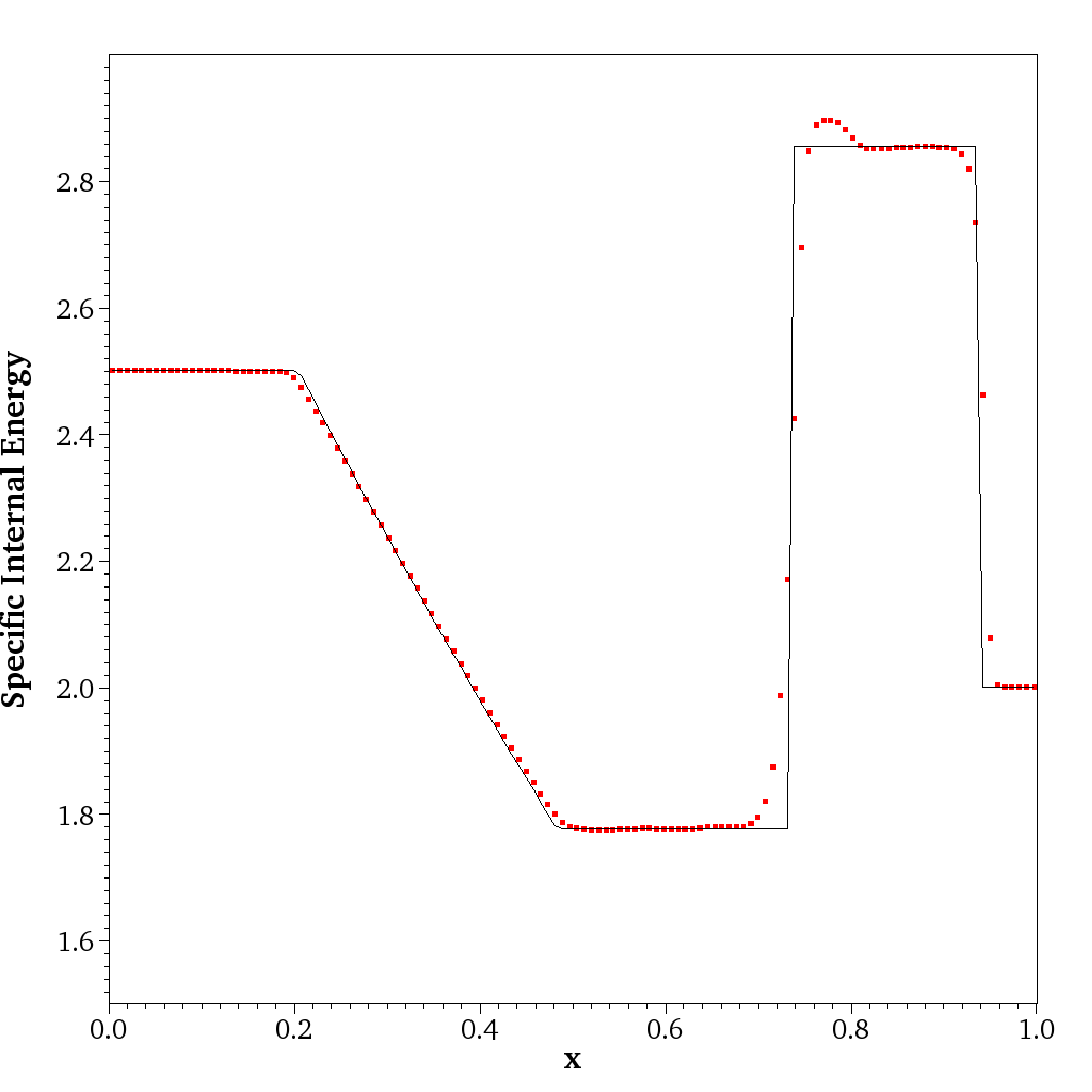}
          {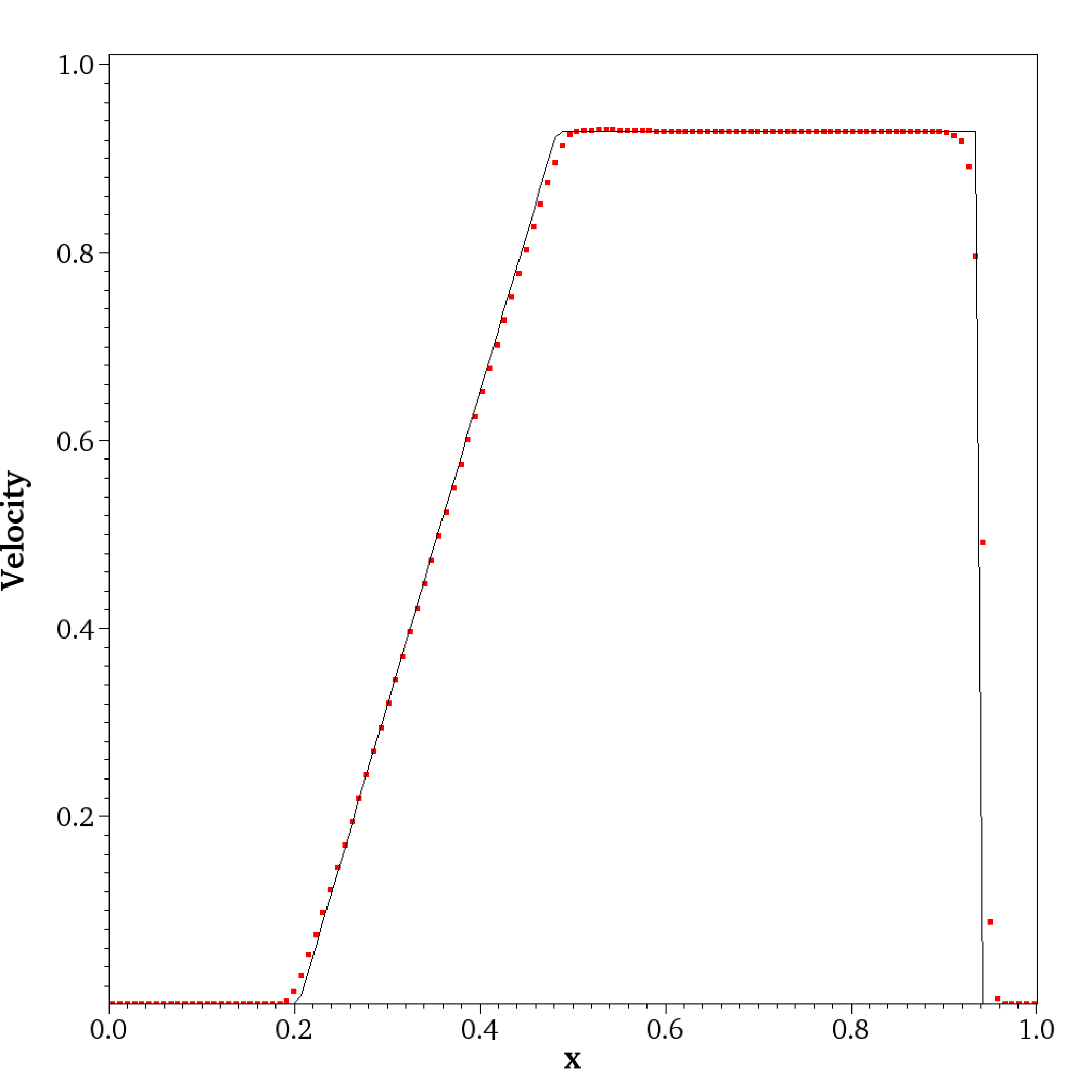}
  \caption{Results from running Sod's shocktube test at $t=0.25$, using 128 cells and the HLLC Riemann solver.  
  The mass density (left) and pressure (right) are plotted in the upper panels, and the specific internal energy ($e/\rho$; left) and the velocity ($v_{x}$; right) are plotted in the lower panels.  
  The analytic solution is shown as a black line in each panel.  }
  \label{fig:SodShockTube_1D}
\end{figure}

\begin{figure}
  \epsscale{1.0}
  \plotone{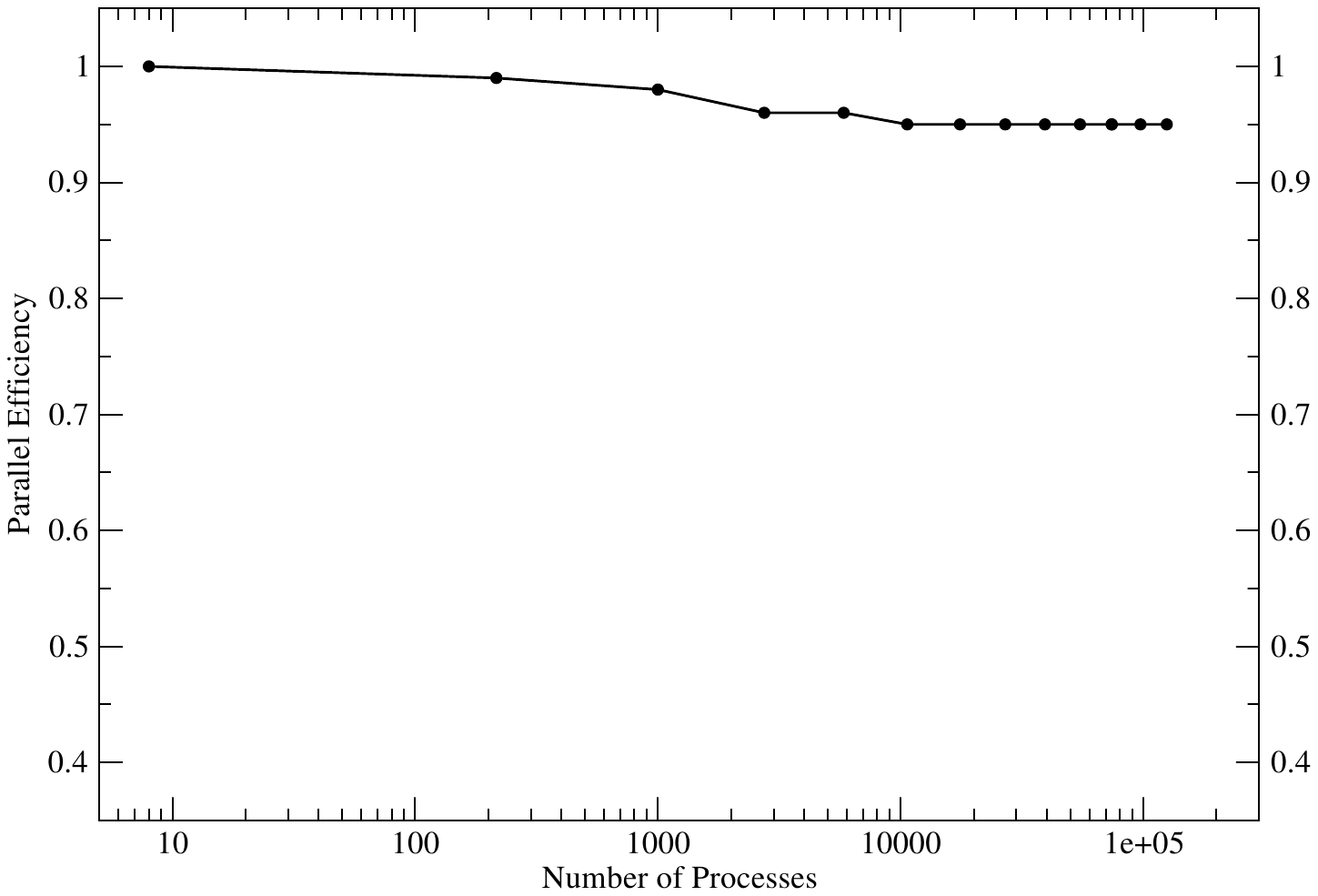}
  \caption{MPI weak scaling of the hydrodynamics algorithms in \genasis.  
  Sod's shocktube test in 3D was run on a single-level cell-by-cell mesh, keeping the number of computational cells per MPI task fixed to $48^{3}$. The wall time per time step for the first data point is 3.10~s.}
  \label{fig:scaling}
\end{figure}

\begin{figure}
  \epsscale{1.0}
  \plottwo{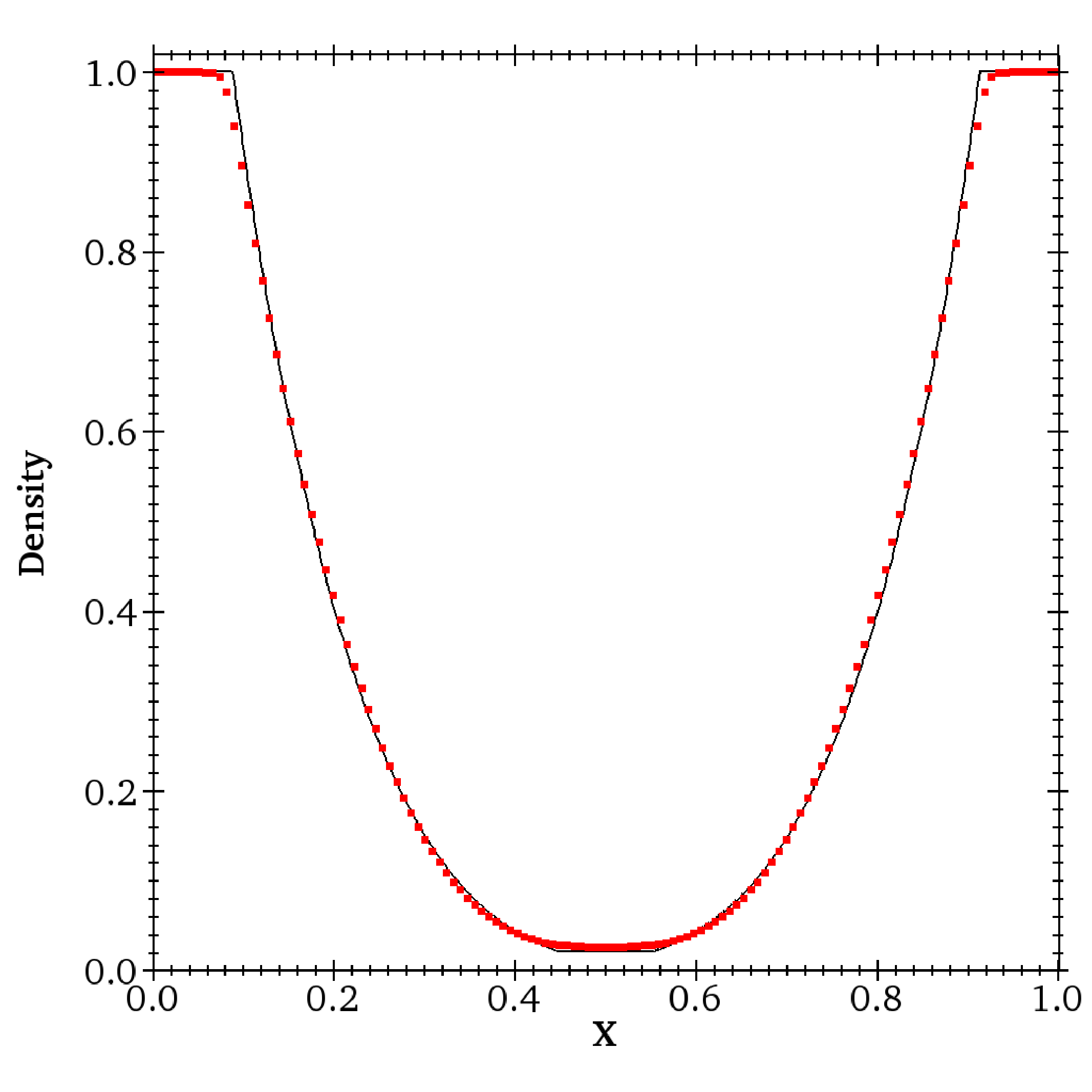}
          {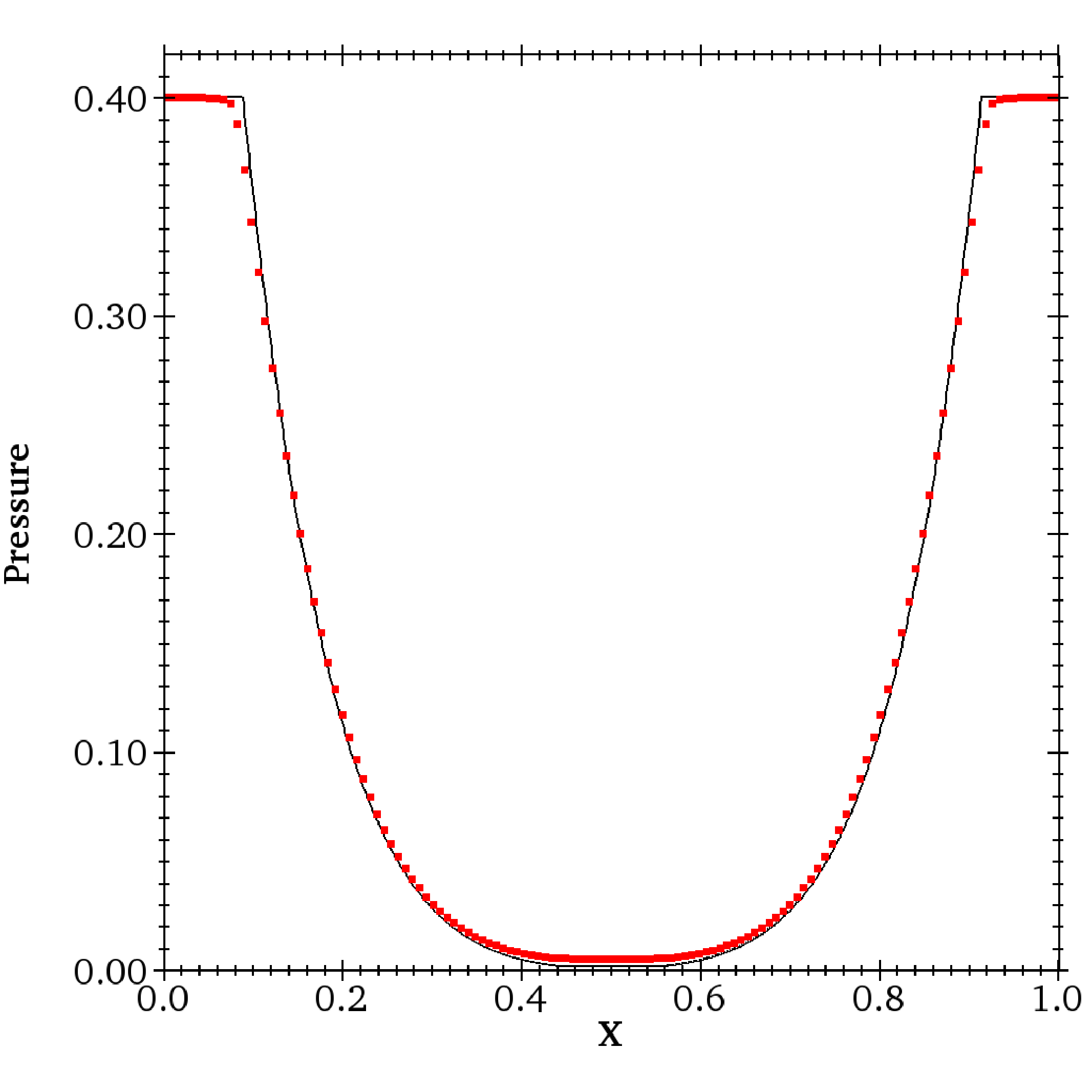} \\
  \plottwo{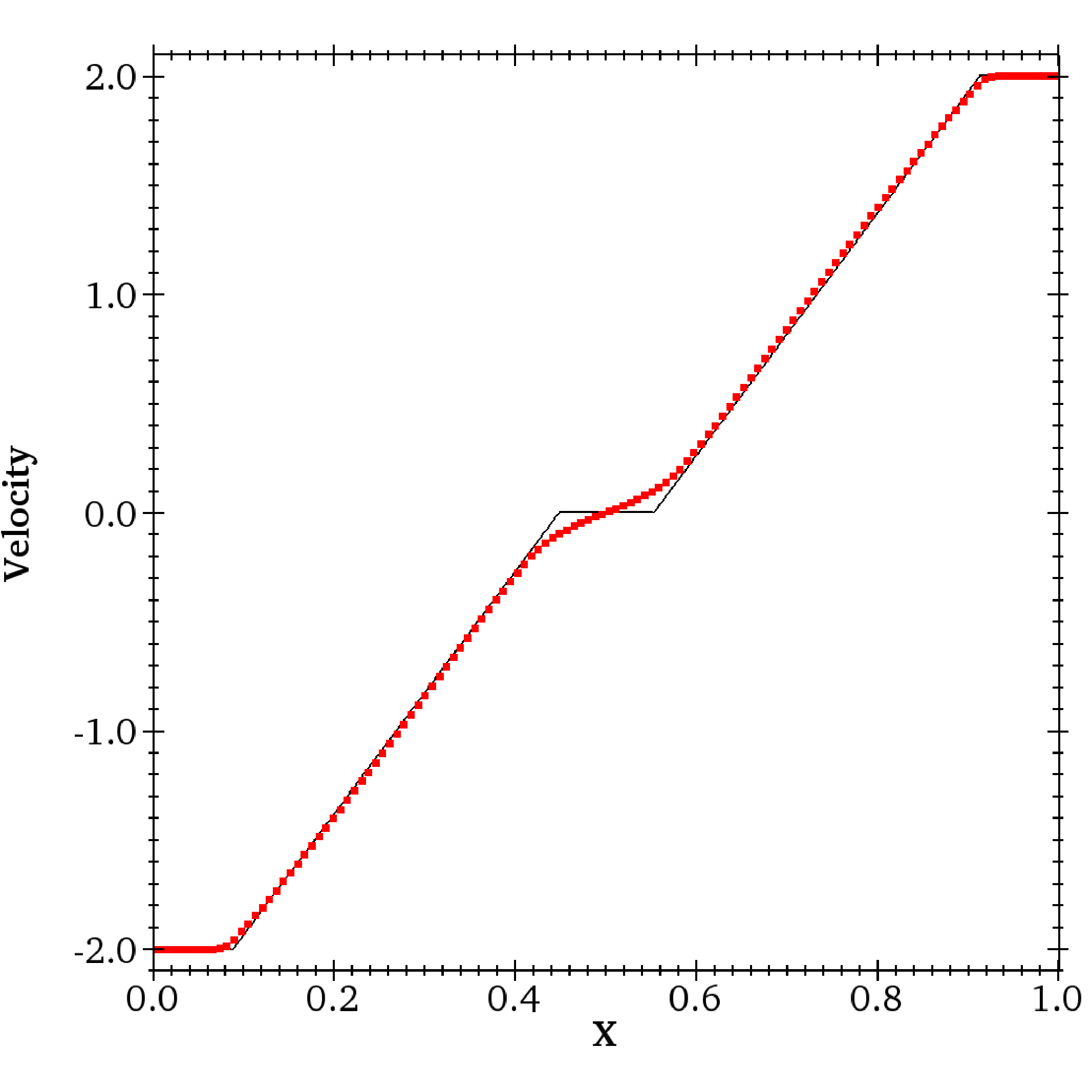}
          {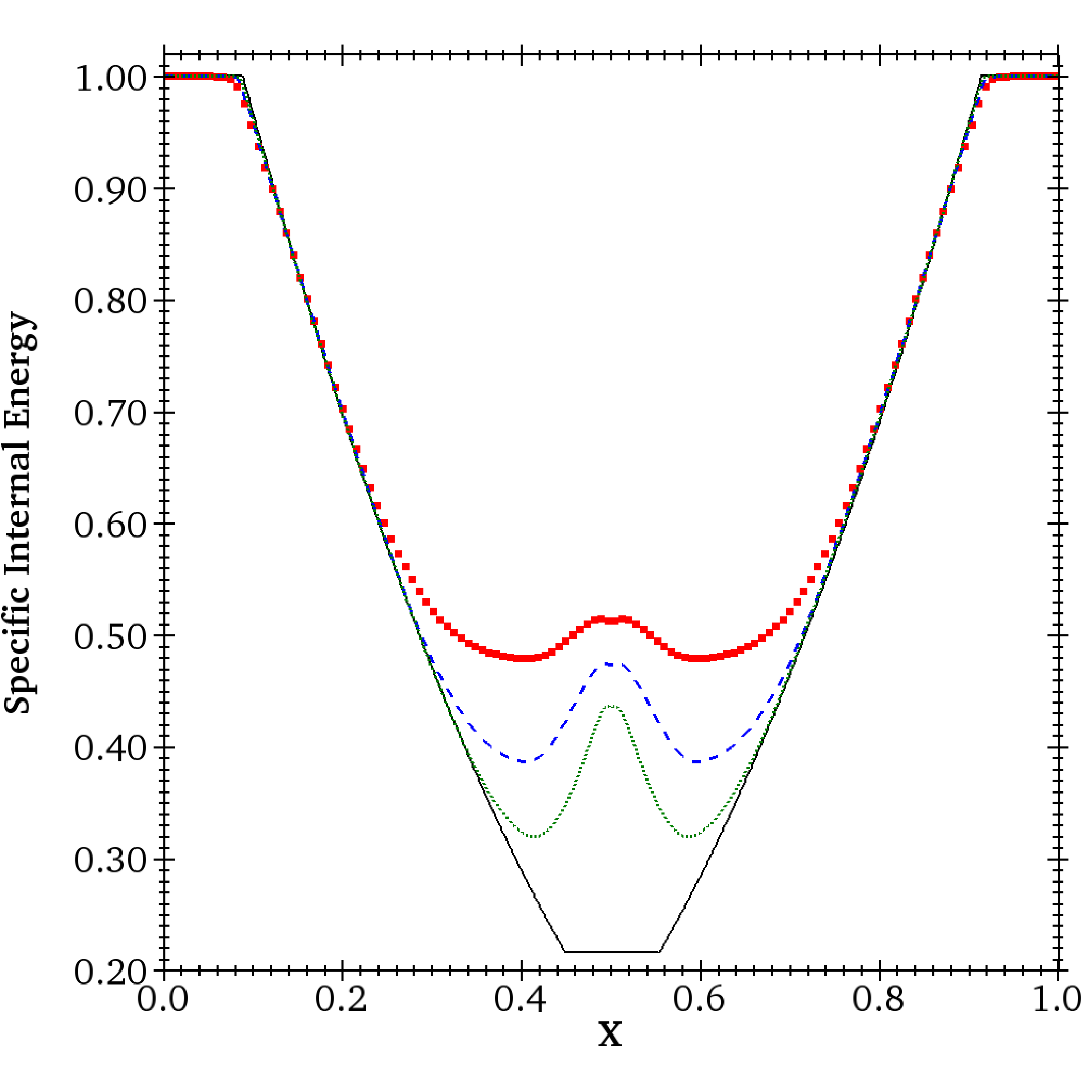}
  \caption{Results from the double rarefaction problem at $t = 0.15$, computed with $128$ cells and using the HLLC  Riemann solver.  
  The density (left) and pressure (right) are plotted in the upper panels, and the velocity ($v_{x}$; left) and specific internal energy ($e/\rho$; right) are plotted in the lower panels.  
  The solid black line is a reference solution obtained with an exact Riemann solver using $1000$ cells.  
  For the specific internal energy, results from runs with $256$ cells (blue dashed line) and $512$ cells (green dotted line) are also plotted.}
\label{fig:DoubleRarefaction_1D}
\end{figure}

\begin{figure}
  \epsscale{1.0}
  \plottwo{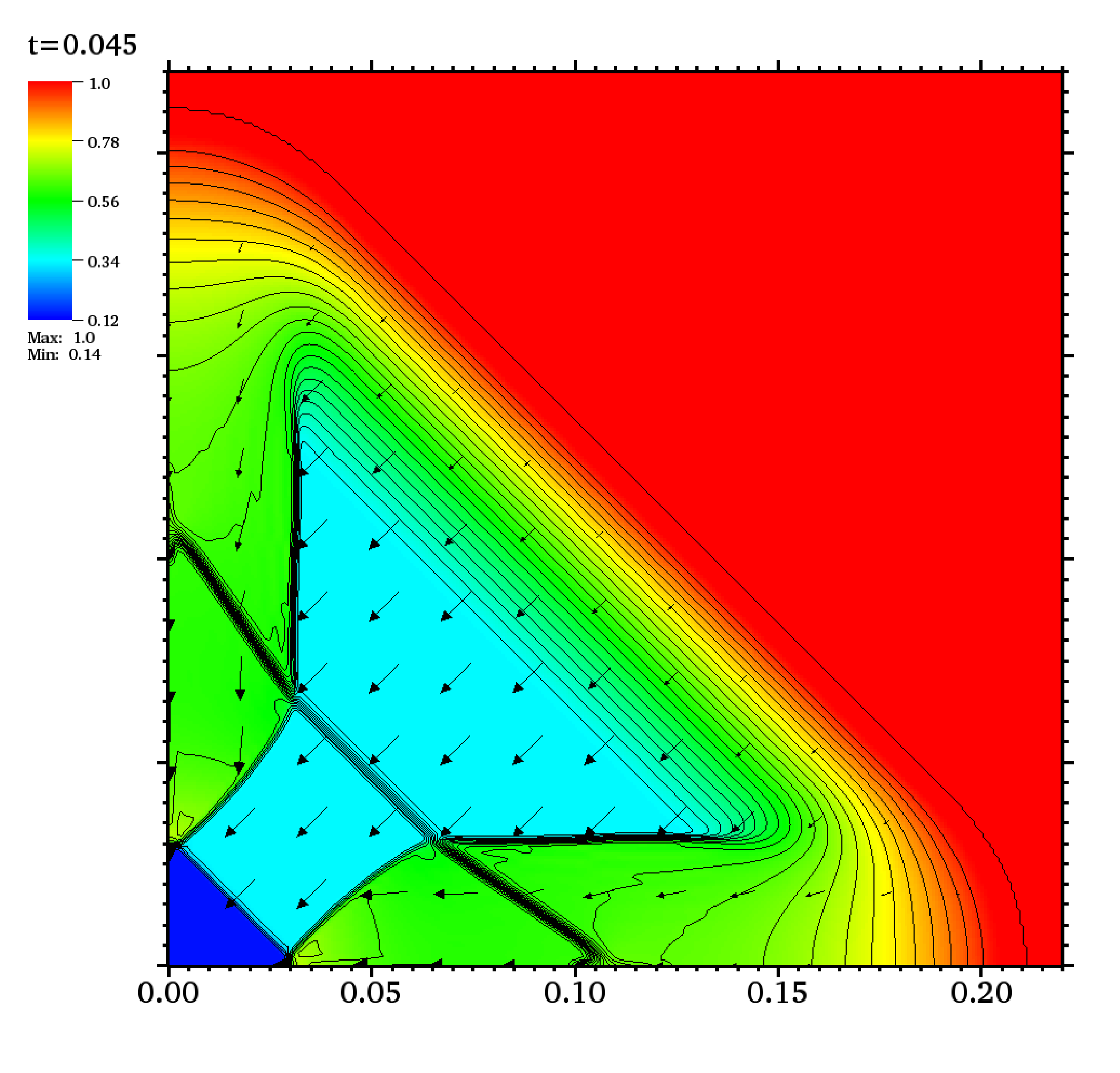}
                {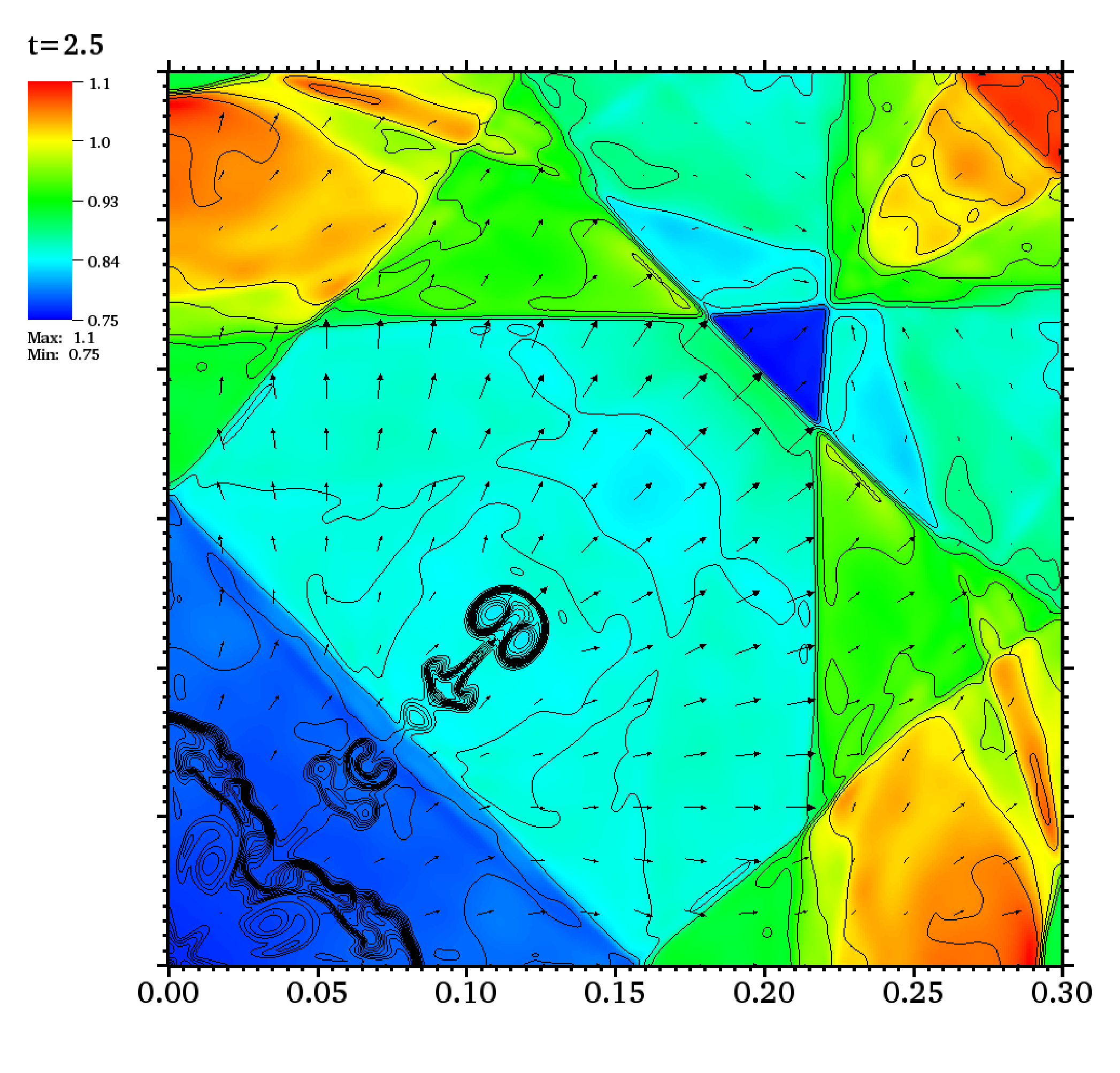}
  \caption{Results from running the implosion problem with \genasis\ using $400\times400$ cells and the HLLC Riemann solver.  
  The left and right panels can be compared directly with Figures 4.10 and 4.11 of \citet{Liska2003}.  
  The color maps represent the fluid pressure.  
  The density is plotted with contours.  
  Left panel: 36 contours from 0.125 to 1 (in steps of 0.025).  
  Right panel: 31 contours from 0.35 to 1.1 (in steps of 0.025).  
  The arrows indicate the flow velocity.}
  \label{fig:implosionProblem}
\end{figure}

\begin{figure}
  \epsscale{1.0}
  \plottwo{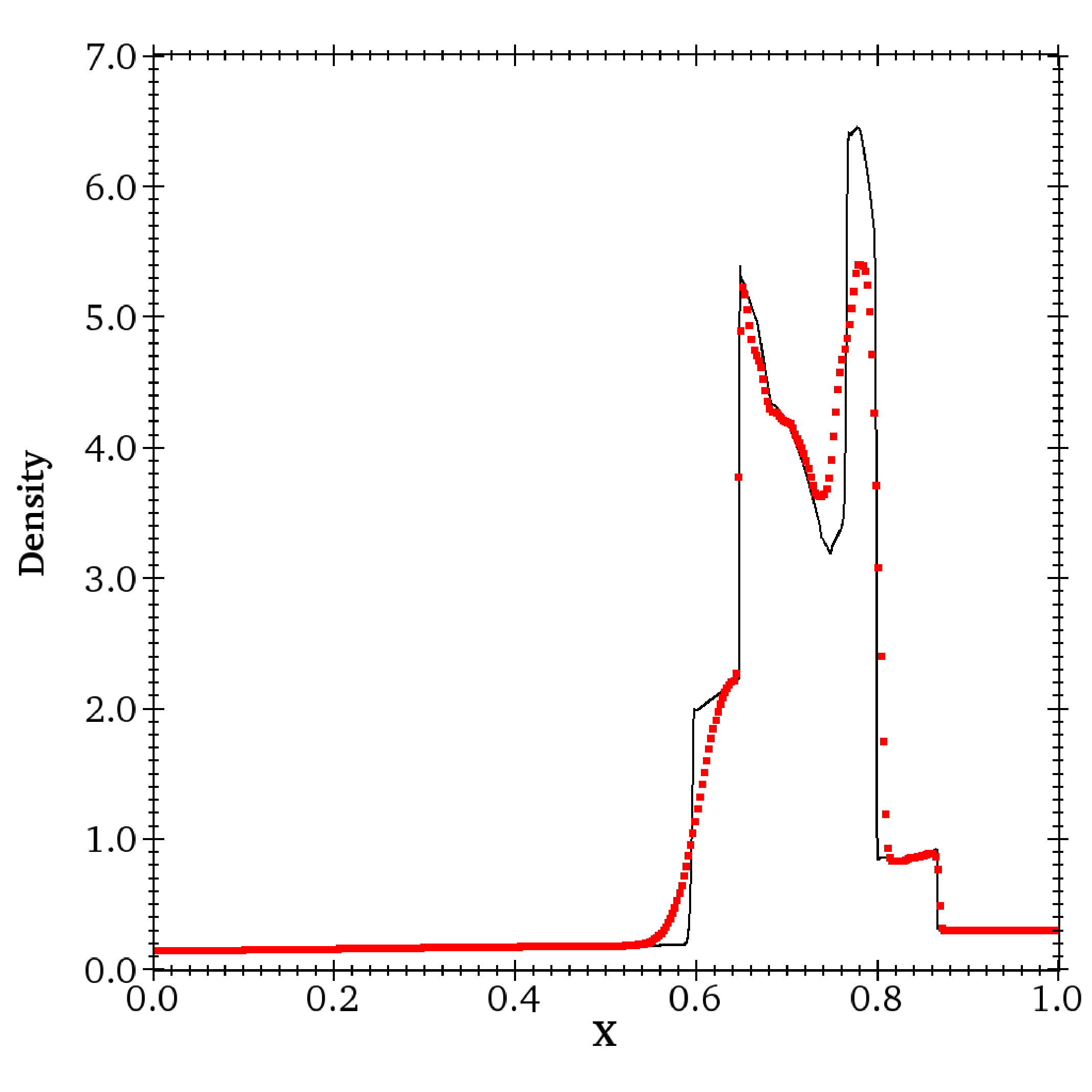}
          {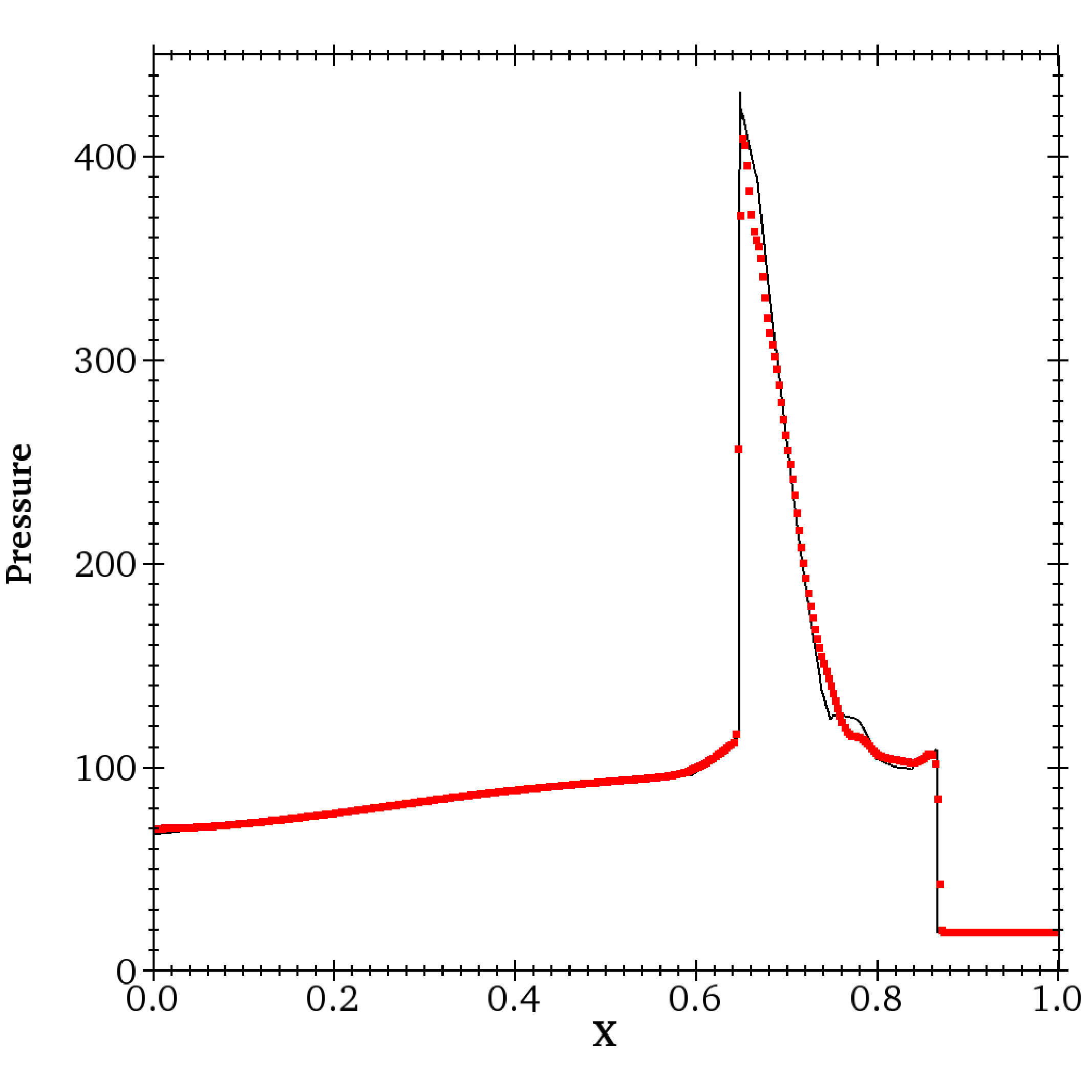}
  \caption{Density (left) and pressure (right) from the two-interacting-blast-waves test problem at $t = 0.038$, computed using $400$ cells and the HLLC Riemann solver in \genasis. 
  The solid black line is a high-resolution reference solution obtained by using $10^{4}$ cells.}
\label{fig:InteractingBlastWave_1D}
\end{figure}

\begin{figure}
  \epsscale{1.0}
  \plottwo{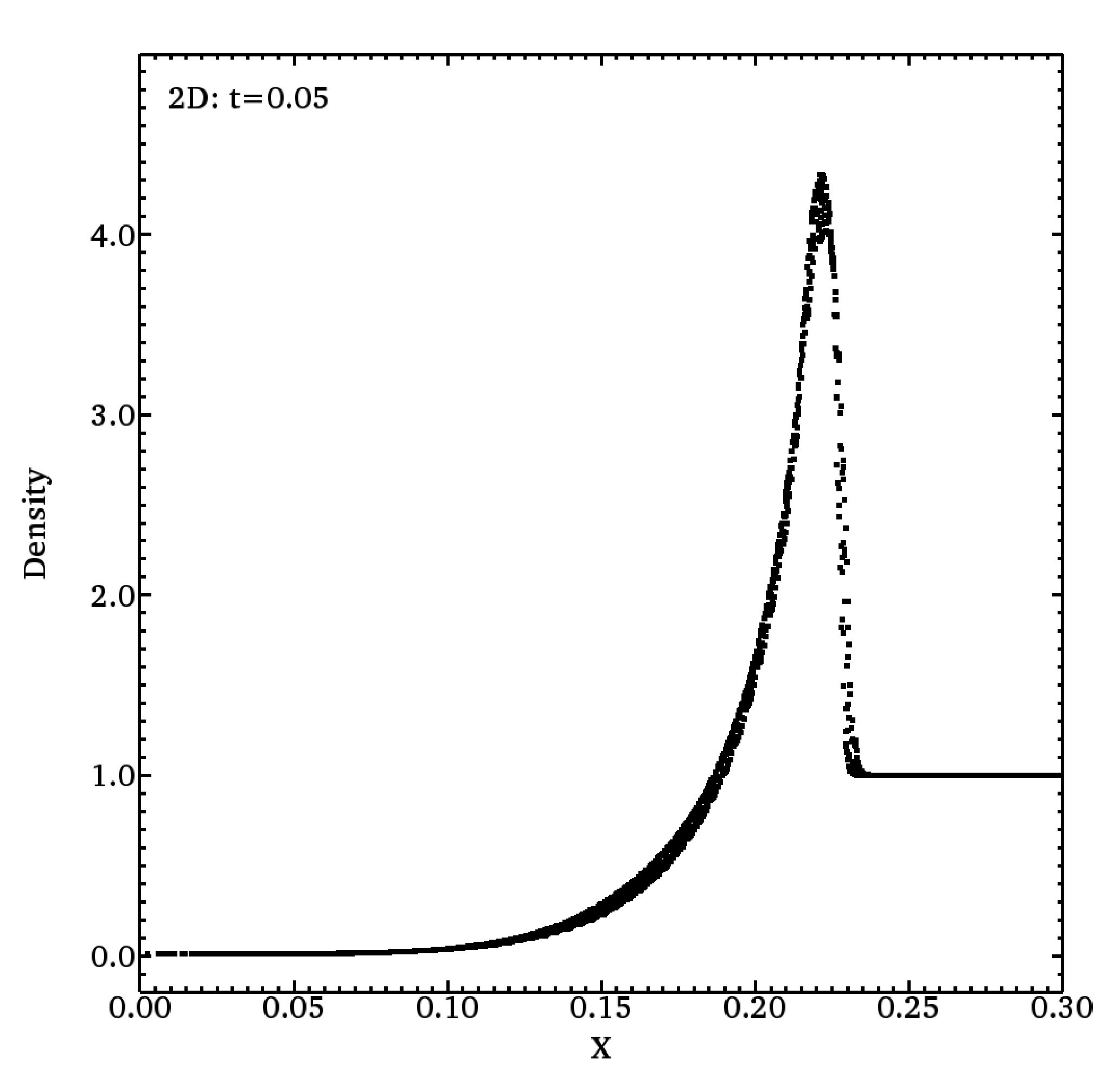}
                {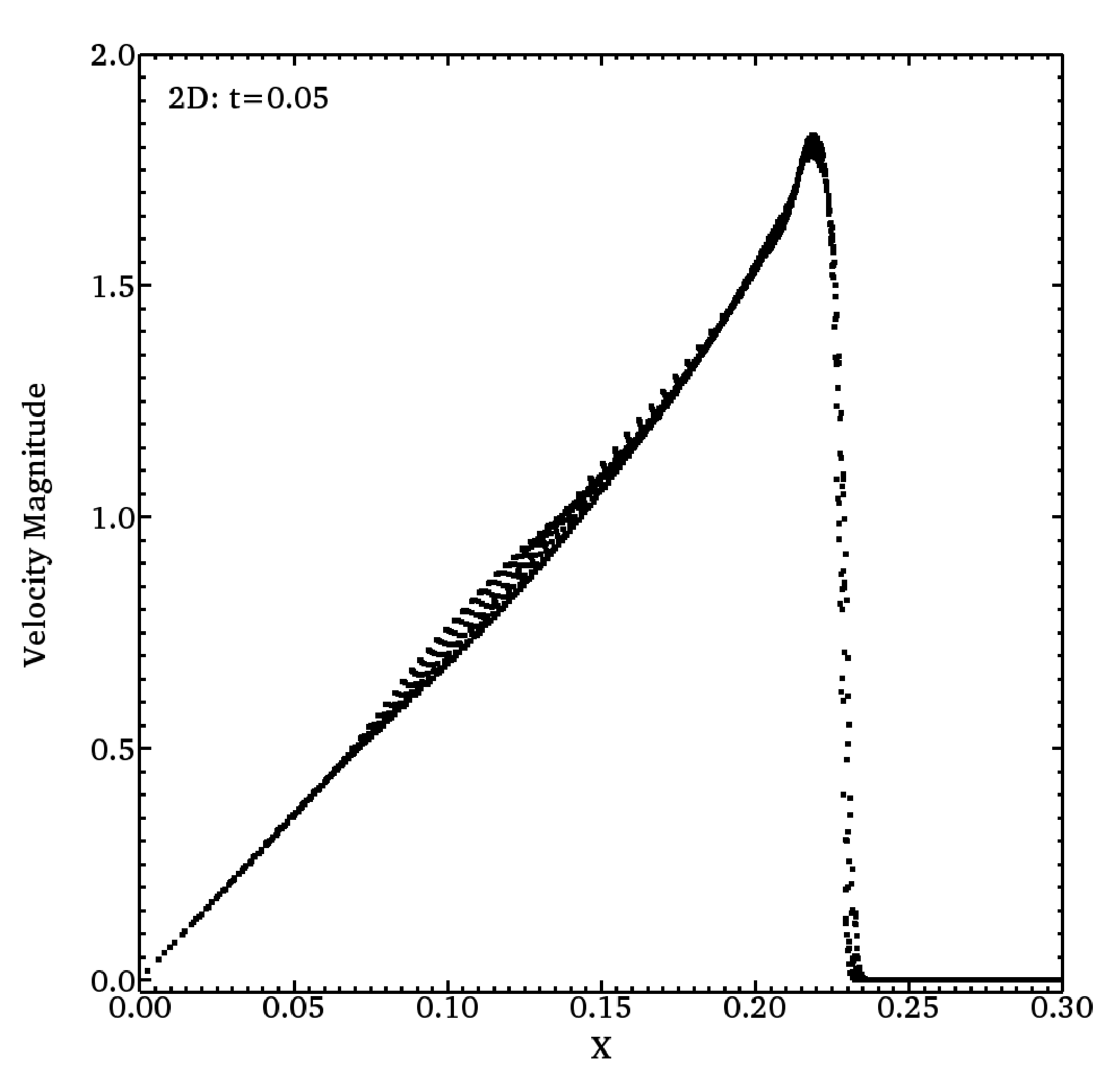} \\
  \plottwo{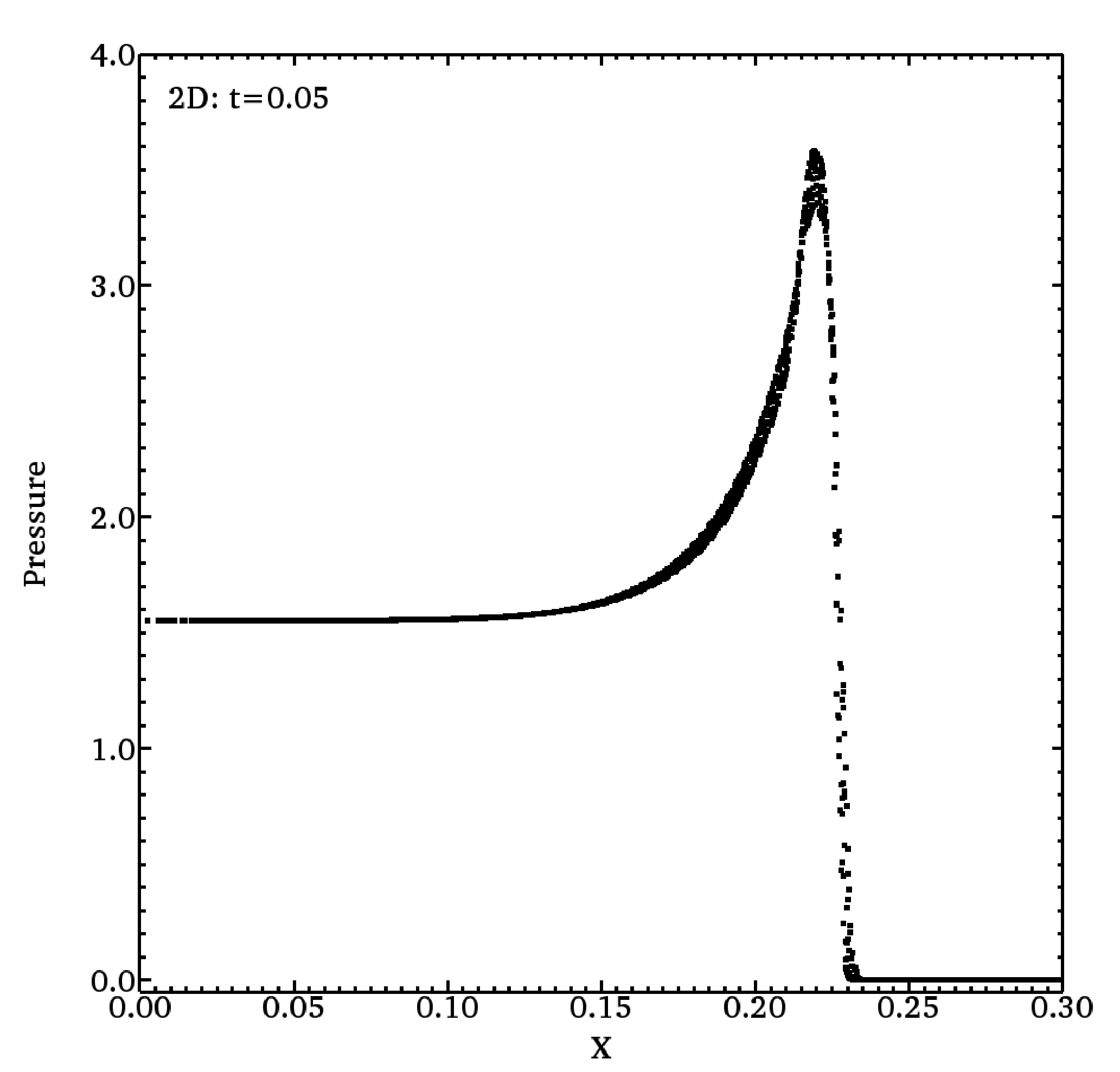}
                {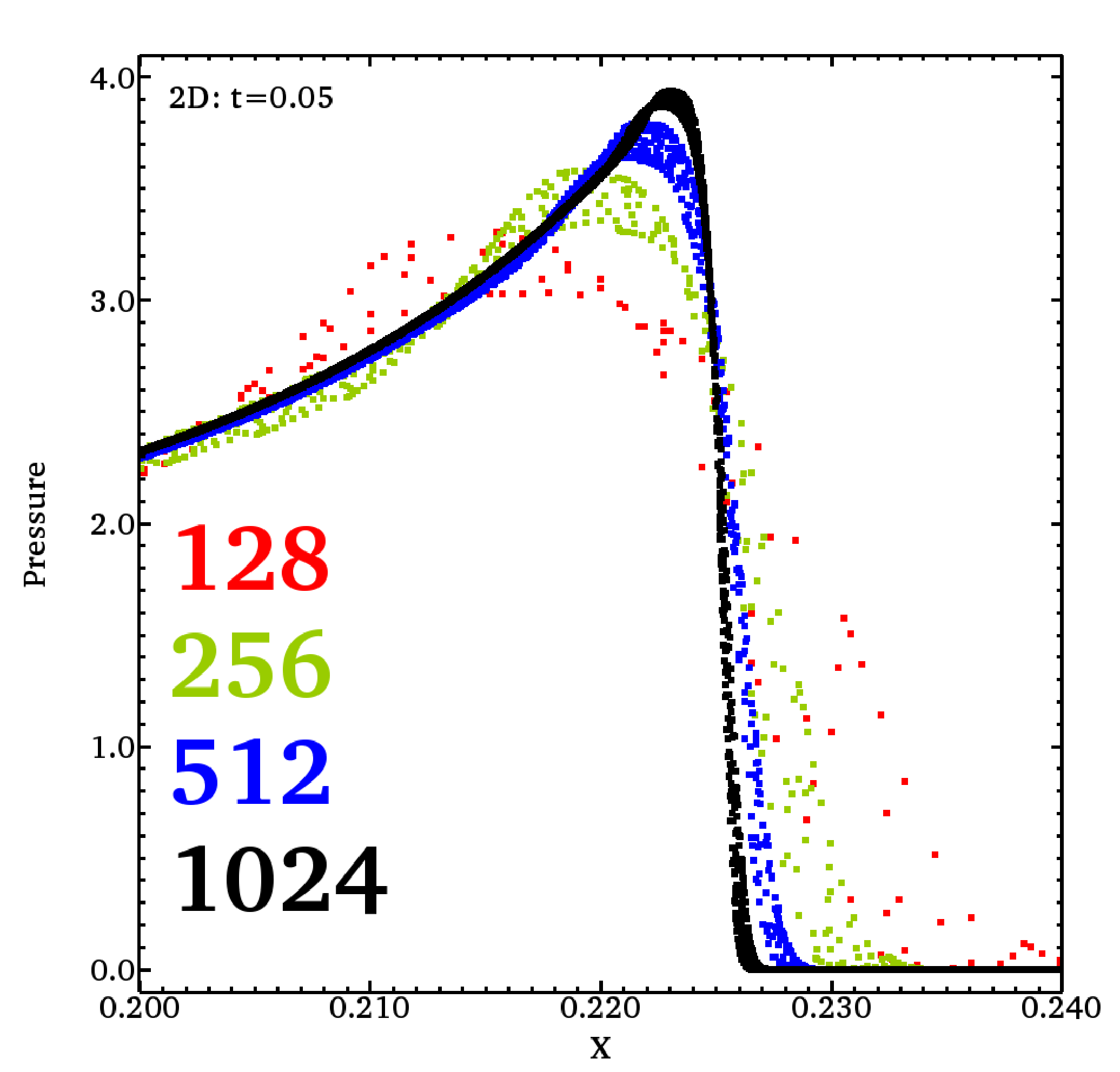}
  \caption{Scatter plots of the 2D Sedov-Taylor blast wave problem computed with the HLL Riemann solver.}
  \label{fig:SedovTaylor2D}
\end{figure}

\begin{figure}
  \epsscale{1.0}
  \plottwo{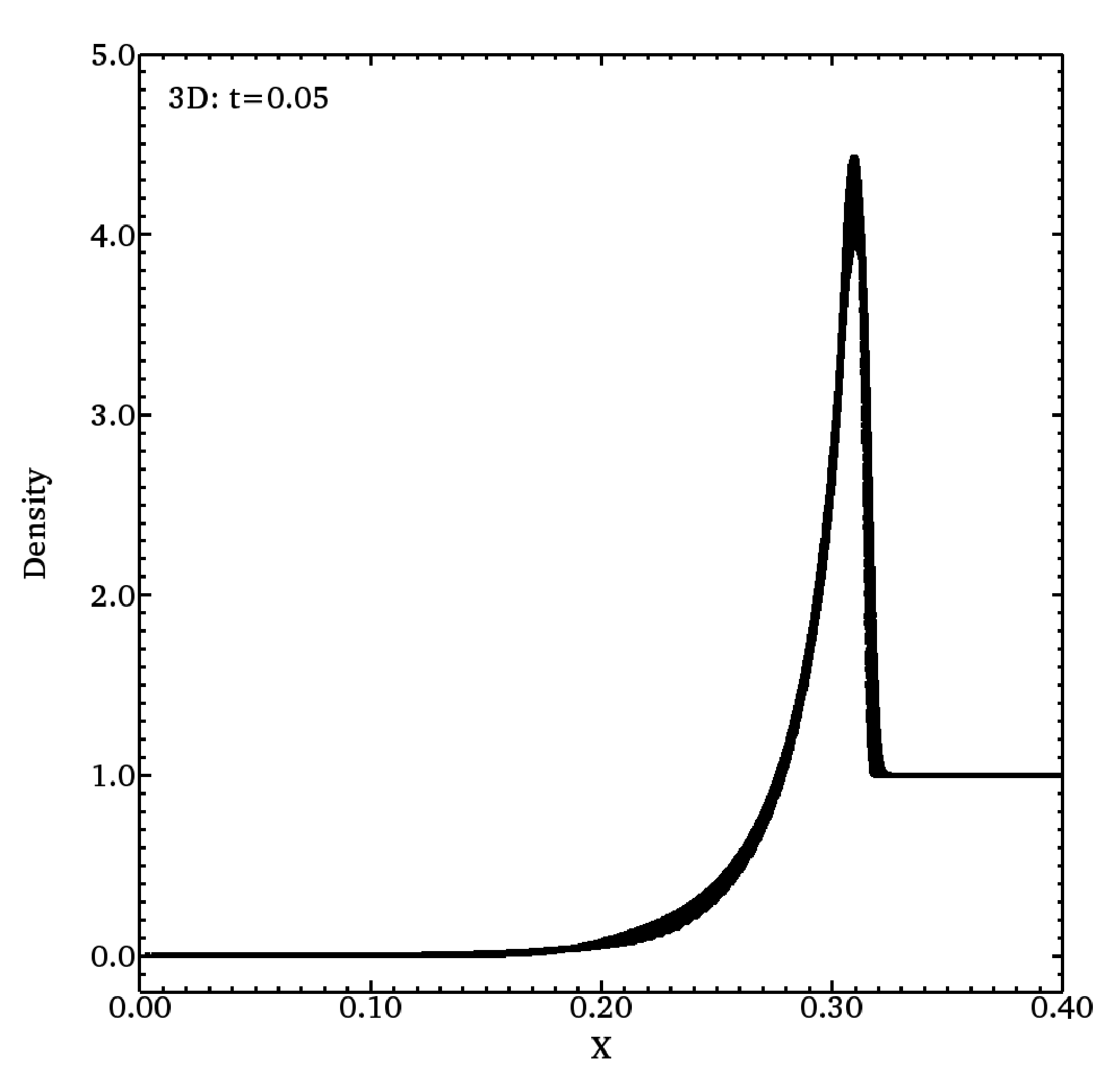}
                {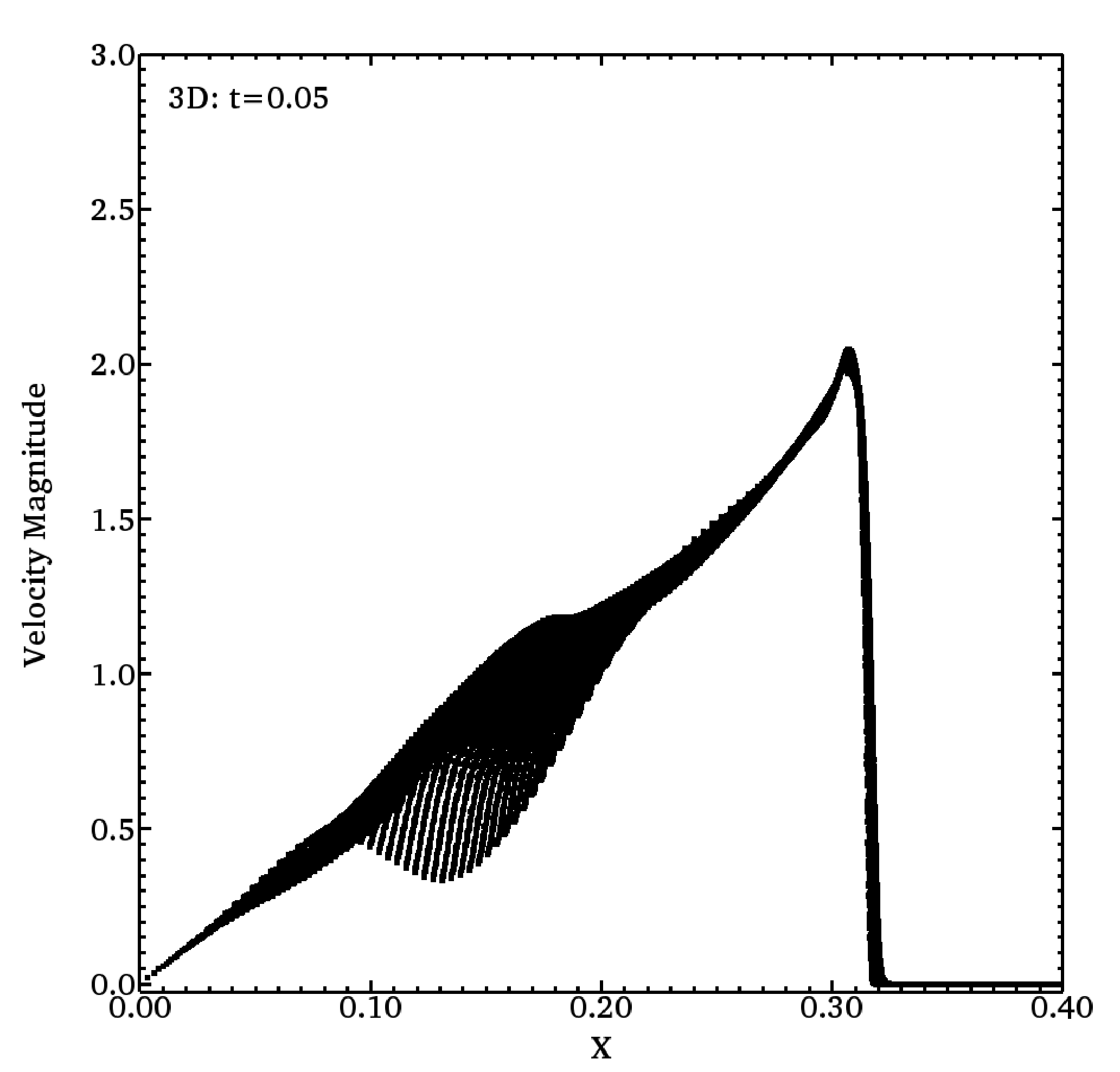} \\
  \plottwo{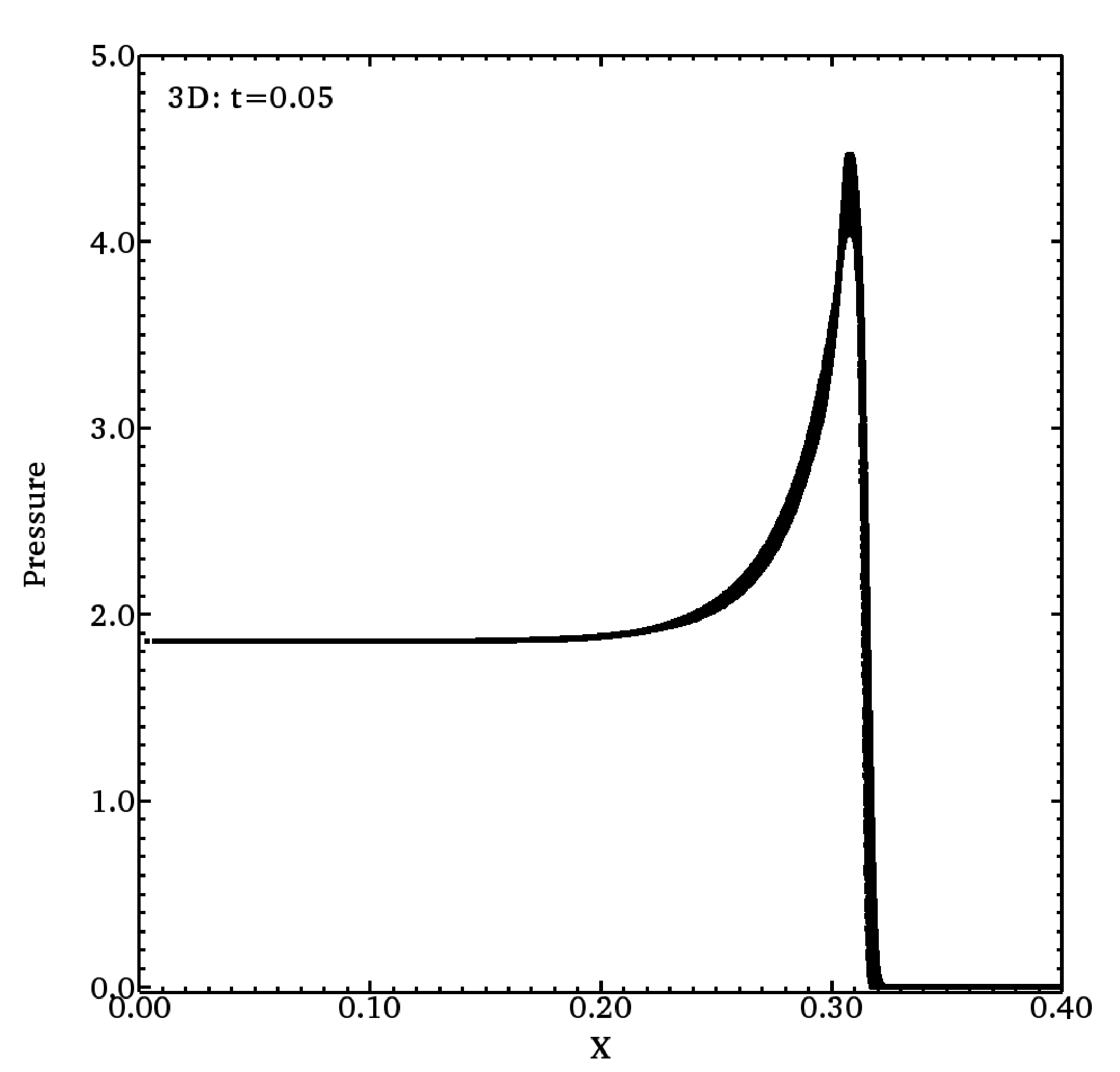}
                {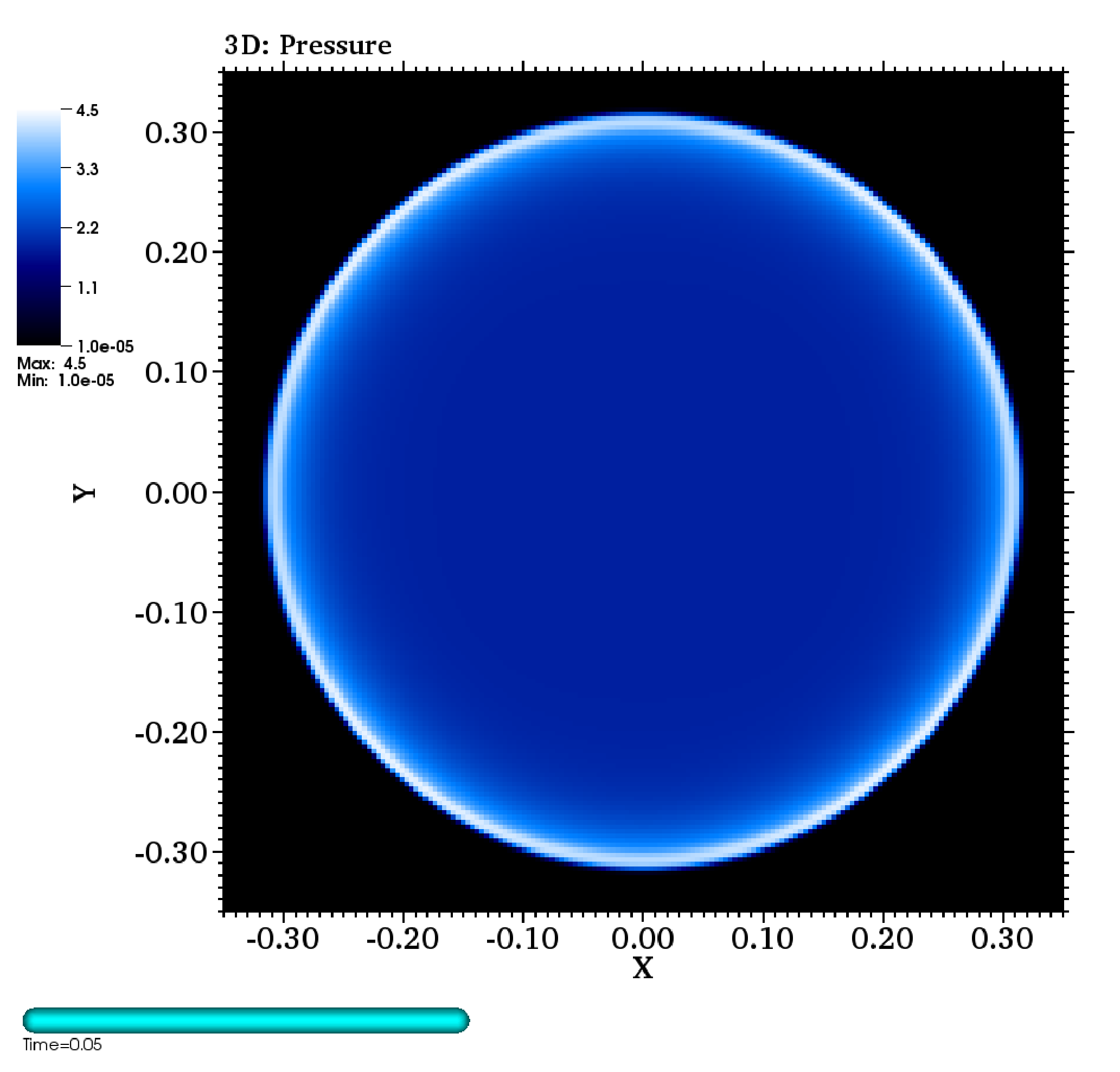}
  \caption{Plots of the 3D Sedov-Taylor blast wave problem computed with the HLL Riemann solver with $256^{3}$ zones.}
  \label{fig:SedovTaylor3D}
\end{figure}

\begin{figure}
  \epsscale{1.0}
  \plotone{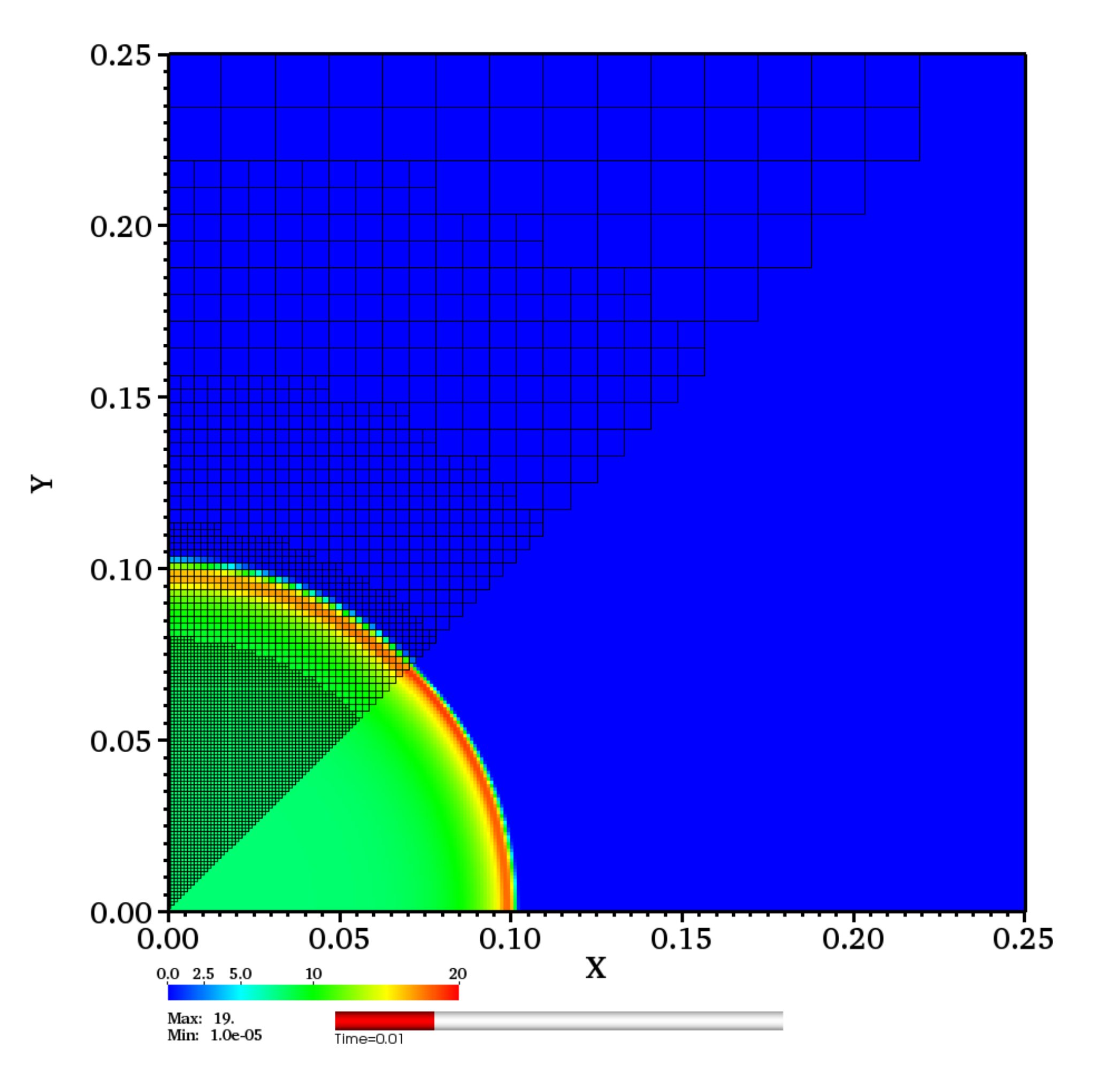}
  \caption{Pressure at $t=0.01$ in the 2D Sedov-Taylor blast wave problem computed with the HLL Riemann solver.  
  Results obtained with the multilevel mesh solver (using seven mesh levels) are shown above the diagonal $y=x$ and compared with single-level results shown below the diagonal.  
  (The multilevel mesh is also shown for $y>x$.)}
  \label{fig:SedovTaylor2D_FMR}
\end{figure}

\begin{figure}
  \epsscale{1.0}
  \plotone{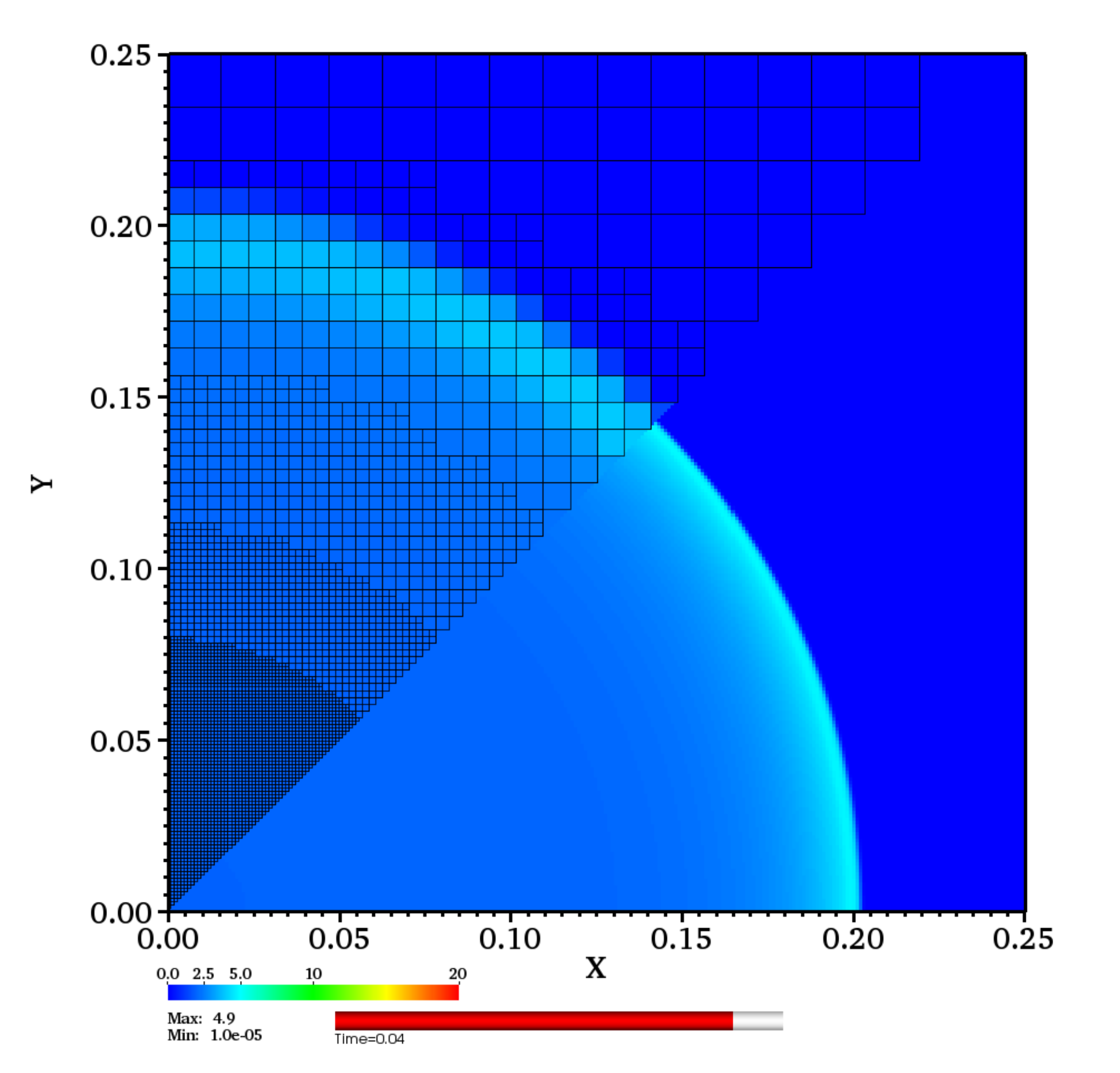}
  \caption{Same as Figure~\ref{fig:SedovTaylor2D_FMR}, but for $t=0.04$.}
  \label{fig:SedovTaylor2D_FMR_2}
\end{figure}

\begin{figure}
  \epsscale{1.0}
  \plotone{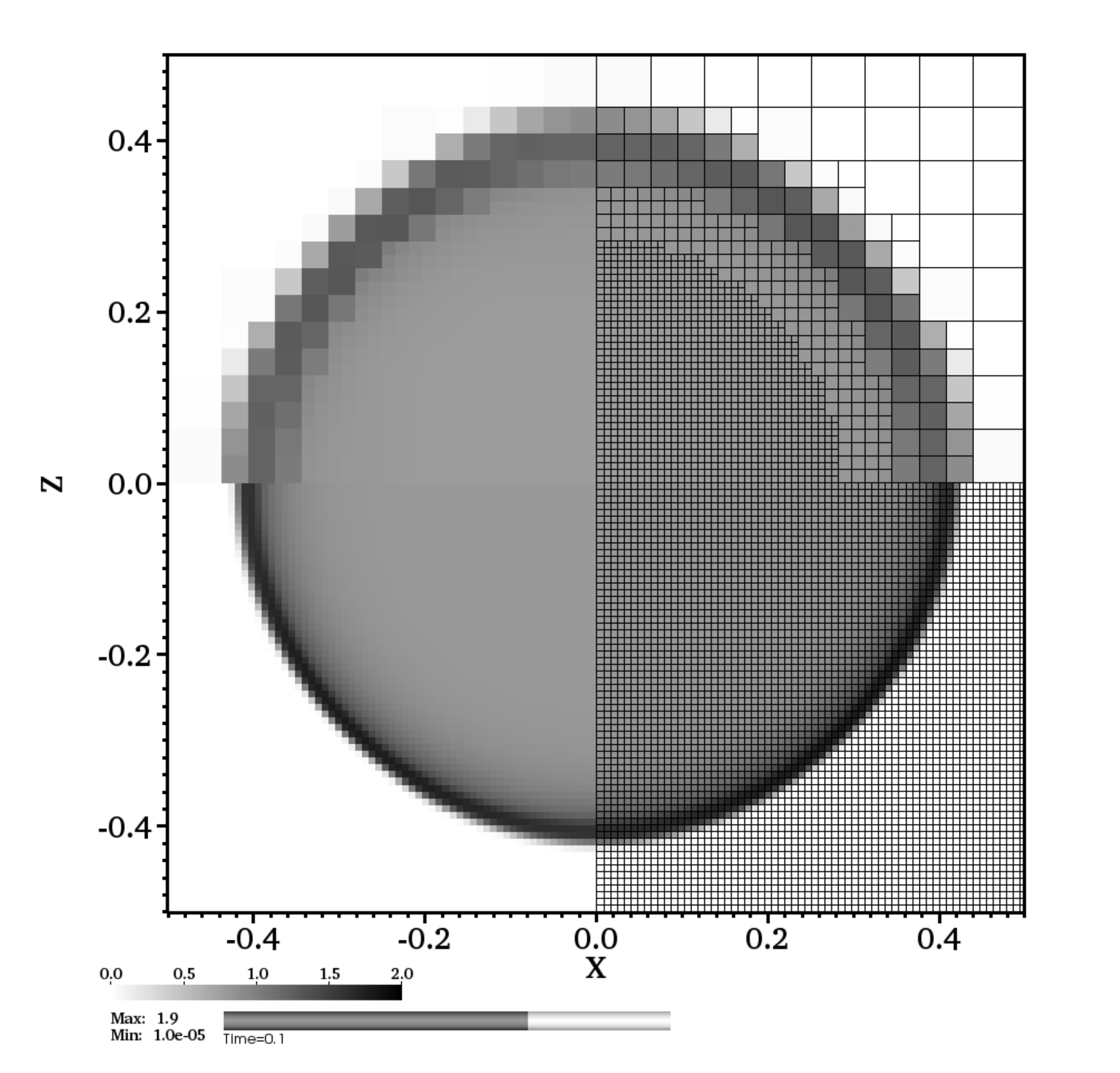}
   \caption{Plots of the pressure in the $xz$-plane ($y=0$) from the 3D Sedov-Taylor blast wave problem at $t=0.1$, computed with the HLL Riemann solver.  
  Results obtained with the multilevel mesh solver (using four mesh levels) are shown for $z>0$, and compared with unigrid results ($z\le0$).  
  (The mesh is shown for $x>0$.)}
  \label{fig:SedovTaylor3D_FMR}
\end{figure}

\begin{figure}
\epsscale{1.0}
\plottwo{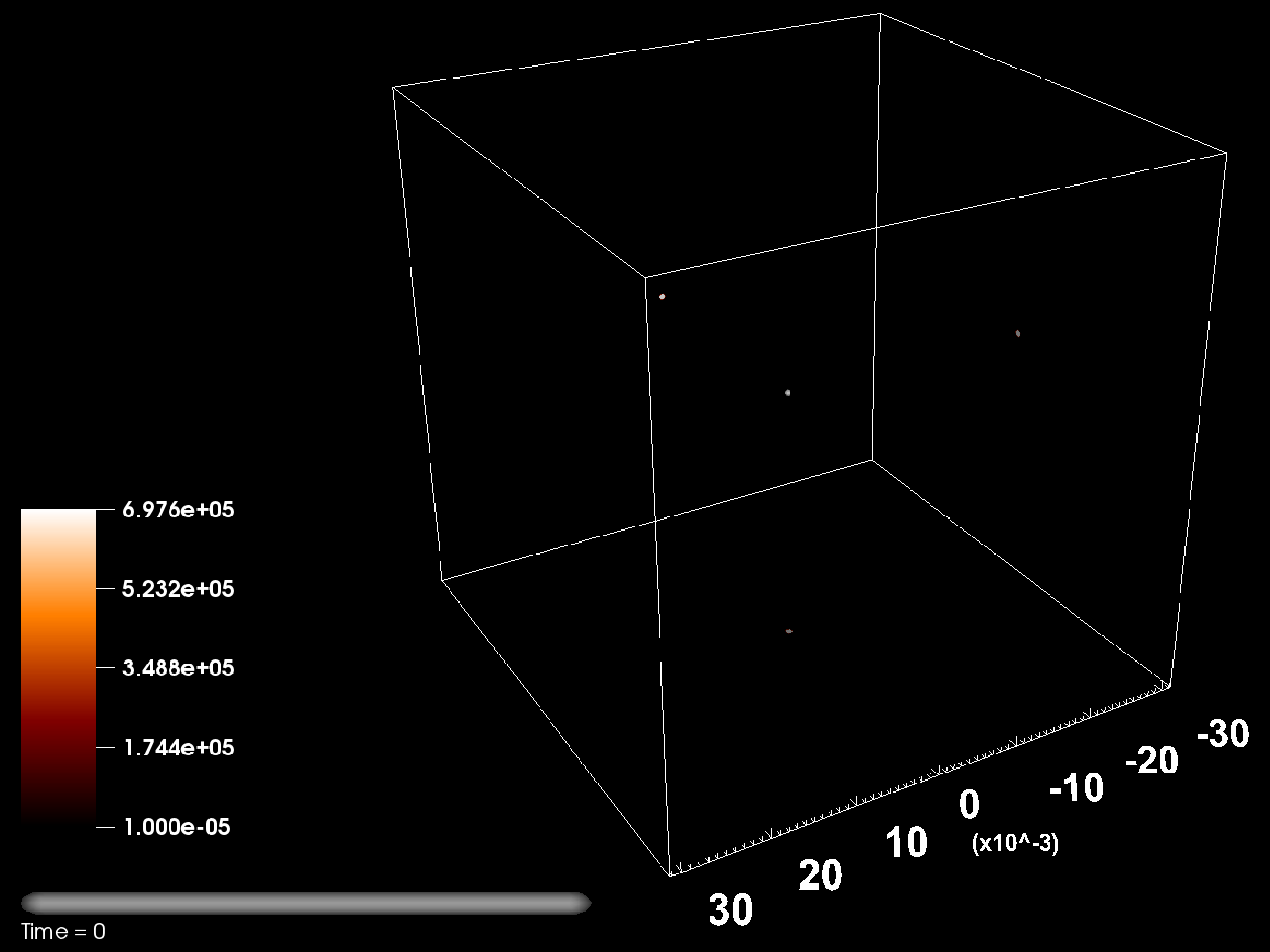}{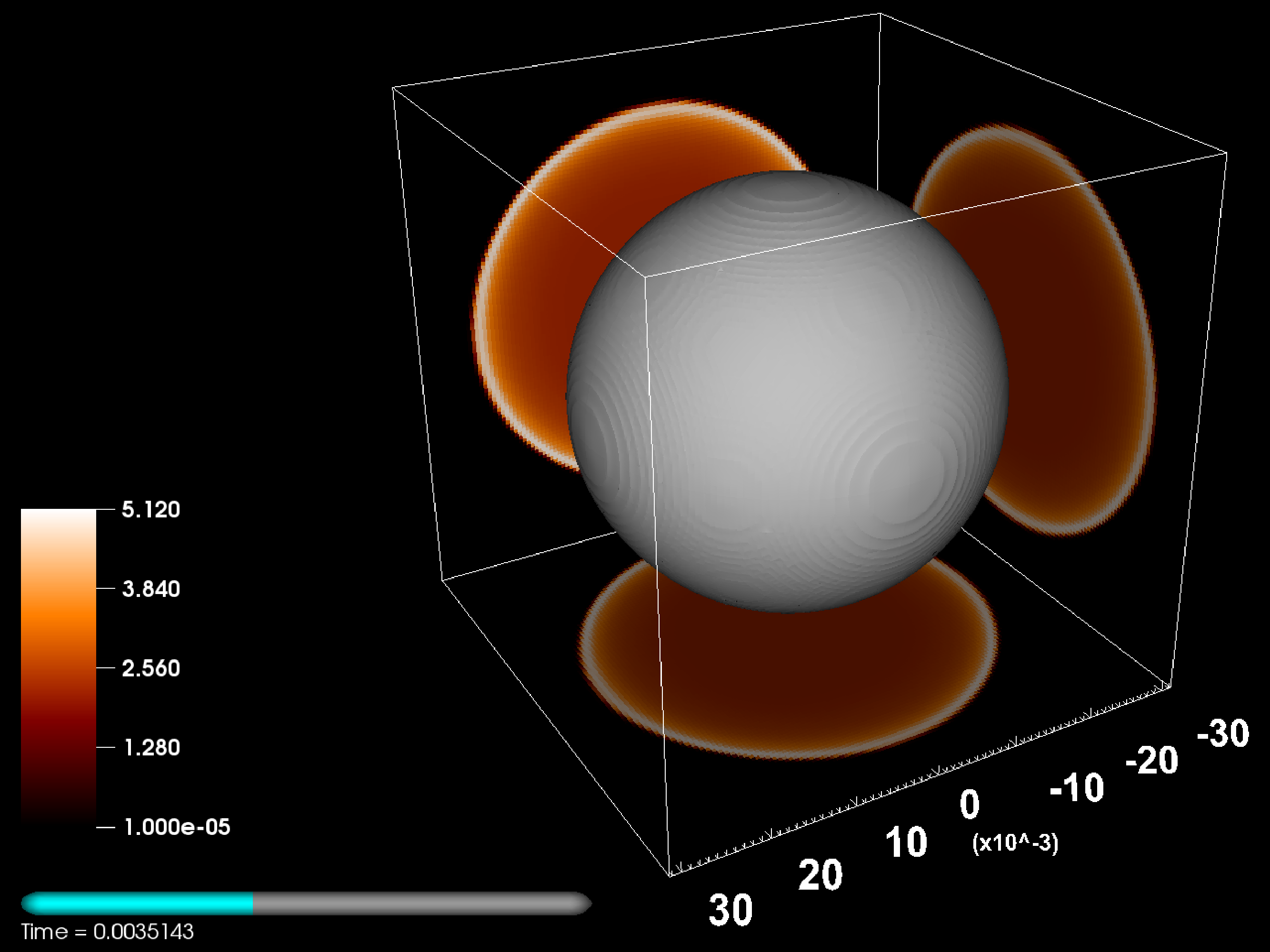}
\plottwo{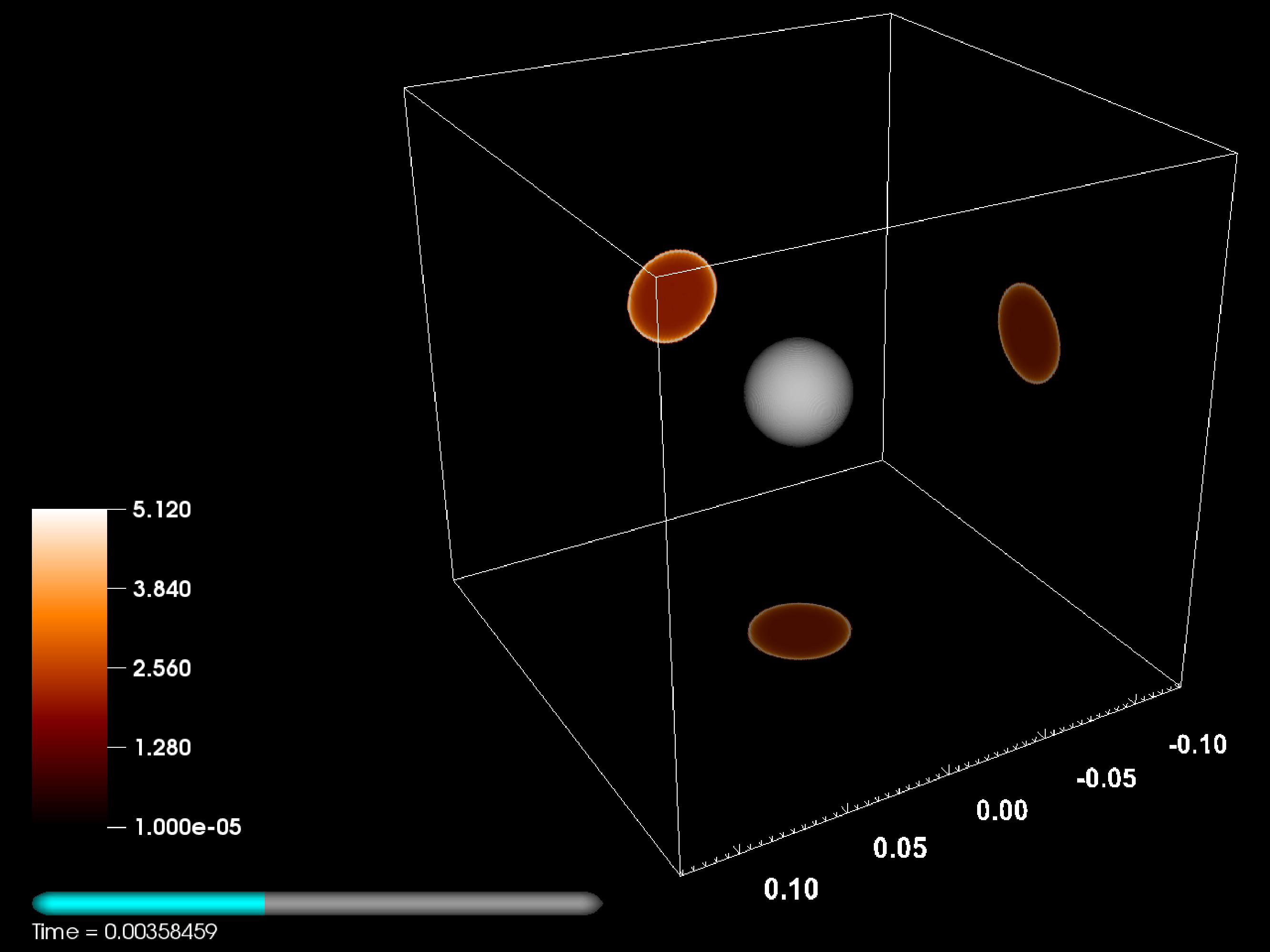}{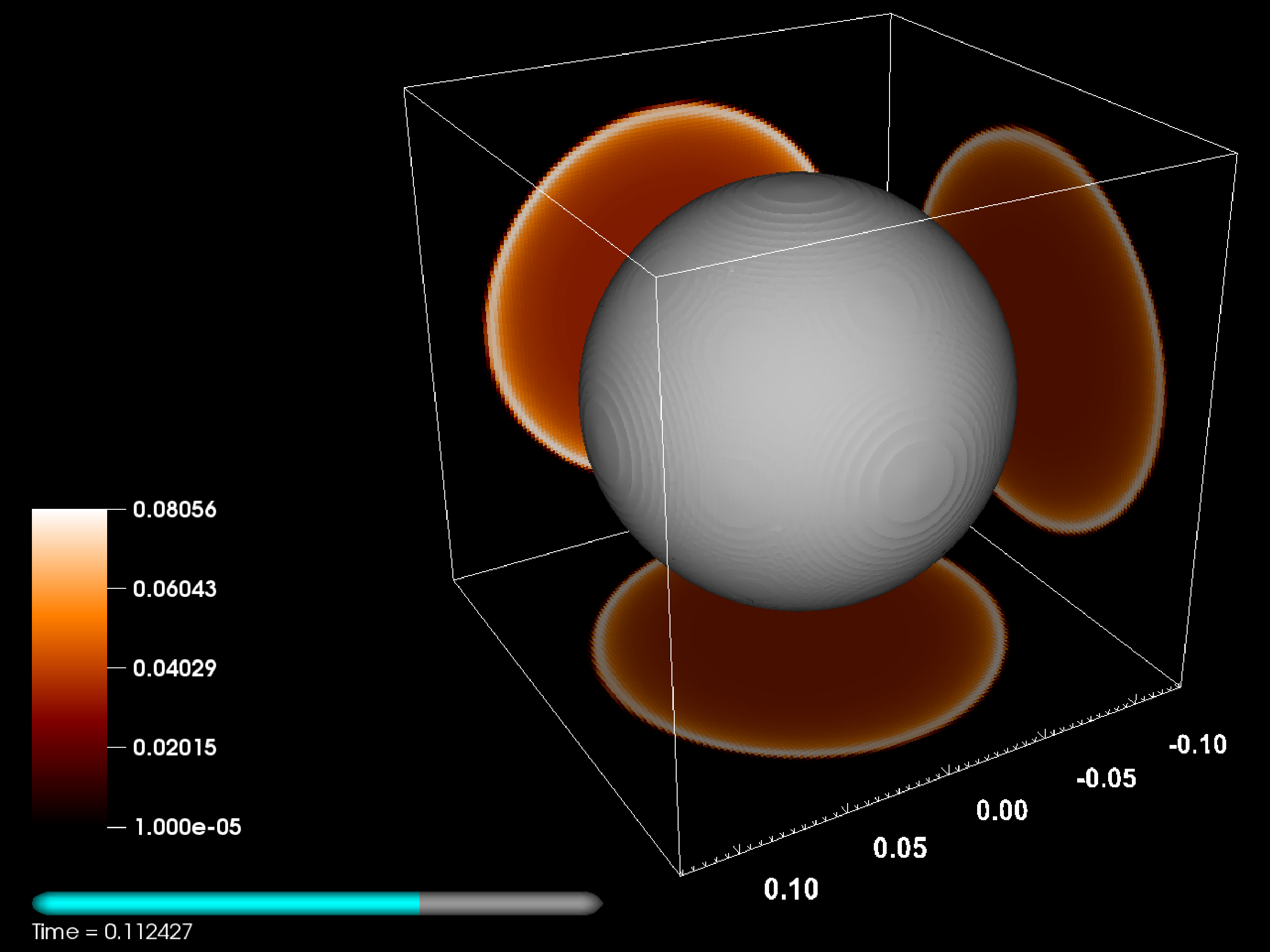}
\plottwo{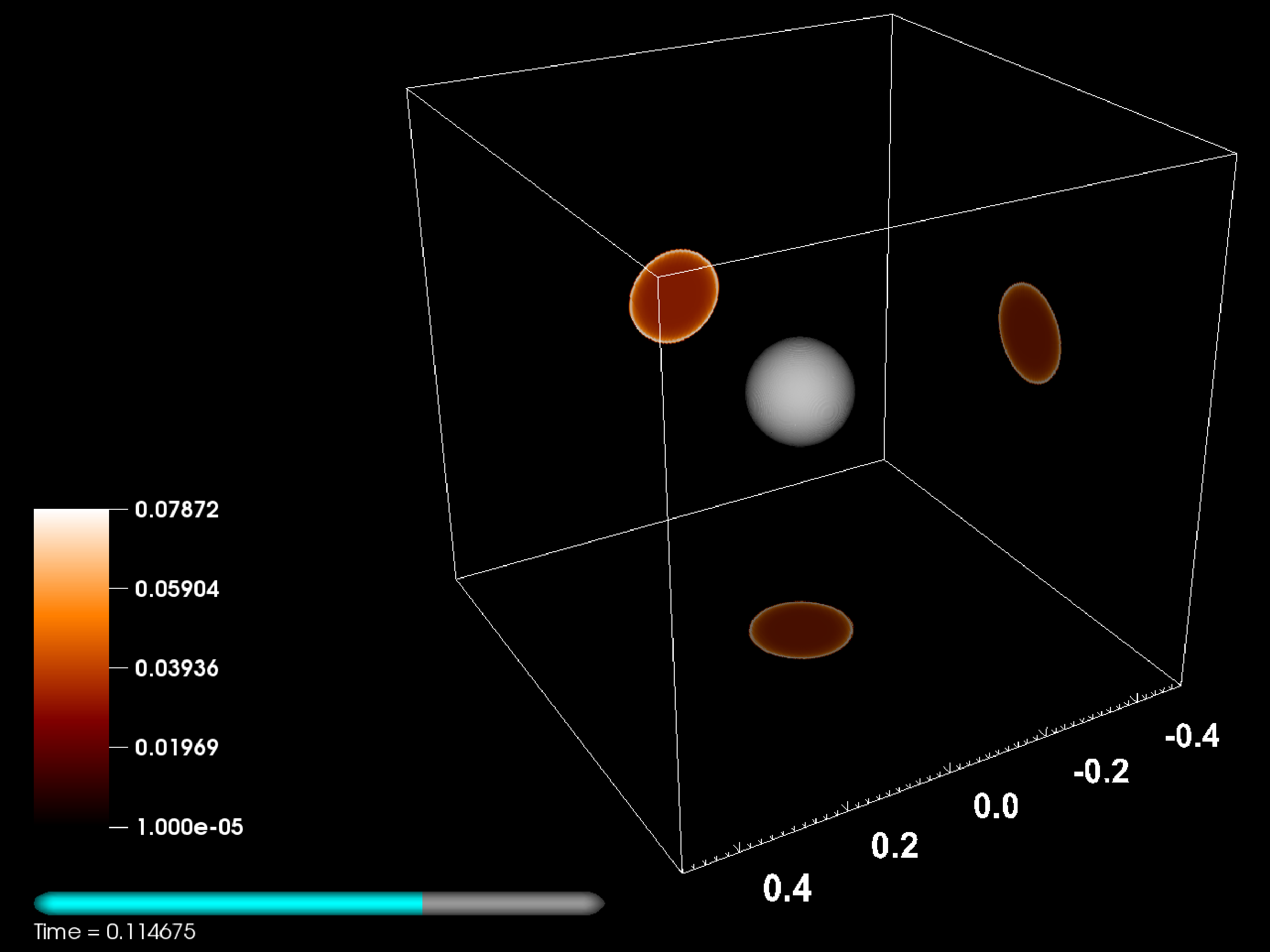}{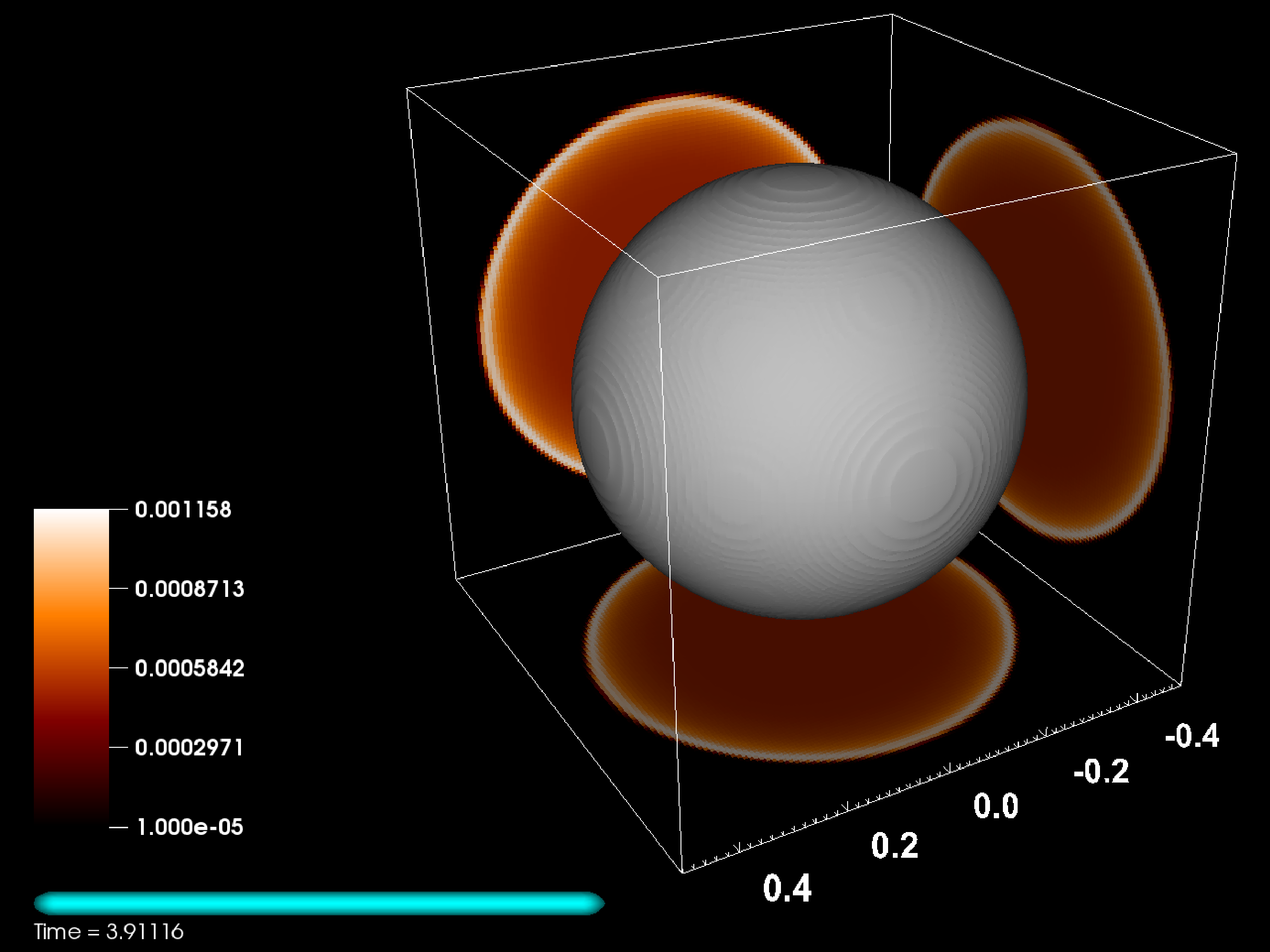}
\caption{Snapshots of a larger-scale 3D Sedov-Taylor blast wave simulation with seven mesh levels and $128^3$ cells on the coarsest level.
Pictured in each panel are a grey shock surface, and cross sections through the origin of the pressure, projected onto the rear walls of the visualization box.
The visualization boxes in the two panels in a given row have the same spatial scale;
this scale changes by a factor of four between each successive row. Animated version online includes episodic depictions of the multilevel mesh.}
\label{fig:Sedov_3D_Big}
\end{figure}

\begin{figure}
\epsscale{1.0}
\plotone{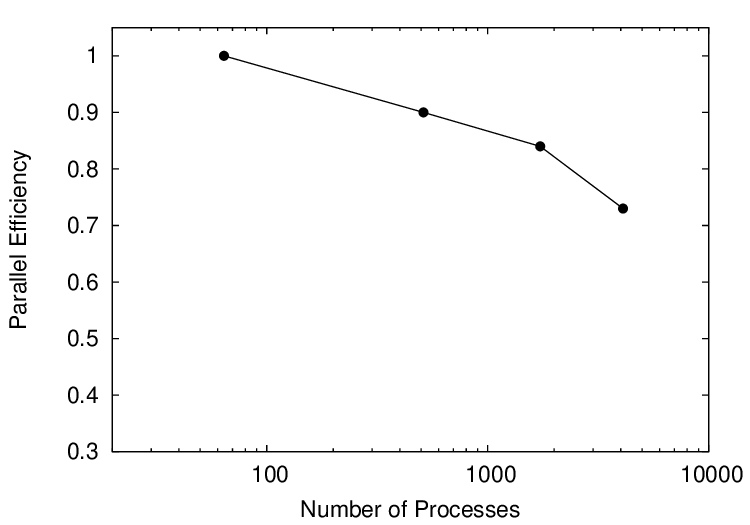}
\caption{Weak scaling of hydrodynamics evolution on a centrally refined mesh with five levels. 
On the coarsest level, the number of computational cells per MPI task is $32^3$.
The data points are for coarsest-level resolutions of $128^3$, $256^3$, $384^3$, and $512^3$, corresponding to effective finest-level resolutions of $2048^3$, $4096^3$, $6144^3$, and $8192^3$.
The wall time per time step for the first data point is 2.59~s.}
\label{fig:MultilevelScaling}
\end{figure}

\begin{figure}
  \epsscale{1.0}
  \plottwo{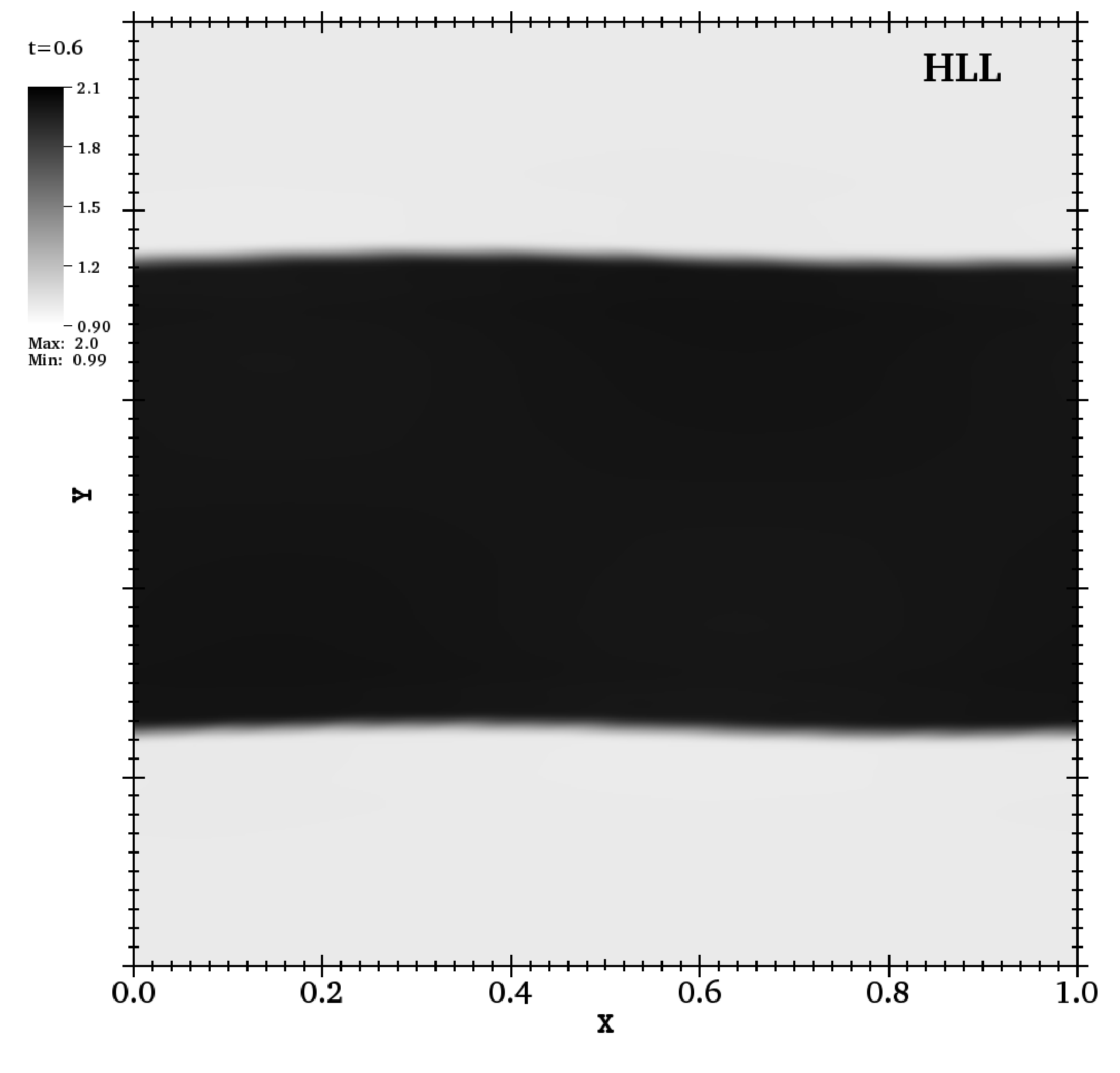}
                {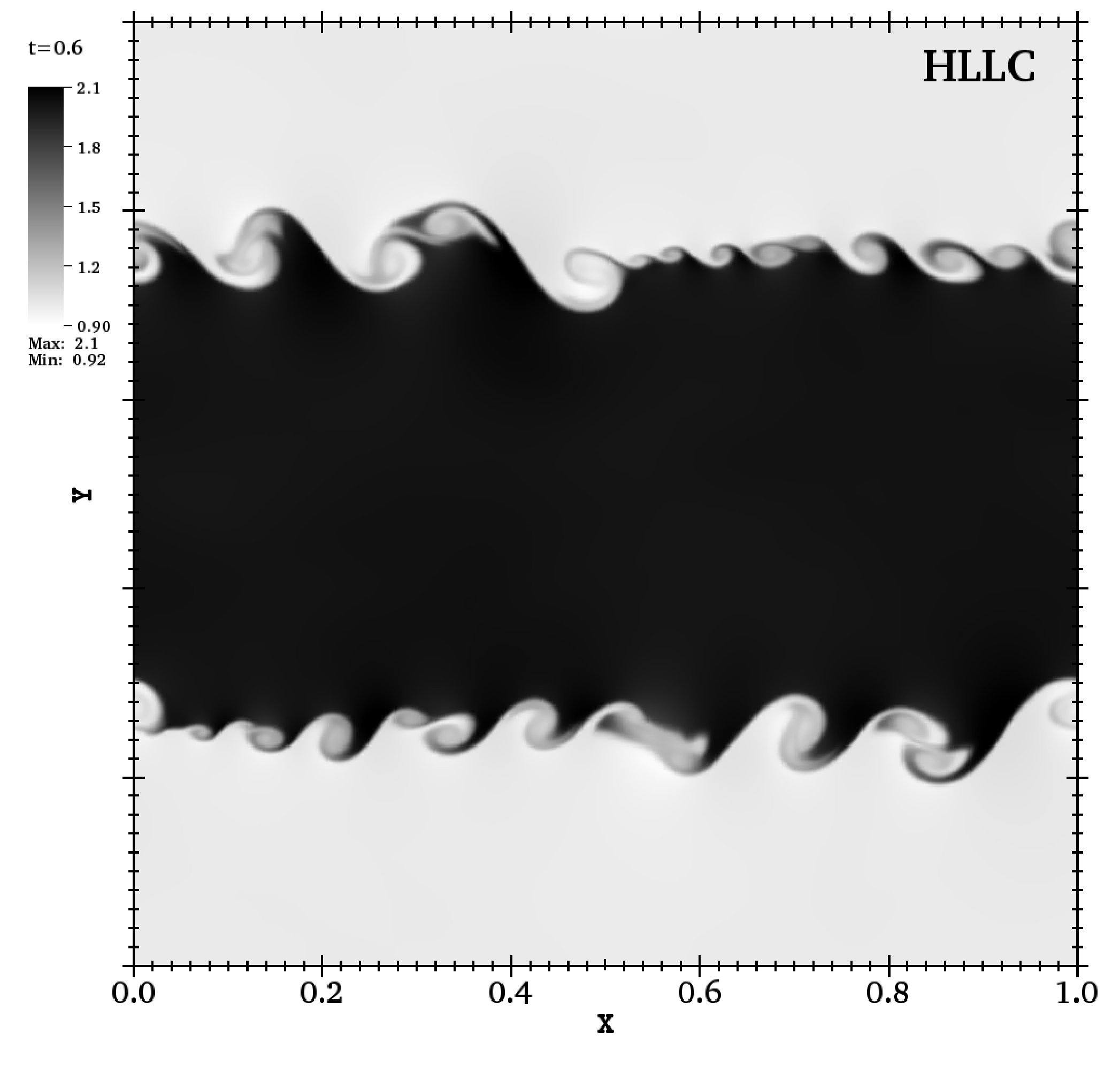}
  \caption{Plots of density from the Kelvin-Helmholtz instability test at $t=0.6$, computed with $512\times512$ zones.  
  The left and right panels show results computed with the HLL and HLLC Riemann solver, respectively.  }
  \label{fig:kelvinHelmholtz1}
\end{figure}

\begin{figure}
  \epsscale{1.0}
  \plottwo{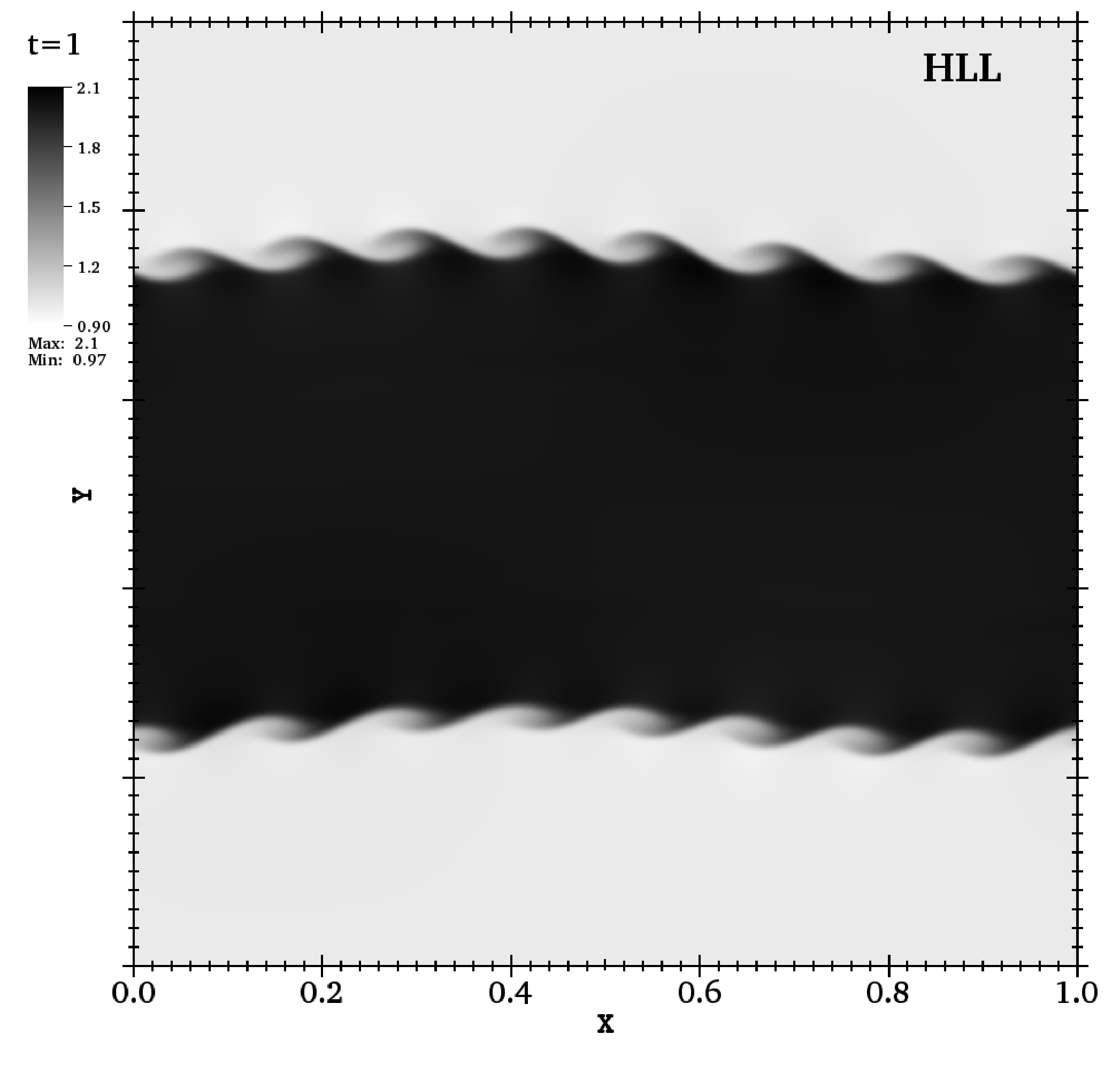}
                {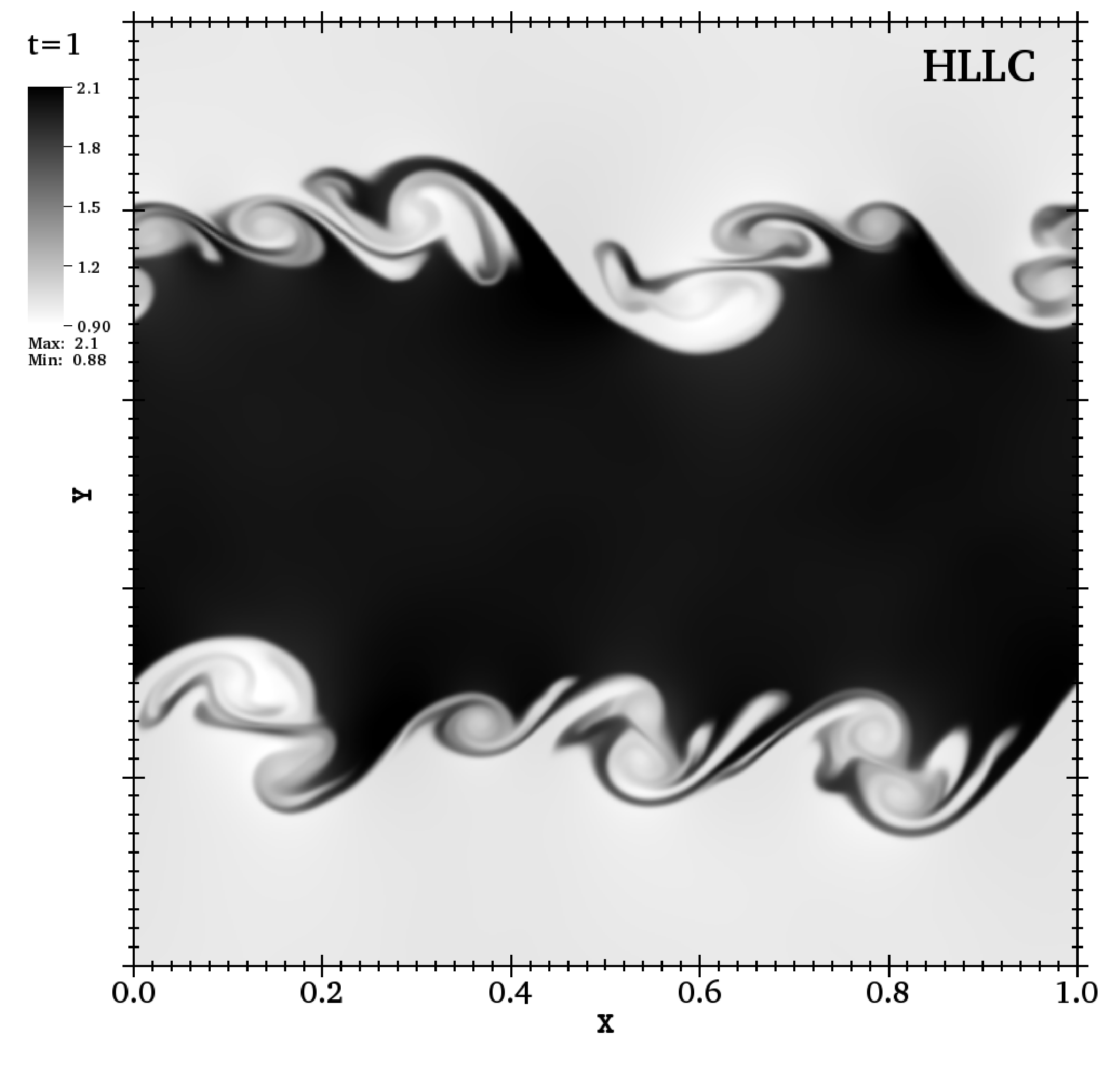}
  \caption{Same as Figure \ref{fig:kelvinHelmholtz1}, but for $t=1.0$.}
  \label{fig:kelvinHelmholtz2}
\end{figure}

\begin{figure}
  \epsscale{1.0}
  \plottwo{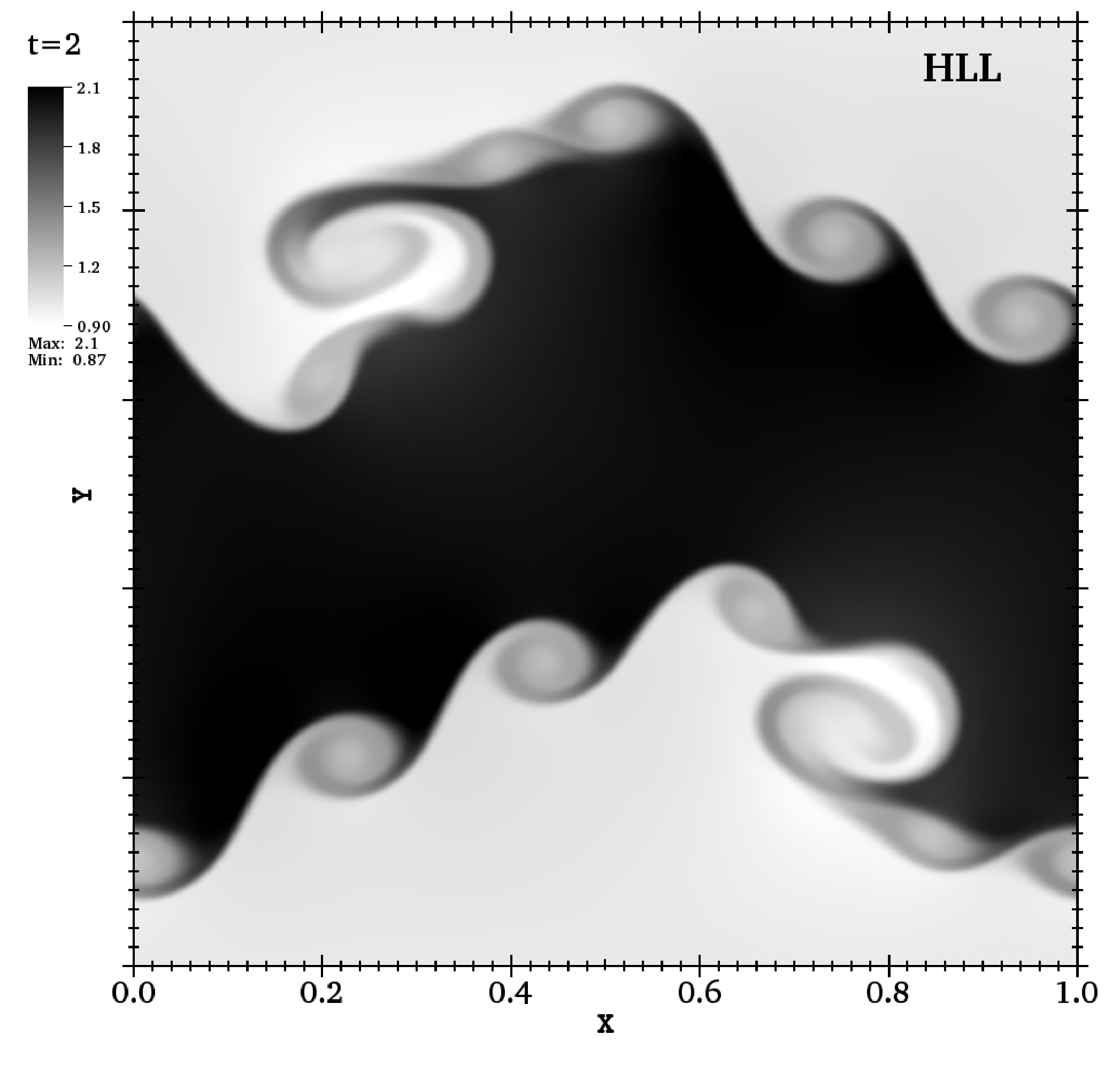}
                {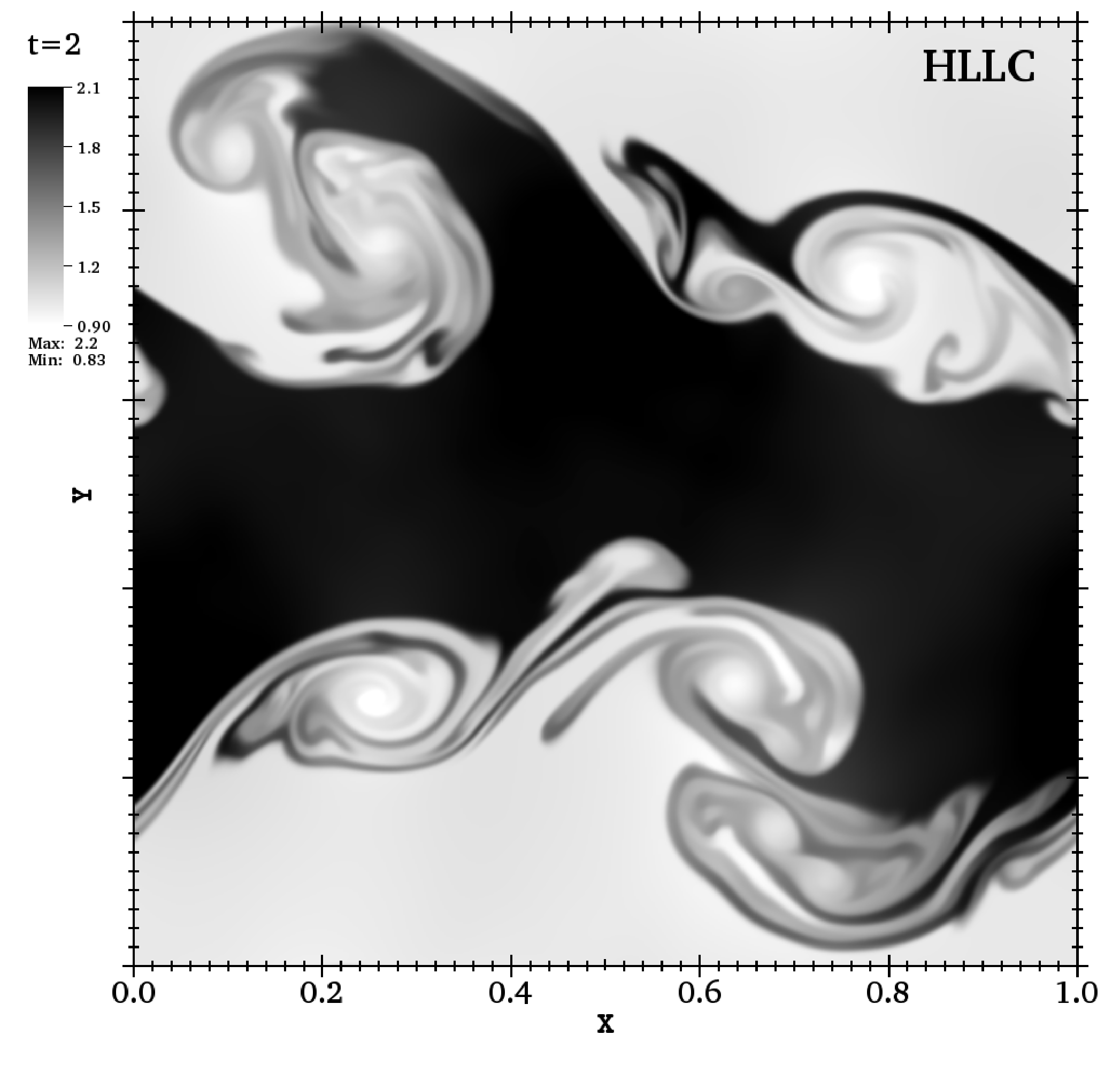}
  \caption{Same as Figure \ref{fig:kelvinHelmholtz1}, but for $t=2.0$.}
  \label{fig:kelvinHelmholtz3}
\end{figure}

\begin{figure}
  \epsscale{1.0}
  \plottwo{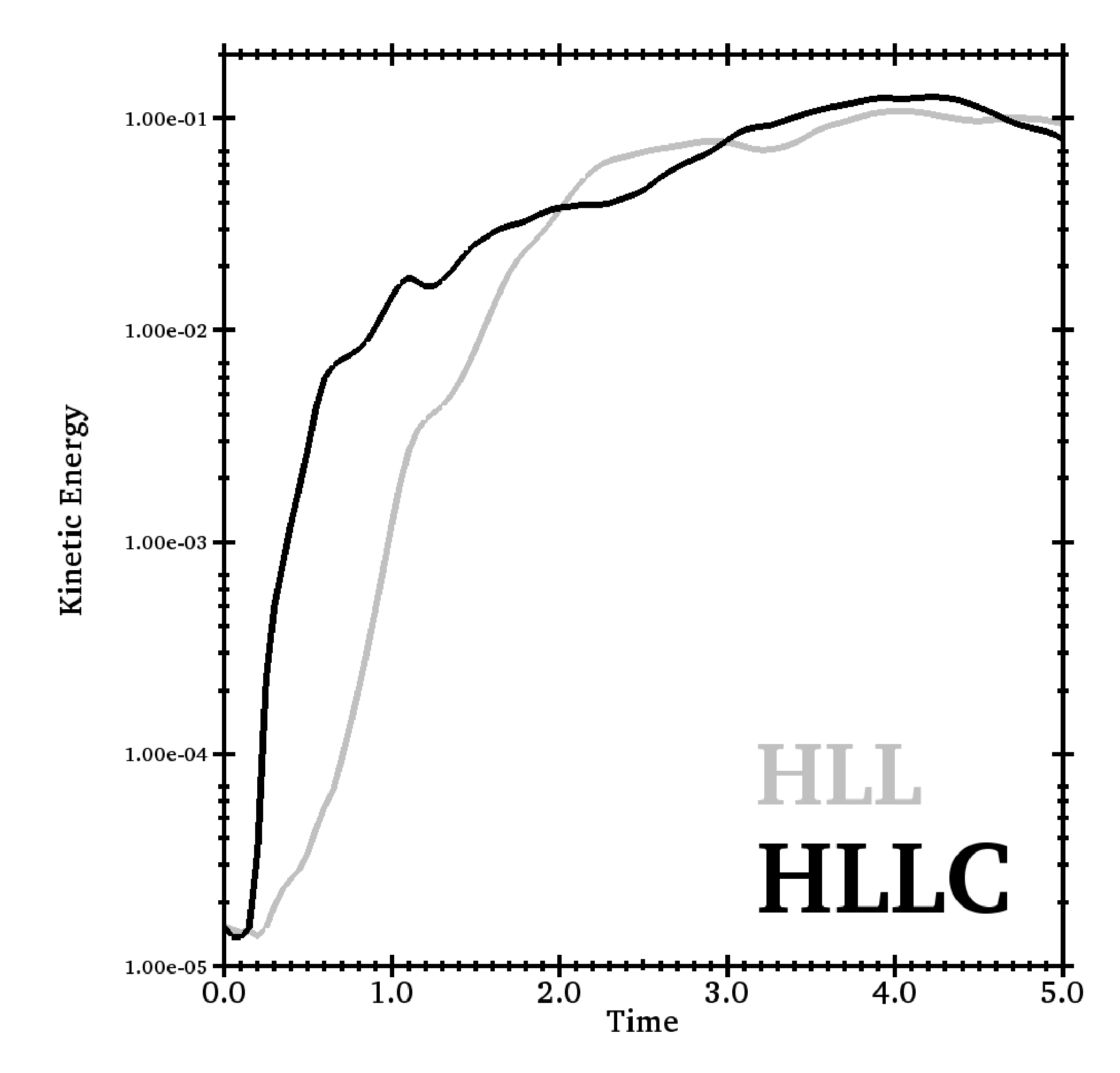}
                {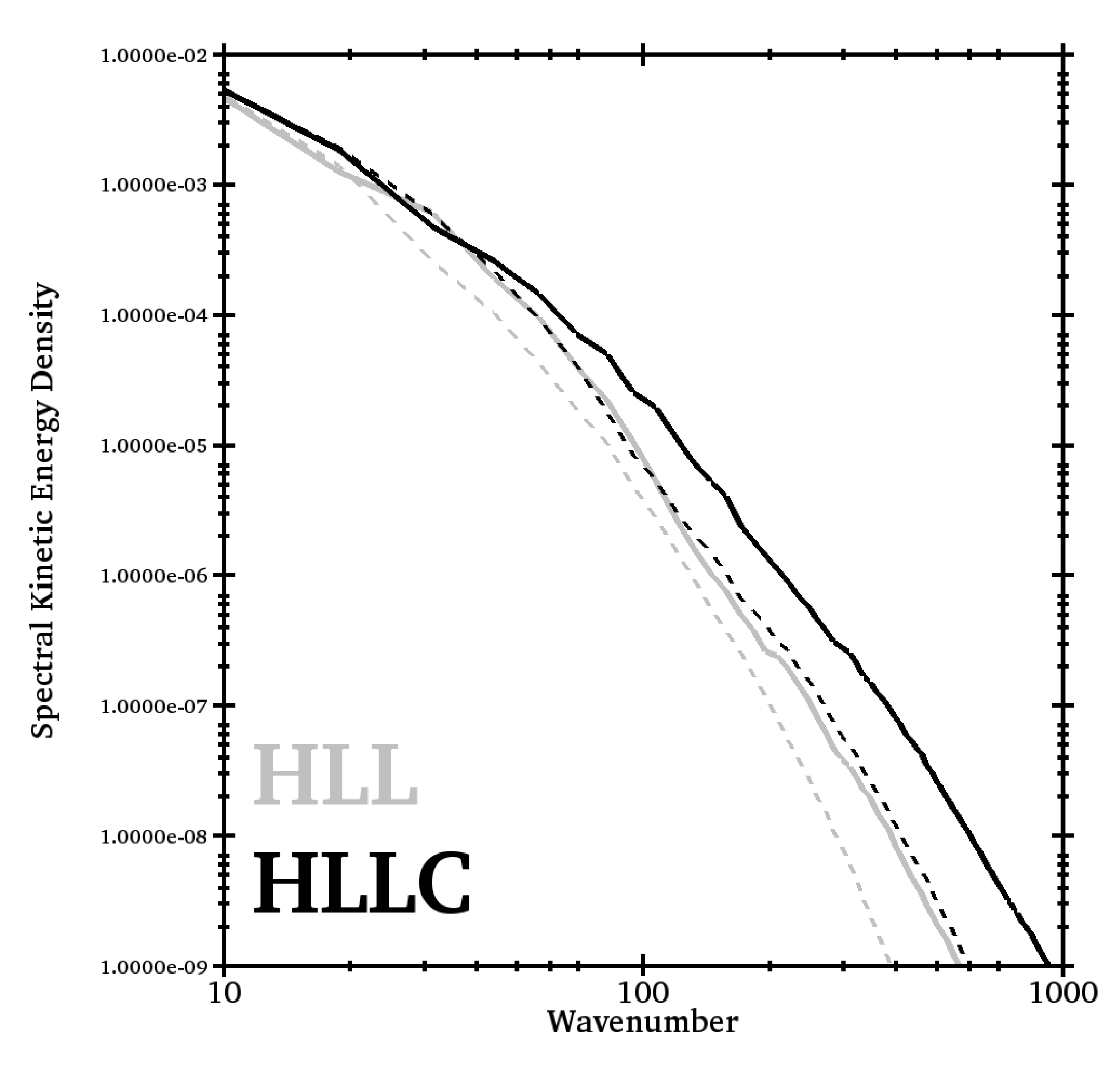}
  \caption{Some quantitative results from the Kelvin-Helmholtz instability test: $y$-component of the kinetic energy versus time (left panel) and the spectral kinetic energy density (right panel).  
  Results obtained with the HLLC and HLL Riemann solvers are represented with black and grey lines, respectively.  
  Solid lines represent results obtained with a $512\times512$ grid, while dashed lines represent results obtained with a $256\times256$ grid.  
  The energy spectra in the right panel are time-averaged over the time period extending from $t=1.9$ to $t=2.1$.}
  \label{fig:kelvinHelmholtzKineticEnergy}
\end{figure}

\begin{figure}
  \epsscale{0.95}
  \plottwo{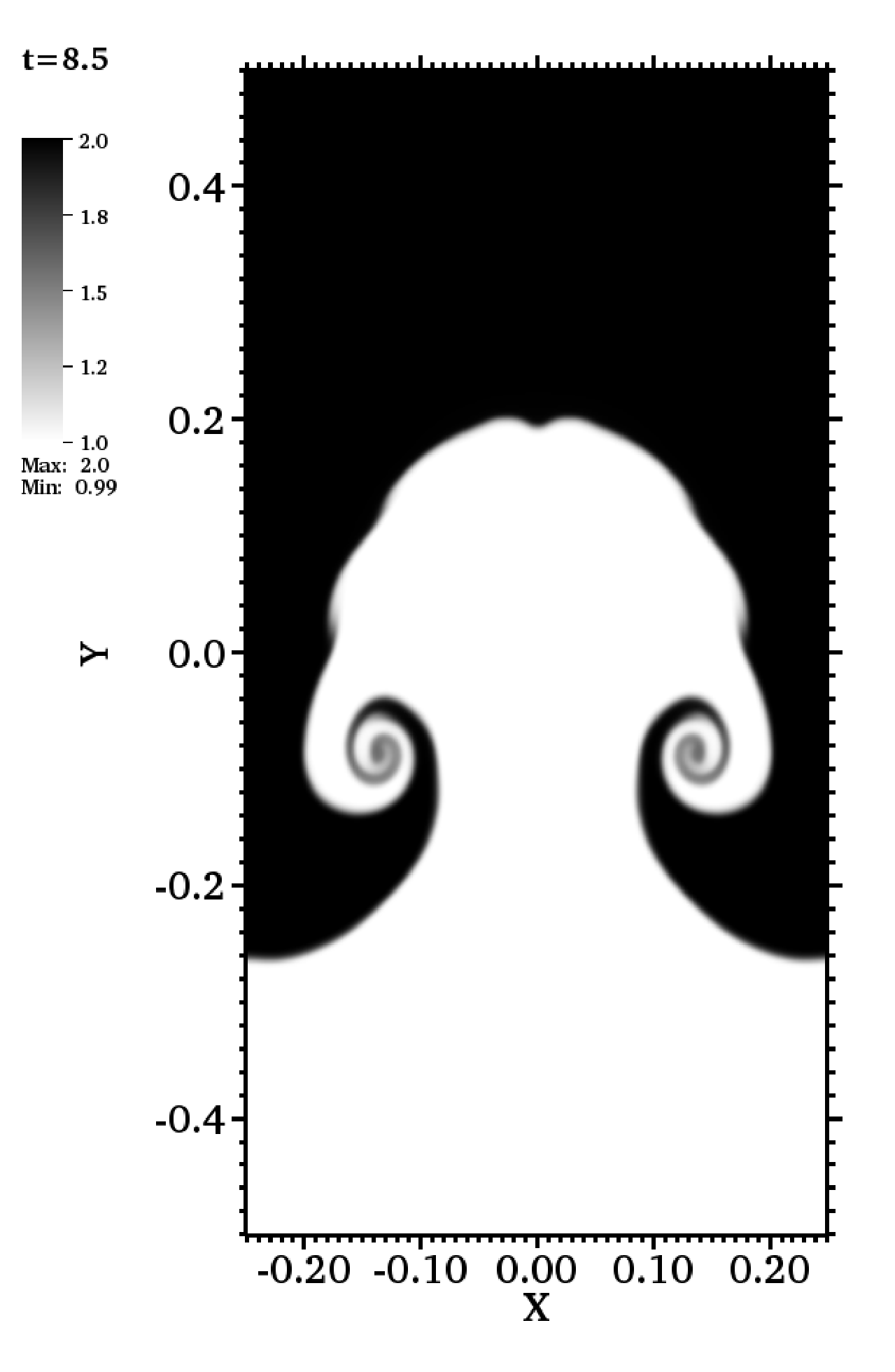}
                {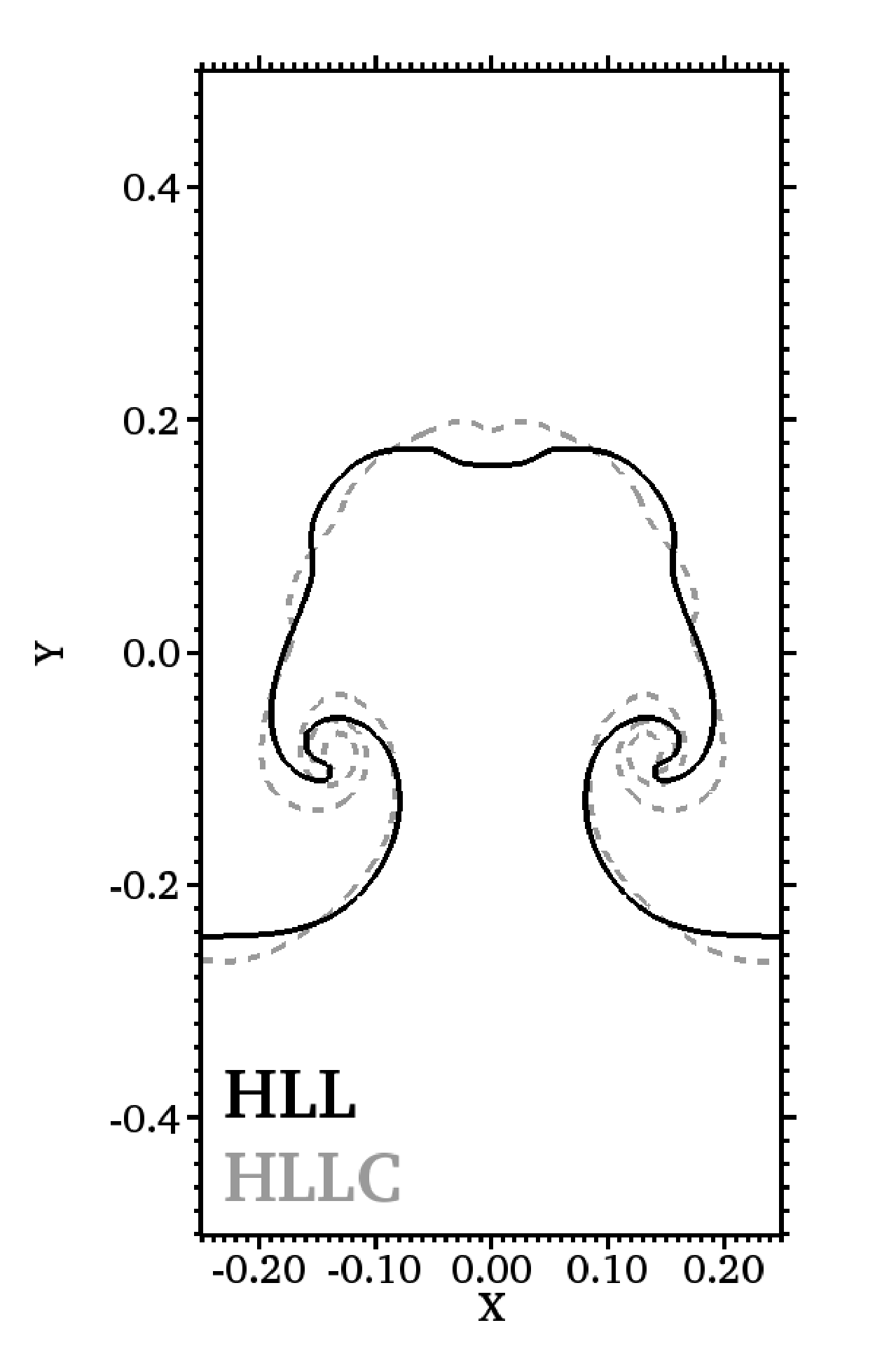}
  \caption{Plots of the density from the Rayleigh-Taylor instability test at $t=8.5$, computed with $256\times768$ zones.  
  The left panel shows results obtained the HLLC Riemann solver.  
  In the right panel we plot contours of constant density $\rho=1.25$: computations using the HLL (solid black) and HLLC (dashed grey) Riemann solvers are compared.  }
  \label{fig:rayleighTaylor}
\end{figure}

We use the hyperbolic balance equation solver on several test problems involving a nonrelativistic polytropic fluid. 
Included are smooth fluid tests that enable us to determine the order of accuracy of the hydrodynamics solvers;
discontinuous fluid tests that test the ability of GenASiS to handle shocks and other discontinuities; 
and fluid instability tests that demonstrate how two instabilities of astrophysical relevance grow and develop in GenASiS simulations.
On the whole, these tests illustrate the basic competence of the classes relevant to nonrelativistic hydrodynamics; 
demonstrate second-order convergence for problems with smooth analytic solutions, and first-order convergence for problems with discontinuous solutions;
and produce results that are comparable to and consistent with those presented by other authors.  
The superiority of the HLLC Riemann solver over the HLL Riemann solver is especially apparent in the contact discontinuity test (Figure~\ref{fig:contactWaves}) and tests of the Kelvin-Helmholtz (Figures~\ref{fig:kelvinHelmholtz1}-\ref{fig:kelvinHelmholtzKineticEnergy}) and Rayleigh-Taylor (Figure~\ref{fig:rayleighTaylor}) instabilities.
The Sedov-Taylor blast wave exercises the functionality of our explicit multi-level time stepping algorithm for the evolution of balance equations with cell-by-cell fixed-mesh refinement (Figures~\ref{fig:SedovTaylor2D_FMR}--\ref{fig:Sedov_3D_Big});
that large dynamic ranges in time, space, and pressure (or energy density) have been handled successfully with modest resources suggests that the multilevel hydrodynamics solver in GenASiS will be well suited to the hydrodynamics needs of core-collapse supernova simulations.
As we continue the development of \genasis, the test problems presented here will form the basis for a comprehensive regression test suite in Bellerophon \citep{Lingerfelt2011}, in order to ensure continued reliable performance and guard against the unintentional introduction of code bugs.  

Subsequent papers in this series will document additional functionality that will make \genasis\ suitable for multiphysics simulations of core-collapse supernovae and other astrophysical problems.  
Development of additional solvers and physics is already underway.  
A nuclear equation of state is one obvious requirement.  
We have implemented magnetohydrodynamics (MHD) capabilities in a `draft version' of \genasis\ \citep{Endeve2012b}, which has been used to study standing accretion shock instability (SASI)-driven magnetic field amplification \citep{Endeve2010,Endeve2012}.  
In the current version of \genasis\ we plan to implement MHD capabilities based on the HLLD solver \citep{Miyoshi2005,Mignone2009}.  
For Newtonian gravity, a multi-level Poisson solver will use a distributed FFT solver \citep{Budiardja2011} for the coarsest level, in conjunction with finite-difference solves on individual levels, in a multigrid approach.  
Work on general relativistic gravity---a major undertaking---is also underway  in \genasis\ \citep{Tsatsin2011}.  
The greatest challenge of all is neutrino transport.  
Building on our group's past experience \citep{Liebendorfer2004,Bruenn2009,Bruenn2013}, theoretical developments \citep{Cardall2003,Cardall2005b}, and initial forays into transport in the earliest version of \genasis\ \citep{Cardall2005,Cardall2009}, we plan to deploy a multigrid approach here as well, using a multigroup Variable Eddington Tensor formulation \citep{Cardall2013a,Endeve2012c} with closures of increasing sophistication, ultimately culminating in a full `Boltzmann solver' \citep{Cardall2013b}.  

In conjunction with additional physics capabilities---particularly radiation transport---future work will also report further progress on the goal of more fully exploiting the evolving nature of the world's leading capability supercomputers with \genasis.
We address the architectural features of distributed memory and distributed processing capacity with MPI (Message Passing Interface) in the work presented here.
But parallelism through MPI alone will become untenable as many-core architectures become more prevalent;  
additional or alternative possibilities for parallelism must be identified and implemented through `threading' (for instance, with OpenMP) in order to exploit the possibilities of many-core processing units.
Heterogeneous processing capacity, in which co-processors or accelerators provide (in principle) large flop counts with low power consumption, may also become a common feature of capability machines; 
exploitation of these architectures will require reconsideration of algorithms in light of, for example, memory and data transfer issues associated with use of such accelerators.
We expect to address many-core and accelerator architectures more directly as the capabilities for radiation transport are further developed, for it is here that memory and processing requirements will increase greatly beyond the requirements of hydrodynamics reported here.

Finally, the suitability of exascale machines for cell-by-cell AMR remains to be determined.
The cost of data movement on such machines is expected to grow faster than the cost of flops on these machines, which may be a disadvantage for cell-by-cell AMR. 
On the other hand, the amount of memory may not grow as quickly as the processing capacity, and cell-by-cell AMR may prove advantageous in memory-constrained environments.

\acknowledgments

This research was supported by the Office of Advanced Scientific Computing Research and the Office of Nuclear Physics, U.S. Department of Energy.  
This research used resources of the Oak Ridge Leadership Computing Facility at the Oak Ridge National Laboratory provided through the INCITE program.

\bibliographystyle{apj}
\bibliography{references}


\clearpage

\clearpage

\begin{table}
  \begin{center}
  \caption{$L_{1}$ error norm and convergence rate for 1D advection test with flow Mach number $\mbox{Ma}=0.6$.  \label{tab:convergenceResultsAdvection1DMach06}}
  \begin{tabular}{ccccccccc}
    \midrule
    & \multicolumn{2}{c}{HLL ($t=1$)} & 
    \multicolumn{2}{c}{HLLC ($t=1$)} &
    \multicolumn{2}{c}{HLL ($t=5$)} &
    \multicolumn{2}{c}{HLLC ($t=5$)} \\
    \cmidrule(r){2-3} \cmidrule(r){4-5} \cmidrule(r){6-7} \cmidrule(r){8-9}
    $N_{x}$ & $L_{1}(\rho)$ & Rate & $L_{1}(\rho)$ & Rate & $L_{1}(\rho)$ & Rate & $L_{1}(\rho)$ & Rate \\
    \midrule
    16   & $3.443\times10^{-2}$ &         $-$ & $2.810\times10^{-2}$ &         $-$ & $1.306\times10^{-1}$ &         $-$ & $9.729\times10^{-2}$ &         $-$ \\
    32   & $1.227\times10^{-2}$ & $-1.49$ & $1.043\times10^{-2}$ & $-1.43$ & $2.831\times10^{-2}$ & $-2.21$ & $2.653\times10^{-2}$ & $-1.87$ \\
    64   & $3.294\times10^{-3}$ & $-1.90$ & $2.770\times10^{-3}$ & $-1.91$ & $1.046\times10^{-2}$ & $-1.44$ & $9.421\times10^{-3}$ & $-1.49$ \\
    128 & $8.062\times10^{-4}$ & $-2.03$ & $6.942\times10^{-4}$ & $-2.00$ & $2.936\times10^{-3}$ & $-1.83$ & $2.717\times10^{-3}$ & $-1.79$ \\
    256 & $1.900\times10^{-4}$ & $-2.09$ & $1.667\times10^{-4}$ & $-2.06$ & $7.600\times10^{-4}$ & $-1.95$ & $7.170\times10^{-4}$ & $-1.92$ \\
    512 & $4.460\times10^{-5}$ & $-2.09$ & $4.074\times10^{-5}$ & $-2.03$ & $1.900\times10^{-4}$ & $-2.00$ & $1.827\times10^{-4}$ & $-1.97$ \\
    \midrule
  \end{tabular}
  \end{center}
\end{table}

\begin{table}
  \begin{center}
  \caption{$L_{1}$ error norm and convergence rate for 1D advection test with flow Mach number $\mbox{Ma}=0.1$.  \label{tab:convergenceResultsAdvection1DMach01}}
    \begin{tabular}{ccccccccc}
      \midrule
      & \multicolumn{2}{c}{HLL ($t=1$)} & 
      \multicolumn{2}{c}{HLLC ($t=1$)} &
      \multicolumn{2}{c}{HLL ($t=5$)} &
      \multicolumn{2}{c}{HLLC ($t=5$)} \\
      \cmidrule(r){2-3} \cmidrule(r){4-5} \cmidrule(r){6-7} \cmidrule(r){8-9}
      $N_{x}$ & $L_{1}(\rho)$ & Rate & $L_{1}(\rho)$ & Rate & $L_{1}(\rho)$ & Rate & $L_{1}(\rho)$ & Rate \\
      \midrule
      16   & $1.377\times10^{-1}$ &         $-$ & $2.777\times10^{-2}$ &         $-$ & $1.995\times10^{-1}$ &         $-$ & $9.616\times10^{-2}$ &         $-$ \\
      32   & $2.591\times10^{-2}$ & $-2.41$ & $1.026\times10^{-2}$ & $-1.44$ & $1.044\times10^{-1}$ & $-0.93$ & $2.621\times10^{-2}$ & $-1.88$ \\
      64   & $8.095\times10^{-3}$ & $-1.68$ & $2.709\times10^{-3}$ & $-1.92$ & $1.779\times10^{-2}$ & $-2.55$ & $9.176\times10^{-3}$ & $-1.51$ \\
      128 & $1.805\times10^{-3}$ & $-2.17$ & $6.758\times10^{-4}$ & $-2.00$ & $4.769\times10^{-3}$ & $-1.90$ & $2.628\times10^{-3}$ & $-1.80$ \\
      256 & $3.894\times10^{-4}$ & $-2.21$ & $1.616\times10^{-4}$ & $-2.06$ & $1.119\times10^{-3}$ & $-2.09$ & $6.893\times10^{-4}$ & $-1.93$ \\
      512 & $8.117\times10^{-5}$ & $-2.26$ & $3.804\times10^{-5}$ & $-2.09$ & $2.525\times10^{-4}$ & $-2.15$ & $1.750\times10^{-4}$ & $-1.98$ \\
      \midrule
    \end{tabular}
  \end{center}
\end{table}

\begin{table}
  \begin{center}
  \caption{$L_{1}$ error norm and convergence rate for 2D advection test with flow Mach number $\mbox{Ma}=0.6$.  \label{tab:convergenceResultsAdvection2DMach06}}
    \begin{tabular}{ccccccccc}
      \midrule
      & \multicolumn{2}{c}{HLL ($t=1$)} & 
      \multicolumn{2}{c}{HLLC ($t=1$)} &
      \multicolumn{2}{c}{HLL ($t=5$)} &
      \multicolumn{2}{c}{HLLC ($t=5$)} \\
      \cmidrule(r){2-3} \cmidrule(r){4-5} \cmidrule(r){6-7} \cmidrule(r){8-9}
      $N_{x}$ & $L_{1}(\rho)$ & Rate & $L_{1}(\rho)$ & Rate & $L_{1}(\rho)$ & Rate & $L_{1}(\rho)$ & Rate \\
      \midrule
      32   & $1.019\times10^{-2}$ &         $-$ & $8.028\times10^{-3}$ &         $-$ & $3.944\times10^{-2}$ &         $-$ & $2.722\times10^{-2}$ &         $-$ \\
      64   & $3.577\times10^{-3}$ & $-1.51$ & $2.897\times10^{-3}$ & $-1.47$ & $8.376\times10^{-3}$ & $-2.24$ & $7.892\times10^{-3}$ & $-1.79$ \\
      128 & $9.485\times10^{-4}$ & $-1.92$ & $7.816\times10^{-4}$ & $-1.89$ & $3.014\times10^{-3}$ & $-1.47$ & $2.698\times10^{-3}$ & $-1.55$ \\
      256 & $2.311\times10^{-4}$ & $-2.04$ & $1.933\times10^{-4}$ & $-2.02$ & $8.419\times10^{-4}$ & $-1.84$ & $7.715\times10^{-4}$ & $-1.81$ \\
      512 & $5.421\times10^{-5}$ & $-2.09$ & $4.688\times10^{-5}$ & $-2.04$ & $2.167\times10^{-4}$ & $-1.96$ & $2.026\times10^{-4}$ & $-1.93$ \\
      \midrule
    \end{tabular}
  \end{center}
\end{table}

\clearpage

\begin{table}[ht]
\centering
\caption{$L_1$ error norm and convergence rate for Sod's shocktube problem.  \label{table:SodShockTube_1D_Convergence}}
\begin{tabular}{ c c c c c c c c c }
\hline
        & \multicolumn{4}{c}{HLL}                                               & \multicolumn{4}{c}{HLLC}                                              \\ \cmidrule(r){2-5} \cmidrule(r){6-9}
$N_x$   & $L_1 (\rho)$           & Rate     & $L_1 (p)$              & Rate     & $L_1 (\rho)$           & Rate     & $L_1 (p)$              & Rate     \\ \hline
$32$    & $2.982 \times 10^{-2}$ & $-$	    & $2.567 \times 10^{-2}$ & $-$      & $2.809 \times 10^{-2}$ & $-$      & $2.553 \times 10^{-2}$ & $-$      \\
$64$    & $1.529 \times 10^{-2}$ & $-0.96$ & $1.293 \times 10^{-2}$ & $-0.99$ & $1.452 \times 10^{-2}$ & $-0.95$ & $1.269 \times 10^{-2}$ & $-1.01$ \\
$128$   & $8.098 \times 10^{-3}$ & $-0.92$ & $6.500 \times 10^{-3}$ & $-0.99$ & $7.803 \times 10^{-3}$ & $-0.90$ & $6.388 \times 10^{-3}$ & $-0.99$ \\
$256$   & $4.383 \times 10^{-3}$ & $-0.89$ & $3.289 \times 10^{-3}$ & $-0.98$ & $4.256 \times 10^{-3}$ & $-0.88$ & $3.233 \times 10^{-3}$ & $-0.98$ \\
$512$   & $2.434 \times 10^{-3}$ & $-0.85$ & $1.692 \times 10^{-3}$ & $-0.96$ & $2.362 \times 10^{-3}$ & $-0.85$ & $1.658 \times 10^{-3}$ & $-0.96$ \\
$1024$  & $1.311 \times 10^{-3}$ & $-0.89$ & $7.911 \times 10^{-4}$ & $-1.10$ & $1.279 \times 10^{-3}$ & $-0.89$ & $7.748 \times 10^{-4}$ & $-1.10$ \\
$2048$  & $7.495 \times 10^{-4}$ & $-0.81$ & $4.095 \times 10^{-4}$ & $-0.95$ & $7.322 \times 10^{-4}$ & $-0.81$ & $4.013 \times 10^{-4}$ & $-0.95$ \\ \hline
\end{tabular}
\end{table}

\clearpage

\clearpage

\clearpage

\clearpage

\end{document}